\documentclass[3p, times]{elsarticle}
\usepackage[cmex10]{amsmath}
\usepackage[normalem]{ulem} 
\usepackage[mathscr]{euscript}
\usepackage[utf8]{inputenc}
\usepackage{%
	algorithm,%
	algorithmic,%
	amssymb,%
	amsthm,%
	array,%
	booktabs,%
	bbm,%
	dsfont,%
	graphicx,%
	multirow,%
	paralist,%
	pgf,%
	pgfgantt,%
	pgfplots,%
	siunitx,%
	subcaption,%
	subfiles,%
	tabulary,%
	tikz,
	tikzscale,%
	url,%
	xcolor,%
}
\usepackage{scalerel,stackengine}

\graphicspath{{Figures/}{../Figures/}}

\usepgflibrary{shapes}
\usetikzlibrary{%
	arrows,%
	automata,%
	calc,%
	chains,%
	patterns,%
	positioning,%
	shapes,%
	shapes.callouts,%
	shapes.multipart,%
	shapes.symbols,%
}

\renewcommand{\le}{\leqslant}
\renewcommand{\ge}{\geqslant}

\newcommand{\EQ}[1]{\begin{equation*}#1\end{equation*}}
\newcommand{\EQN}[1]{\begin{equation}#1\end{equation}}
\newcommand{\eq}[1]{\begin{align*}#1\end{align*}}
\newcommand{\eqn}[1]{\begin{align}#1\end{align}}

\newcommand{\set}[1]{\left\{#1\right\}}

\newcommand{\SetIn}[1]{\mathbbm{1}_{\set{#1}}}
\newcommand{\abs}[1]{\left\lvert #1\right\rvert}

\newcommand{\given}{\Big\lvert}

\newcommand{\iid}{\emph{i.i.d.~}}

\newcommand{\E}{\mathbb{E}}

\newcommand{\N}{\mathbb{N}}
\newcommand{\R}{\mathbb{R}}
\newcommand{\Z}{\mathbb{Z}}

\newcommand{\bF}{\bar{F}}
\newcommand{\bG}{\bar{G}}
\newcommand{\bH}{\bar{H}}
\newcommand{\bl}{\bar{\lambda}}
\newcommand{\bxi}{\bar{\xi}}

\newcommand{\cH}{\mathcal{H}}

\newcommand{\cB}{\mathcal{B}}

\newcommand{\expect}[1]{\E\left[#1\right]}

\newcommand{\nix}[1]{}

\theoremstyle{plain}
\newtheorem{theorem}{Theorem}
\newtheorem{corollary}[theorem]{Corollary}
\newtheorem{lemma}[theorem]{Lemma}
\newtheorem{proposition}[theorem]{Proposition}

\theoremstyle{definition}
\newtheorem{definition}[theorem]{Definition}

 \newtheorem{innercustomgeneric}{\customgenericname}
\providecommand{\customgenericname}{}
\newcommand{\newcustomtheorem}[2]{%
  \newenvironment{#1}[1]
  {%
   \renewcommand\customgenericname{#2}%
   \renewcommand\theinnercustomgeneric{##1}%
   \innercustomgeneric
  }
  {\endinnercustomgeneric}
}

\newcustomtheorem{conj2}{Conjecture}
\newcustomtheorem{exmp}{Example}

\theoremstyle{remark}
\newtheorem{remark}{Remark}

\pgfplotsset{compat=1.5}

\begin{document}
\begin{frontmatter}

\title{Load balancing policies without feedback using timed replicas}

\author[label1]{Rooji~Jinan\corref{cor1}}
\ead{roojijinan@iisc.ac.in} 
\author[label2]{Ajay~Badita}
\ead{ajaybadita@iisc.ac.in} 
\author[label3]{Tejas~Bodas}
\ead{tejas.bodas@iiit.ac.in}
\author[label2]{Parimal~Parag}
\ead{parimal@iisc.ac.in}

\cortext[cor1]{Corresponding author}

\address[label1]{Department of cyber-physical systems, IISc, Bangalore, KA 560012, India.}
\address[label2]{Department of electrical communication engineering, IISc, Bangalore, KA 560012, India.}
\address[label3]{Computer systems group at IIIT Hyderabad, TS 500032, India.}

\begin{abstract}
Dispatching policies such as the join shortest queue (JSQ), join smallest work (JSW) and their power of two variants are used in load balancing systems where the instantaneous queue length or workload information at all queues or a subset of them can be queried. 
In situations where the dispatcher has an associated memory, 
one can minimize this query overhead by maintaining a list of idle servers to which jobs can be dispatched.  
Recent alternative approaches that do not require querying such information include the cancel on start and cancel on complete based replication policies. 
The downside of such policies however is that the servers must communicate the start or completion of each service to the dispatcher and must allow cancellation of redundant copies. 
In practice, the requirements of query messaging, memory, and replica cancellation pose challenges in their implementation and their advantages are not clear. 
In this work, we consider load balancing policies that do not query load information, 
do not have a memory, and do not cancel replicas. 
Surprisingly, we were able to identify operating regimes where such policies have better performance when compared to some of the popular policies that utilize server feedback information.
Our policies allow the dispatcher to append a timer to each job or its replica. 
A job or a replica is discarded if its timer expires before it starts receiving service. 
We analyze several variants of this policy which are novel, simple to implement, and also have remarkably good performance in some operating regimes, 
despite no feedback from servers to the dispatcher. 
\end{abstract}


\end{frontmatter}

\section{Introduction}
\label{sec:Intro}
Load balancing policies play a vital role in latency reduction in distributed systems such as large data centers and cloud computing. 
A typical load balancing system comprises of a large number of homogeneous servers and a dispatcher that routes arriving jobs to the queue of these servers. 
When the instantaneous queue length of different servers is known, 
an obvious approach would be to use the join-shortest-queue (JSQ) policy~\cite{Winston77}. 
If instead of queue length, the workload i.e., the pending amount of work at each server is known, 
the optimal policy is the join smallest work queue (JSW). 
Unfortunately, in most practical systems, 
the number of servers is large and therefore obtaining the instantaneous queue lengths or workloads from all servers is difficult. 
\par
A popular remedy for this is to consider the power of $d$ choice variant of JSQ and JSW. 
In a JSQ($d$) policy, 
the dispatcher samples $d$ servers uniformly at random and queries their queue lengths. 
The job is then routed to a sampled server with the least number of waiting jobs. 
Implementing such a policy requires $2d$ messages per job 
and was shown to have very good performance characteristics~\cite{Mitzenmacher01,Vvedenskaya96}. 
The equivalent workload based policy JSW($d$) also has a $2d$ query overhead per job 
and was analyzed recently~\cite{Hellemans18, Ayesta19}.   
For many systems, a $2d$ query exchange is a considerable overhead, 
especially when $d$ is large or when the timescale for message exchange is comparable to the actual service requirement of a job.

Recent efforts have therefore been directed towards bringing down this overhead using smart feedback techniques~\cite{Van19,Van20}. 
The authors of~\cite{Van19} consider a hyper-scalable dispatching scheme where the dispatcher maintains queue length estimates for the different queues and sends an arriving job to the server with the least estimated queue length.
Each server occasionally updates the dispatcher about its true queue length and this enables the dispatcher to synchronize its estimates with reality.  
The authors of~\cite{Van20} introduce the join-open-queue scheme where servers send busy alerts to the dispatcher at predetermined times. When a server is idle, it does not send the alert and thus the dispatcher can infer idle servers without considerable message exchanges.
In such cases, there is some feedback communicated by the servers to the dispatcher, and this can be non negligible in some settings.  
Furthermore, the dispatcher operates under noisy queue/workload information, that affects the system performance. 
It is well known that for correlated processes, there is a tradeoff between the estimated accuracy and the frequency of updates~\cite{Jinan2020MDPI}. 
Another policy that works under sparse communication and approximate state information can be found at \cite{Mendelson2022Arxiv}.

The feedback communication overhead and the noisiness of estimates get exacerbated for the case of multiple dispatchers, 
which is common for modern data centers comprising of a huge number of servers.  
A load balancing system with multiple dispatchers is analyzed in~\cite{Vargaftik20,Zhou2021PER}, 
where the authors consider policies that require infrequent communication between servers and dispatchers. 
In these policies, the dispatchers perform load balancing based on a local estimate of the queue length. 
It is observed that in such systems, jobs could be concurrently dispatched by different dispatchers to the same server which might drive the system to instability.

An alternative low feedback policy that uses memory, is join the idle queue (JIQ) policy. 
In this policy, idle queues willingly inform the dispatcher about their idleness and the dispatcher lists this in an associated memory. 
This policy records accurate information of idleness of all queues, and has very good performance characteristics~\cite{Lu11}.
An arriving job is sent to an idle queue selected randomly from the list if it is non-empty and therefore this policy has an overhead of a single feedback message in each busy period per server.  
Some recent load balancing policies that make use of memory in their dispatching decisions appear in~\cite{Gamarnik18, Hellemans20}. 

An alternative way to achieve good performance without querying instantaneous queue length or workload information is to use redundancy based load balancing policies.
Two popular variants of redundancy-$d$ based load balancing are cancel on start (c.o.s.)~\cite{Ayesta18} 
and cancel on complete (c.o.c.)~\cite{Gardner17}. 
In these policies, independent replicas of an arriving job are sent to $d$ randomly chosen servers. 
In c.o.s. (resp. c.o.c.), when one of the copy starts receiving service (resp. completes service), 
the $d-1$ replicas are canceled. 
Such policies also have superior delay performance and are quite amenable to analysis. 
A detailed product form analysis characterizing the delay performance for both variants is presented in~\cite{Gardner2020QS,Ayesta2021OR}.  
However, a major implementation problem with replication based policies is the synchronized cancellation of the redundant replicas. 
The sophistication required for implementing such an approach in fact may even be non-trivial. 
Further, depending on the operating scenario, 
instantaneous cancellation may not always be feasible, 
thereby adding an overhead on the system
~\cite{Lee17b, Hellemans19}.
In many redundancy based practical systems, 
replica cancellation is an undesirable overhead that is often avoided. 
The authors of~\cite{Qiu2017TDS, Perez2018CC} discuss applications where delay due to request cancellations cannot be tolerated and describe systems where it is difficult to incorporate a functionality to terminate requests while being executed.
The authors of~\cite{Vulimiri2012ACM, Anantha2013USENIX, Primorac2021USENIX} study systems where replication is implemented without cancellation (r.w.c.). 
In particular, the authors of~\cite{Vulimiri2012ACM} suggest that there is a threshold system load above which replication can be detrimental and cannot offer any improvement in terms of latency. 
This provides the motivation for designing a policy that would intelligently replicate only when the workload conditions are favorable.
The idea of replication without cancellation has also been used in multipath routing in networks~\cite{Liu2018TSC, Maxemchuk1975ICC}.

Besides replication based policies for latency reduction, latency in distributed storage systems with MDS coded data have been widely studied in literature~\cite{Joshi2014JSAC, Joshi2017ACM,Lee2017TIT,Badita2019TIT,Bodas2021ISIT}. 
Although they have superior delay performance, such schemes have additional decoding costs and scalability issues besides the cancellation costs.
Also, there are efficient replication based strategies that have competitive performance with that of MDS coded systems~\cite{Jinan2021ISIT,Jinan2021TIT} and we do not discuss them in this article.
In addition, load balancing policies which consider different cost functions like throughput~\cite{Joshi2021TNET}, server utilization~\cite{Badita2020INFOCOM, Badita2020TNET} etc. have also been studied. 
We do not delve in to these details.


Note that except for the r.w.c. policy, the load balancing policies discussed earlier either involve (a) communication of messages, 
or (b) require a memory, or (c) require replication with cancellation. 
Such policies therefore always have an element of feedback from the server to the dispatcher. 
In this work, we aim to characterize the impact of such communication/memory/coordinated replica-cancellation in load balancing by comparing their performance with policies that do not make use of server feedback information. 
We focus on static load balancing policies that do not demand queue length information or memory at the dispatcher. 
Static load balancing approach is very similar to the forward error correction~\cite[Chapter 1]{Shu2011ECC} in communication, where the message redundancy is designed in advance without any receiver feedback. 
Analogously, the traditional server feedback based load balancing approaches are similar to adaptive coding policies~\cite[Chapter 22]{Shu2011ECC}.

While we allow the dispatcher to possibly replicate jobs to $d$ different servers, 
we assume that the dispatcher does not have any server state information and does not send any state dependent cancellation message. 
This is in line with some of the practical policies discussed in the preceding paragraph. 
While the random routing policy $(d=1)$ is an obvious choice for such static load balancing, its performance is known to be poor and is therefore of limited interest. 
Replicate without cancellation (r.w.c.) is an alternative candidate, but the impact on the system load due to uncanceled replicas is not clear. 
At this point, an imminent question is can we add some functionality to the system (without incurring much overhead) and achieve much better performance as compared to random routing  or r.w.c.-like policy of~\cite{Vulimiri2012ACM, Anantha2013USENIX, Primorac2021USENIX}? 
Further, would it be possible to have comparable or even better performance as compared to policies like JSQ($d$) or JSW($d$) that make use of the server state information? 
An affirmative answer to the latter question even under say a restricted parameter setting, 
may go a long way in establishing the true value of feedback information in load balancing.


In this paper, we propose a policy where the dispatcher has the ability to append a \emph{server-side cancellation criteria} to each job or its replica. Before picking any job or its replica for service, 
each server will check if the appended criteria is satisfied or not. 
If the criteria is met, then the replica is served or else it is dropped. 
We consider a criteria that depends on the waiting time of the replica in a queue. 
For example, the criteria that we consider is to serve the replica only if it has waited in the queue no more than a fixed preset amount of time.
More formally, we assume for each arriving job referred to as the primary replica, 
the dispatcher creates $d-1$ secondary replicas. 
The servers where the replicas are sent are chosen randomly. 
Associated with the primary and secondary replicas are non-negative \emph{discard thresholds} $T_1$ and $T_2$. 
A replica is discarded by the server if the waiting time experienced by the replica is more than its discard threshold and we label our load balancing policy by $\pi(d,T_1,T_2)$.
Such a criteria is easy for the server to validate, 
and can be achieved by logging the arrival time information of each job/replica.
Furthermore, our policy can even be implemented in a multiple dispatcher setting without incurring delay overhead and can be designed to not cause an instability. 
The key essence of our approach is to exploit possible gains from replication of jobs, 
but at the same time prevent overloading the system due to extra replicas by preemptively performing server-side cancellation of potentially wasteful replicas.
We compare this policy against policies with access to side information on the status of the system and show that the proposed load balancing policy in certain regimes provides a superior latency performance as compared to most of the popular policies.
This begs two important questions:
\begin{compactenum}
\item Are the load balancing policies with server feedback information utilizing the available information optimally? 
\item Can a load balancing policy without server feedback perform as well as the ones with server feedback in certain operating regimes?
\end{compactenum}
One needs to answer these two questions to know whether the performance gains derived from using server feedback information are worth the cost structure imposed by such information gathering in the system. 
We note that  answering the first question will need a more thorough study on the characterization of value of information in such systems and can be an independent study of its own.
However, our work provides the evidence to show that the utilization of the server feedback information is sub-optimal in many of the prominent load balancing policies with information feedback. 
Ideally if the information utilization is optimal, then the load balancing policies with extra feedback information is never supposed to perform worse than policies without feedback with the same amount of redundancy and under similar system settings.
In this work, we have been successful in designing a policy without server feedback which is shown to outperform the policies with feedback in certain load regimes. 

We observe that when $T_1$ and $T_2$ are both finite, 
arriving jobs could potentially be lost without service. 
Keeping this in mind, the two key performance metrics that we consider are the conditional mean response time of jobs admitted into the system and the loss probability of an arriving job. Note that for systems where loss cannot be tolerated, we can set $T_1 = \infty$ and adapt suitably.
To analyze our policy, we make use of the cavity process method of~\cite{Bramson10, Bramson12} along with an assumption on the asymptotic independence of the stationary workloads at the different queues as the number of servers $N \to \infty$. While we prove that the queues are asymptotically independent over any finite time horizon, the absence of monotonicity arguments makes it difficult to extend this result to time-stationary regimes when thresholds $T_1, T_2$ are finite. 
When both the discard thresholds are infinite, the workload monotonicity of queues continues to hold, and the asymptotic independence for stationary workloads is easy to prove. 
Note that asymptotic independence is difficult to prove in general, and proofs are available only under specific service disciplines, load balancing policies, and assumptions on service distributions~\cite{Bramson10, Bramson12, Shneer2021QS}. 
Having said that, the use of this assumption as a conjecture is widespread~\cite{Vasantam2019Thesis, Bramson10, Hellemans19} and supported by extensive numerical evidence.

\subsection{Contributions}
\label{subsec:contributions}
We have listed our key contributions below. 
\begin{compactenum}  
\item We propose a distributed load balancing policy $\pi(d,T_1,T_2)$, where the dispatcher needs no feedback from the servers. 
Further, replicas are discarded at a server if the waiting time exceeds the discard threshold.

\item 
We show that the workloads in the various queues in the system are asymptotically independent over a finite time horizon.
We empirically verify that the independence assumption on the limiting marginal workload distribution is a good approximation even for a finite number of servers.

\item We obtain expressions for loss probability in Lemma~\ref{lem:LossProb} and conditional mean response time of admitted jobs in Theorem~\ref{thm:MeanResponseTime} in terms of the loss probability, the limiting marginal workload distribution, and the service time distribution.

\item We obtain the moment generating function (MGF) for limiting workload of an arbitrary queue under the policy $\pi(d,T_1, T_2)$. 
We invert this function for the exponential service time distribution, 
to obtain the limiting workload distribution in Corollary~\ref{cor:GeneralCase}. 

\item We provide some design guidelines on choice of number of replicas $d$, and the corresponding cancellation thresholds $T_1, T_2$ for the proposed policy. 

\item We analytically show in Lemma~\ref{lem:RRProposedComparison} that the proposed policy $\pi(d, \infty, 0)$ always outperforms the random routing policy under exponential service times. 
We also provide an analytical comparison it 
with c.o.c.($d$) policy when service times are exponential in Proposition~\ref{pro:cocProposedComparison}.

\item We conduct numerical experiments to show that the $\pi(d, \infty, 0)$ policy can outperform the replication-$d$
c.o.s., JSQ($d$), and JIQ($d$), in a low arrival rate regime. 
This policy converges to replication-1 c.o.s. policy in the high arrival rate regime.  
We observe that the arrival rate threshold for this regime switch increases with redundancy $d$.

\item We also provide the performance comparison of our policy with other server feedback based policies for general service time distributions and observe similar performance improvement as seen under exponential service times.
\end{compactenum}

\subsection{Organization}
We introduce the system model and notations in Section~\ref{sec:SystemModel}. 
This is followed by a discussion on the cavity process method and its application to our problem along with the discussion on the asymptotic independence of the workloads at different queues. 
In Section~\ref{sec:Analysis}, 
we compute the performance metrics for the proposed policy $\pi(d,T_1, T_2)$ for a general service time distribution, 
in terms of the limiting marginal workload distribution.  
In Section~\ref{sec:Exponential}, we find the closed-form expression for marginal workload distribution when the service time distribution is exponential. 
We also compute the conditional mean of response time for admitted jobs, for some special cases of $\pi(d,T_1,T_2)$ policy. 
We provide comparison of our policy with policies with feedback for various service time distributions in Section~\ref{sec:NumericalComparison}.
We conclude with a summary of our work and future directions in Section~\ref{sec:discussion}.
\section{System Model and Preliminaries }
\label{sec:SystemModel}

We consider a load balancing system with $N$ servers, 
where jobs arrive according to a Poisson process of rate $\lambda N$. 
There is a dispatcher associated with this system whose objective is to minimize the response time experienced by each job by suitably balancing the workload across different servers. 
Owing to the popularity of redundancy based load balancing policies, 
we assume that the dispatcher has the ability to replicate an arriving job across multiple servers.

Throughout this article, we denote 
the set of first $n$ consecutive positive integers as $[n] \triangleq \set{1, \dots, n}$, 
the set of non-negative integers as $\Z_+$, 
the set of positive integers as $\N$, 
the set of non-negative reals as $\R_+$
and the set of positive reals as $\R^+$.
We also use the notation $x \wedge y \triangleq \min\set{x,y}$.

\subsection{Service}
We denote the service time for $n$th arriving job at $i$th server by $X_{n,i} \in \R_+$.  
We assume that the job service time sequence $(X_{n,i} \in \R_+: n \in \N, i \in [N])$ is random and independent and identically distributed (\emph{i.i.d.}) with the common distribution $G: \R_+ \to [0,1]$ and the common mean $\frac{1}{\mu}$. 

A more generalized service model is the S\&X model ~\cite{Gardner2017TON} where the service time of $n$th job at server $i$ is defined to be the product random variable $Y_{n,i} \triangleq S_i \cdot X_n$ where $S_i$ is the slowdown factor at the server $i$ and $X_n$ denotes the random size of the incoming job $n$. 
The random slowdown factor $S_i$ is assumed to have a mean greater than or equal to $1$. 
The random job size sequence is assumed to be \iid across the jobs.
Owing to the difficulty in the analysis posed by this model (also noted in~\cite{Gardner2017TON, Hellemans19}), we focus on
\iid service time model in this work.
%
This can also be considered as a special case of the S\&X model where the slow down factor $S$ is \iid exponential with unit rate across servers
and the service times of job $X_i$ have a constant size of $\frac{1}{\mu}$.
Another interesting special case is when the slowdown factor is deterministic and the job sizes are \iid. 
This special case has been discussed in detail in ~\ref{App:DeterministicSlowdown}.

In the main results discussed in the paper, we assume that the service time for each replica of the job is  \iid according to the same distribution $G$. 
Even if we consider all servers to be identical in terms of configuration and compute power, there could be
some uncertainties in the time taken to service a job at any server due to other background processes~\cite{Cheng2014IMC,Koole2008JS}.
The randomness assumption accomodates these uncertainties.
Further, we also assume the service times to be exponentially distributed.
Recent studies suggests that the service times in distributed computing systems can be modelled to have two components; a constant startup delay and a random memoryless component~\cite{Lee2018TIT,Bitar2017ISIT,Abbasi2018TNET}.
Although, it is the shifted exponential model that best fits this profile, whenever the startup time is negligibe the service time distribution can be approximated by an exponential distribution.
This along with analytical tractability motivates us to assume that the service times follow \iid exponential distribution with rate $\mu$.
We denote the tail distribution of the service time or the complementary service time distribution by $\bar{G} \triangleq 1 - G$. 
When we focus on a single queue~$i$, we will drop the subscript $i$ for brevity. 

\subsection{Threshold based cancellation}
\label{subsection:ThresholdCancel}

We assume that the dispatcher has limited functionality and that it cannot cancel redundant copies when one of the replica has received (or starts receiving) service. 
Instead, we assume that the dispatcher can append \emph{discard instruction} along with each replica.  
Before selecting a job/replica for service, 
each server will read the \emph{discard instruction} and possibly discard the replica based on the instruction. 
We call this as a redundancy based approach with server side cancellation of replicas. 
For ease of exposition, 
we assume that the \emph{instruction} is almost identical for all copies in the system and hence the overhead of implementing this approach is minimal. 
In this article, we restrict to \emph{instructions} that are characterized by a threshold $T \in [0, \infty)$.
To elaborate, we assume that the server serves a replica if it is chosen for service within $T$ units of its arrival or else discards the replica. 
We call $T$ as the \emph{discard threshold} for brevity.
\subsubsection{Primary replica and discard threshold}
We consider the following dispatching policy based on the above idea of a \emph{discard threshold}. 
When a job arrives, the dispatcher samples a single \emph{primary} server uniformly at random and sends a primary replica of the job to the server along with the \emph{primary discard threshold} $T_1$. 

\subsubsection{Secondary replicas and discard thresholds} 
For each job arrival, the dispatcher creates $d-1$ secondary replicas.
Then, it samples $d-1$ other servers uniformly at random and sends \iid replicas of the same job to the sampled $d-1$ servers after appending each replica with a \emph{secondary discard threshold} of $T_2$ where $T_2 \le T_1.$ 
We choose the secondary discard threshold to be smaller than the primary discard threshold
as a smaller value for secondary threshold will ensure that the secondary replicas will not overload the system when the current workloads at the queues are high.
Furthermore, note that we expect the secondary replicas to be helpful only if the primary is delayed.

Since our policy is parametrized by number of replicas $d$, primary discard threshold $T_1$, 
and secondary discard threshold $T_2$, 
we shall henceforth denote it by $\pi(d,T_1,T_2)$ for simplicity.  
Following are some special cases of our \emph{discard threshold} based redundancy-$d$ policy that we analyze in this article.
\begin{enumerate}
\item Replication with identical thresholds ($\pi(d,T,T)$): In this policy, each job is replicated $d$ times and assigned to $d$ servers chosen at random.
Each job replica will have a threshold of $T$ time units which can possibly result in loss of jobs. 
When $T = \infty$, the policy reduces to that of a simple replication-$d$ policy without cancellation.
\item Replication with no loss ($\pi(d,\infty,T_2)$): 
Under this policy, as primary threshold $T_1 = \infty,$ each primary replica of the job is definitely served. 
The advantage of this policy is that no jobs are lost.
\item Replication on idle secondary servers ($\pi(d,\infty,0)$): This is special case of replication policy with minimal redundancy addition, 
since secondary replicas only join idle queues. 
\end{enumerate}

\subsection{Server}
We assume that each server has an infinite sized buffer where arriving job replicas can wait for service, 
on a first come first served (FCFS) basis. 
We let the random variable $W_{n,i}$ with distribution function $F: \R_+\to [0,1]$, 
denote the waiting time for the $n$th arriving job at server~$i \in[N]$.  
Due to FCFS service, the random variable $W_{n,i}$ is also the effective workload present at server~$i$ that must be served before $n$th job replica can receive service. 
An arriving replica is executed at a server~$i$ if its discard threshold $T$ is larger than the observed workload $W_{n,i}$, 
and is discarded otherwise. 

Each arriving job in the system results in a potential arrival at maximum $d$ randomly sampled queues. 
Depending on the discard threshold $T$ and waiting time $W_{n,i},$ the job either receives service or is discarded. 
If a replica is served, then it results in an actual arrival at the corresponding server queue.   

\subsection{Performance metrics}
\label{sec:PerfMetrics}

We consider the following two performance metrics, 
the mean response time and the loss probability. 
Since our dispatcher replicates each arriving job to at most $d$ servers, 
the response time of an arriving job is the minimum of the sojourn times experienced by its different replicas. 
When both the thresholds $T_1$ and $T_2$ are finite, 
each replica can be discarded without service, leading to a loss.  
For lost jobs, the response time metric is meaningless. 
Hence, we obtain the mean response time of a job, 
conditioned on the event that it is not discarded. 
A job is serviced when at least one of its replicas is not discarded at the servers sampled by the dispatcher, i.e. 
when workload at one of these servers is smaller than or equal to the corresponding discard threshold. 
\begin{definition}
\label{defn:IndQueueSelection}
Let $I_1$ be the singleton set of servers where primary replica is dispatched. 
Let $I_2$ be the candidate set of servers to which the secondary replicas are dispatched. 
For any server $j$, we define the indicators $\gamma^1_j \triangleq \SetIn{j \in I_1}$ and $\gamma^2_j \triangleq \SetIn{j \in I_2}$ which indicates that the queue is selected as a primary or secondary server respectively. 
\end{definition}

\begin{definition}
\label{defn:IndicatorJobDiscard} 
If the replica is dispatched to a server~$j \in I_1\cup I_2$ with current workload $W_j$, 
then we define the indicator that the job is not discarded at this server~$j$ as 
\EQN{
\label{eqn:IndicatorJobDiscardServer}
\xi_j\SetIn{j \in I_1\cup I_2} \triangleq \SetIn{W_j \le T_1} \gamma^1_j + \SetIn{W_j \le T_2} \gamma^2_j. 
} 
We denote the set of servers, 
where the job replicas are not discarded by 
$I \triangleq \set{j \in I_1\cup I_2: \xi_j=1}$.  
A job is not discarded when $I\neq \emptyset$, 
and we denote this by indicator $\xi \triangleq \SetIn{I \neq \emptyset}$. 
We can write this in terms of
the set of servers $I_1, I_2$, 
the indicators $\xi_j$ and $\bxi_j \triangleq 1- \xi_j$ for all $j \in I_1\cup I_2$, 
\EQN{
\label{eqn:IndicatorJobDiscard}
\xi\triangleq 1 -  \prod_{j \in I_1}\bxi_j\prod_{j \in I_2}\bxi_j. 
}
\end{definition}
\begin{definition}
\label{def:LossProbability} 
The loss probability for policy $\pi(d, T_1, T_2)$ is denoted by $P_{L} \triangleq \E\bxi$.
\end{definition} 
\begin{definition}
\label{def:MeanResponseTime}
We denote the response time of any job by $R^{\prime} \in \R_+ \cup {\infty}$ and the response time of an undiscarded job by a random variable $R = \xi  R^{\prime} \in \R_+ $ following the distribution function $H: \R_{+} \to [0,1]$, 
such that the tail distribution $\bH: \R_+ \to [0,1]$ is defined as $\bH(x) \triangleq 1- H(x)$ for all $x \in \R_+$.
We study the conditional mean response time for a job  given that it is not discarded.
Specifically, we define the conditional mean response time as 
\EQN{
\label{eq:mrtDefinition}
\tau \triangleq \frac{\E[R]}{\E[\xi]} = \frac{\int_{x \in \R_+}\bar{H}(x)dx}{1 - P_L}.
}
\end{definition}

In this article, 
we analyze the performance of the $\pi(d,T_1, T_2)$ load balancing policy for different special cases mentioned in Section~\ref{subsection:ThresholdCancel}, 
based on the two performance metrics of conditional mean response time and loss probability. 
Computing the limiting marginal workload distribution at a single queue is straightforward and can be performed by isolating the considered queue from the rest of the system.
However, a job response time is the minimum of response time for all possible job replicas, 
and computation of the conditional mean requires the knowledge of the joint distribution of workloads at all queues with a job replica. 
We would also like to point out that the workloads at different queues are not independent of each other due to the correlated arrivals.
To see the dependence between the workload at different servers, let us consider a simplified scenario.
\begin{exmp}{1}
Consider a system of two servers with initial workloads $W_1(0) = W_2(0) = 0$. 
Suppose the arrivals to the system follows a Poisson distribution with rate $\lambda$ and each arrival brings in a job of constant size $c$. 
The jobs are replicated and sent to both the servers and they are accepted at the servers if their current workloads are smaller than a threshold $T$. 
We denote the inter-arrival times by the random sequence $(T_n \in \R_+: n \in \N)$, 
and the $n$th arrival instant by $S_n \triangleq \sum_{k=1}^nT_k$ for all $k \in \N$. 
Then the workload at server $i$ at the $n$th arrival instant is denoted by $W_{n,i}$, 
and can be written recursively, as 
$
W_{n+1,i} = (W_{n,i}+c-T_{n+1})_+.
$
We observe that $W_{n,1} = W_{n,2}$ for all $n \in \N$. 
Further, we have $W_i(t) = (W_{n,i} -t)_+$ for all $t \in [S_n, S_{n+1})$, 
and hence $W_1(t) = W_2(t)$ for all $t \in \R_+$.
Thus we can see that the workloads in these two queues are not independent of each other.
\end{exmp} 
However, we show that for the proposed load balancing policy $\pi(d,T_1,T_2)$,  
the workload at different queues are independent of each other for any finite time horizon $[0, t]$, 
when the job arrivals are Poisson, the replicas have \iid service time distribution, 
and the number of servers $N$ grows large while keeping the number of replicas $d$ fixed. 
Furthermore, to compute the joint workload distribution, 
we use the cavity process method~\cite{Bramson10,Bramson12,Hellemans18,Hellemans19}) that assumes the asymptotic independence of the limiting marginal workload at server queues.
The next subsection provides a brief discussion on the cavity process method.  
\subsection{Cavity process method}   
Here, we explain the principle of a cavity process method as applied to popular load balancing policies such as least loaded (LL($d$)) or join shortest queue (JSQ($d$)) and then specialize the discussion to our policy $\pi(d,T_1,T_2)$. See \cite{Bramson10,Bramson12,Hellemans18,Hellemans19} for more details about this approach. 
In the LL($d$) (resp. JSQ($d$)) system with $N$ queues and Poisson arrival rate of $\lambda N$,  
$d$ queues are sampled for each arriving job. 
The arriving job is executed on the sampled server with the smallest workload (resp. queue length). 
Let $\set{\cH(t), t \ge 0}$ denote the collection of probability measures on $\R_+$. 
This is called as the environment process. 
We tag one of the queue in the $N$ queue system as the cavity queue and denote the cavity process by $X^{\cH(t)}$ 
which represents the workload process (resp. the queue length process) at the cavity queue under policy LL($d$) (resp. JSQ($d$)). 
The potential arrival rate of jobs to the cavity queue under both policies is $\lambda d$. 
For a potential arrival at the cavity queue at time $t$, 
we compare $d-1$ random variables with law $\cH(t)$ with $X^{\cH(t-)}$. 
The potential arrival becomes an actual arrival  to the cavity queue if the value of $X^{\cH(t-)}$ is lower than the values taken by the $d-1$ other variables, 
else the job is discarded. 
When the job is accepted,  
we have $X^{\cH(t)} = X^{\cH(t-)} + 1$ for the JSQ($d$) policy and $X^{\cH(t)} = X^{\cH(t-)} + x$ for the LL($d$) policy where $x$ is the service requirement of the arriving job. 
When the job is discarded, 
we have $X^{\cH(t)} = X^{\cH(t-)}$. 
For the LL($d$) policy, the workload $X^{\cH(t)}$ at the cavity queue decreases a unit rate, 
and for the JSQ($d$) policy, the queue length $X^{\cH(t)}$ of the cavity queue decreases by one at a unit rate. 
The process $\cH(\cdot)$ is called as the \textit{equilibrium environment process} if $X^{\cH(.)}(t)$ has distribution $\cH(t)$ for all times $t$. 
If $\cH(t) = \cH$ for all $t, $ then $\cH$ is called as \textit{equilibrium environment}.

The cavity process method was used in~\cite{Bramson10,Bramson12} to analyze the LL($d$) and the JSQ($d$) policy. 
A key step in the analysis is to show asymptotic independence between the workloads/queue length random variables at the different queues. 
While the analysis for LL($d$) holds for any service requirement distribution, 
the proof for JSQ($d$) is only known for the case when the service requirement of a job has decreasing hazard rate distribution. 
In~\cite{Hellemans18}, this approach is used further to obtain the functional differential equation for the workload distribution of the cavity queue. 
In~\cite{Hellemans19}, several workload based load balancing policies based on redundancy were considered and the cavity process method was used to identity the workload distribution for a wide range of load balancing policies. 
While the asymptotic independence of the queues was only conjectured,  
this was very recently proved (for most of the policies of~\cite{Hellemans19}) in~\cite{Shneer2021QS} for a variety of such replication based policies.
We prove the asymptotic independence of workloads under our settings for any finite time horizon and provide an empirical validation at time stationarity.

\begin{figure}[h!]%
\centering
		\scalebox{.75}{\tikzset{every picture/.style={line width=0.75pt}} 

\begin{tikzpicture}[x=0.75pt,y=0.75pt,yscale=-1,xscale=1]

\draw  [color={rgb, 255:red, 128; green, 128; blue, 128 }  ,draw opacity=1 ][dash pattern={on 4.5pt off 4.5pt}] (362.35,199.94) -- (553.41,199.94) -- (553.41,250.91) -- (362.35,250.91) -- cycle ;
\draw    (384.32,113.69) -- (517,113.69) ;
\draw    (384.32,143.09) -- (517,142.11) ;
\draw    (517,113.69) -- (517,142.83) ;
\draw   (517.44,128.15) .. controls (517.44,119.35) and (523.87,112.22) .. (531.81,112.22) .. controls (539.75,112.22) and (546.18,119.35) .. (546.18,128.15) .. controls (546.18,136.94) and (539.75,144.07) .. (531.81,144.07) .. controls (523.87,144.07) and (517.44,136.94) .. (517.44,128.15) -- cycle ;
\draw    (500.19,113.69) -- (500.19,142.83) ;
\draw    (509.03,113.69) -- (509.03,142.83) ;
\draw    (384.17,58.8) -- (516.85,58.8) ;
\draw    (384.17,88.2) -- (516.85,87.22) ;
\draw    (516.85,58.8) -- (516.85,87.94) ;
\draw   (517.29,73.26) .. controls (517.29,64.46) and (523.73,57.33) .. (531.66,57.33) .. controls (539.6,57.33) and (546.04,64.46) .. (546.04,73.26) .. controls (546.04,82.05) and (539.6,89.18) .. (531.66,89.18) .. controls (523.73,89.18) and (517.29,82.05) .. (517.29,73.26) -- cycle ;
\draw    (500.04,58.8) -- (500.04,87.94) ;
\draw    (508.89,58.8) -- (508.89,87.94) ;
\draw    (482.35,113.69) -- (482.35,142.83) ;
\draw    (491.2,113.69) -- (491.2,142.83) ;
\draw    (464.66,113.69) -- (464.66,142.83) ;
\draw    (473.51,113.69) -- (473.51,142.83) ;
\draw    (446.97,113.69) -- (446.97,142.83) ;
\draw    (455.82,113.69) -- (455.82,142.83) ;
\draw    (428.4,113.69) -- (428.4,142.83) ;
\draw    (437.24,113.69) -- (437.24,142.83) ;
\draw    (410.7,113.69) -- (410.7,142.83) ;
\draw    (419.55,113.69) -- (419.55,142.83) ;
\draw    (384.17,211.71) -- (516.85,211.71) ;
\draw    (384.17,241.11) -- (516.85,240.13) ;
\draw    (516.85,211.71) -- (516.85,240.85) ;
\draw   (517.29,226.16) .. controls (517.29,217.37) and (523.73,210.24) .. (531.66,210.24) .. controls (539.6,210.24) and (546.04,217.37) .. (546.04,226.16) .. controls (546.04,234.96) and (539.6,242.09) .. (531.66,242.09) .. controls (523.73,242.09) and (517.29,234.96) .. (517.29,226.16) -- cycle ;
\draw    (500.04,211.71) -- (500.04,240.85) ;
\draw    (508.89,211.71) -- (508.89,240.85) ;
\draw    (482.35,211.71) -- (482.35,240.85) ;
\draw    (491.2,211.71) -- (491.2,240.85) ;
\draw    (473.51,211.71) -- (473.51,240.85) ;
\draw  [fill={rgb, 255:red, 0; green, 0; blue, 0 }  ,fill opacity=1 ] (459.06,158.65) .. controls (459.06,157.37) and (460,156.33) .. (461.16,156.33) .. controls (462.32,156.33) and (463.26,157.37) .. (463.26,158.65) .. controls (463.26,159.94) and (462.32,160.98) .. (461.16,160.98) .. controls (460,160.98) and (459.06,159.94) .. (459.06,158.65) -- cycle ;
\draw  [fill={rgb, 255:red, 0; green, 0; blue, 0 }  ,fill opacity=1 ] (459.06,170.42) .. controls (459.06,169.13) and (460,168.09) .. (461.16,168.09) .. controls (462.32,168.09) and (463.26,169.13) .. (463.26,170.42) .. controls (463.26,171.7) and (462.32,172.74) .. (461.16,172.74) .. controls (460,172.74) and (459.06,171.7) .. (459.06,170.42) -- cycle ;
\draw  [fill={rgb, 255:red, 0; green, 0; blue, 0 }  ,fill opacity=1 ] (459.06,182.18) .. controls (459.06,180.89) and (460,179.85) .. (461.16,179.85) .. controls (462.32,179.85) and (463.26,180.89) .. (463.26,182.18) .. controls (463.26,183.46) and (462.32,184.51) .. (461.16,184.51) .. controls (460,184.51) and (459.06,183.46) .. (459.06,182.18) -- cycle ;
\draw [line width=0.75]  [dash pattern={on 0.84pt off 2.51pt}]  (211.98,148.16) -- (388.52,225.67) ;
\draw [line width=0.75]  [dash pattern={on 0.84pt off 2.51pt}]  (211.98,148.16) -- (362.65,128.49) ;
\draw [line width=0.75]  [dash pattern={on 0.84pt off 2.51pt}]  (211.39,147.51) -- (387.71,73.01) ;
\draw [line width=1.5]    (353.58,121.61) -- (366.7,134.46) ;
\draw [line width=1.5]    (365.74,121.61) -- (354.29,134.52) ;
\draw   (147.71,127.9) -- (209.62,127.9) -- (209.62,167.11) -- (147.71,167.11) -- cycle ;
\draw    (117.63,147.51) -- (140.4,147.51) ;
\draw [shift={(142.4,147.51)}, rotate = 180] [color={rgb, 255:red, 0; green, 0; blue, 0 }  ][line width=0.75]    (10.93,-3.29) .. controls (6.95,-1.4) and (3.31,-0.3) .. (0,0) .. controls (3.31,0.3) and (6.95,1.4) .. (10.93,3.29)   ;

\draw (132.14,172.61) node [anchor=north west][inner sep=0.75pt]   [align=left] {{\footnotesize  $\displaystyle d$ potential replicas}\\{\footnotesize  dispatched per job}};
\draw (422.83,34.13) node [anchor=north west][inner sep=0.75pt]   [align=left] {{\footnotesize N servers}};
\draw (151.72,140.52) node [anchor=north west][inner sep=0.75pt]   [align=left] {{\footnotesize dispatcher}};
\draw (278.94,143.4) node [anchor=north west][inner sep=0.75pt]   [align=left] { {\footnotesize replica discarded if }\\{\footnotesize workload $\displaystyle  >$ threshold}};
\draw (360.73,256.63) node [anchor=north west][inner sep=0.75pt]   [align=left] { \ {\footnotesize cavity queue with potential arrival rate $\displaystyle \lambda d$}};
\draw (68.98,129.49) node [anchor=north west][inner sep=0.75pt]   [align=left] {{\footnotesize arrival rate }\\{\footnotesize $\displaystyle \ \ \ \ \ \lambda N$}};

\end{tikzpicture}}
	\caption{\small{
The $N$ server system under $\pi(d,T_1,T_2)$ policy with a job arrival rate $\lambda N$.
The dispatcher dispatches $d$ replicas per job and the potential arrival rate at any cavity queue is $\lambda d$.}}
\label{Fig:Policy}
\end{figure}

For our $\pi(d,T_1,T_2)$ policy, we use this cavity process method along with the conjecture that the workload distribution across any finite subset of queues is asymptotically independent. 
For our policy shown in Fig. \ref{Fig:Policy}, note that the potential arrival rate to the cavity queue is 
$\bl \triangleq \lambda d$. 
If the copy at the cavity queue is a primary replica, 
then $X^{\cH(t)} = X^{\cH(t-)} + x$ if $X^{\cH(t-)} \le T_1$ else the copy is discarded. 
Similarly if the replica at the cavity queue is a secondary one, 
then the replica is served if $X^{\cH(t-)} \le T_2.$ 
Clearly, the potential arrival at the cavity queue becomes an actual arrival based on the workload level at the queue. Remarkably, for our policy there is no influence on the cavity queue of the $d-1$ random variables with law $\cH(\cdot)$.  
With the assumption of the asymptotic independence of the workload at the queue, 
using the cavity process approach, 
we can  view the cavity queue as an $M/G/1$ queue with workload dependent arrival rates. 
The workload distribution of the cavity queue is in fact the equilibrium environment $\cH$ for our system.
 See \cite{Bekker2004QS} for one possible approach to obtain the workload distribution for an $M/G/1$ queue with workload dependent arrival rates. In the following we use a different approach based on the Lindley type recursion and moment generating function (MGF) to obtain the workload distribution for the queue at cavity. We believe that this approach is novel and can be applied to more general load balancing policies beyond this work.

Next, we discuss the conjecture on asymptotic independence.
First, we provide the result on asymptotic independence of workloads over a finite horizon. 
The proof is very similar to the proof of asymptotic independence of queues over a finite time horizon for JSQ($d$) dispatch policy, shown in~\cite[Proposition 7.1]{Bramson12}.
\begin{proposition}[Asymptotic independence over finite time horizon ]
\label{pro:AsymptoticInd} 
Consider an $N$ server system under $\pi(d,T_1,T_2)$ dispatch policy. 
When the number of servers $N$ grows asymptotically large, 
the marginal workload distributions at any finite number of queues are independent over a finite time horizon. 
\end{proposition}
See ~\ref{app:AsymptoticInd} for proof.
\begin{remark}
The above proposition holds true for general \iid service time distributions as well as for the identical service time model discussed in ~\ref{App:DeterministicSlowdown}.
For ease of exposition, we omit the details.
\end{remark}

For power of $d$ variants of dispatch policies, 
once the asymptotic independence is shown for a finite time horizon, 
one can show the asymptotic independence at time stationarity as well when the workloads satisfy certain monotonicity conditions.
(See ~\ref{app:AsymptoticInd} for further details.)
Although we don't have such monotonicity property under $\pi(d,T_1,T_2)$ policy when either of the two thresholds $T_1,T_2$ are finite, we still assume that the asymptotic independence of server workloads continues to hold true at time stationarity, as given in the following conjecture.
\begin{conj2}{1}[Asymptotic Independence at stationarity]
\label{Assum:AsymptoticIndependence}
Consider an $N$ server system under $\pi(d,T_1,T_2)$ dispatch policy. 
When the number of servers $N$ grows asymptotically large, 
the system has a unique equilibrium workload distribution under which any finite number of queues are independent.  
\end{conj2}
See ~\ref{app:ModelValidation} for an empirical validation of this conjecture.
 Please note that Conjecture~\ref{Assum:AsymptoticIndependence} can be proved for a limited regime of arrival rates, by adapting the proof of~\cite[Theorem 2.3]{Bramson12} to our setting.  
However, the empirical evaluations suggest that the asymptotic independence is a valid assumption under all arrival rates under the studied policy.
 
\begin{remark}
We first obtain the MGF for the workload at the cavity queue. 
We then use this to obtain the conditional mean response time for the different policies, under Conjecture~\ref{Assum:AsymptoticIndependence}. 
We illustrate the accuracy of our expressions in ~\ref{app:ModelValidation} by comparing them with simulation experiments for different values of $N$. 
As a validation of the assumption, we see that as $N$ increases, the mean response time from simulations approach the analytical values. 
\end{remark}

\section{Performance Analysis}
\label{sec:Analysis}
As mentioned before, we will be able to compute the identical marginal workload distribution at all $N$ servers for the proposed $\pi(d,T_1,T_2)$ dispatch policy. 
However, we will be able to obtain expressions for both the performance metrics of conditional mean response time and loss probability, 
only under Conjecture~\ref{Assum:AsymptoticIndependence}.
This computation is an approximation for finite number of servers. 
However, we empirically verify that this approximation is quite accurate even for a small number of servers. 

\subsection{Loss probability}
When both primary and secondary thresholds are finite, 
some jobs can be discarded from the system. 
Under Conjecture~\ref{Assum:AsymptoticIndependence}, 
we compute the limiting loss probability of a job being discarded in the following Lemma.
\begin{lemma}
\label{lem:LossProb}
The limiting loss probability of a job under $\pi(d,T_1,T_2)$ dispatch policy with equilibrium workload distribution $F$ and tail distribution of service time $\bG$ is given by
$
P_{L} = \bF(T_1)\bF(T_2)^{d-1}.
$
\end{lemma}
\begin{proof}
From~\eqref{eqn:IndicatorJobDiscard} in Definition~\ref{def:LossProbability}, 
we obtain $P_L = \expect{\prod_{j \in I_1} \bxi_j \prod_{j \in I_2}\bxi_j}.$
The result follows from the independence of the indicators $(\bxi_j : j \in I_1 \cup I_2)$, 
under the assumption of asymptotic independence across the servers $j \in I_1\cup I_2$, 
and the fact that the mean of indicators 
$
\E[\bxi_j\SetIn{j \in I_1\cup I_2}\given I_1,I_2] = \bF(T_1)\gamma^1_j + \bF(T_2)\gamma^2_j.  
$
\end{proof}

\subsection{Conditional mean response time}
Next, we characterize the mean response time for a job under the dispatching policy $\pi(d,T_1,T_2)$. 
Note that when the discard thresholds $T_1, T_2$ are finite, 
then all jobs that arrive at a server with workload $w > T_1$ will be lost. 
For lost jobs, the response time metric is meaningless. 
Hence, we obtain the conditional mean response time given that the job is not discarded. 
A job is serviced when at least one of its replicas is not discarded at the servers sampled by the dispatcher, 
i.e. 
when workload at one of these servers is smaller than or equal to the corresponding discard threshold. 

\begin{theorem}
\label{thm:MeanResponseTime}
The conditional mean response time of an undiscarded job under $\pi(d,T_1,T_2)$ policy with equilibrium workload distribution $F$ and tail distribution of service time $\bG$ is given by
\EQ{
\tau
=  \frac{1}{1 - P_L}\int_x\Big[ \Big\lbrace(\bF(T_1) + k(x,T_1))( \bF(T_2)+k(x,T_2))^{d-1}  
- \bF(T_1)\bF(T_2)^{d-1} \Big\rbrace\Big]dx,
}
where $k(x,T) \triangleq \expect{\bG(x-W)\SetIn{W \le T}}$. 
\end{theorem}
\begin{proof}
Refer ~\ref{App:ProofMeanResponseTime}.
\end{proof}
\begin{remark} 
From the non-negativity of distribution functions, 
we can exchange two integrals using Monotone convergence theorem. 
Therefore, we have 
$
\int_{x\in\R_+}k(x,T)dx =  \frac{F(T)}{\mu} +\expect{W\SetIn{W \le T}}.
$
Defining $k(x) \triangleq \lim_{T\to \infty}k(x,T)$, we observe that $\int_{x\in\R_+}k(x)dx = \E W + \frac{1}{\mu}$.  
\end{remark}
\begin{remark}
 When the thresholds $T_1$ and $T_2$ are infinity, 
 we see the tail workload distributions $\bF(T_1) = \bF(T_2) = 0$ and we have $k(x) = k(x, \infty) = \E\bG(x-W)$. 
It follows that the tail distribution of response time is 
$\bH(x) = k(x)^d $.
\end{remark}
\section{Workload distribution and conditional mean response time under exponential service times}
\label{sec:Exponential}
In this section, we evaluate the workload distribution $F$ in the cavity queue under various load balancing polices discussed in section~\ref{subsection:ThresholdCancel} when the service times of each job is independent and follows an identical exponential distribution with rate $\mu$. 
We choose the service times to be exponentially distributed as they are amenable to analytical computations, due to their memoryless property. 
Let us first introduce some preliminary definitions  prior to introducing the results.

We denote the indicator that the $j$th server is selected by $n$th job as a primary or secondary server by $\gamma^1_{n,j}$ and $\gamma^2_{n,j}$ respectively. 
Recall that the workload seen by the $n$th job arrival at server $j$ is $W_{n,j}$ and the service time for $n$th job if it joins server $j$ is given by $X_{n,j}$. 
Since we are interested in a single cavity queue~$j$, we drop the subscript $j$ in the following. 
For $T_2 \le T_1$, 
we can use Lindley's recursion to write the single queue workload sequence
$(W_n: n \in \N)$ in terms of random service time sequence $(X_n: n \in \N)$, 
inter-arrival time sequence $(T_n: n \in \N)$, as
\EQN{
\label{eq:LindleysRecursion}
W_{n+1} = (W_n+ X_n((\gamma^1_n+ \gamma^2_n)\SetIn{W_n \in [0,T_2]} 
+ \gamma^1_n\SetIn{W_n \in (T_2,T_1]})-T_{n+1})_+,\quad n \in \Z_+.
}
That is, 
we have 
\eqn{
\label{eq:WorkloadRecursionGeneral}
W_{n+1} &=
\begin{cases}
(W_n-T_{n+1})_+,& W_n \in (T_1,\infty),\\ 
(1 - \gamma^1_n)(W_n-T_{n+1})_+ + \gamma^1_n(W_n+X_n-T_{n+1})_+,& W_n \in (T_2,T_1],\\ 
(1-\gamma^1_n-\gamma^2_n)(W_n-T_{n+1})_+ + (\gamma^1_n+\gamma^2_n)(W_n+X_n-T_{n+1})_+, & W_n \in [0,T_2].
\end{cases}
}
In order to derive the workload distribution in the cavity queue, 
we make use of the moment generating function of the workload.
\begin{definition}
\label{def:mgf}
The moment generating function of the limiting workload $W$ in a single queue, 
restricted to different workload regimes is defined as
\begin{xalignat*}{3}
&\Phi_W(\theta) \triangleq \expect{e^{-\theta W}},& 
&\Phi_2(\theta) \triangleq \expect{e^{-\theta W}\SetIn{W > T_2}},&
&\Phi_1(\theta) \triangleq \expect{e^{-\theta W}\SetIn{W > T_1}}.
\end{xalignat*}
\end{definition}

\begin{theorem} 
\label{thm:mgfGeneral}
For an $N$ server system with \iid exponential service times of rate $\mu$ and Poisson arrivals of rate $N\lambda$, 
the moment generating function $\Phi_W(\theta)$ for the waiting time of admitted jobs at any queue under $\pi(d,T_1, T_2)$ policy is given by 
\eqn{
\label{eqn:MGFGeneral}
&F(0)(1 + \frac{\bl}{\theta + \mu - \bl}) + 
\big((\mu - \lambda)\bF(T_2) + \lambda \bF(T_1)\big)
\big[\frac{e^{-\theta T_2}}{\theta + \mu - \lambda} - \frac{e^{-\theta T_2}}{\theta + \mu - \bl}\big] 
-\mu\bF(T_1)\big[\frac{e^{-\theta T_1}}{\theta + \mu - \lambda} - \frac{e^{-\theta T_1}}{\theta + \mu} \big],  
}
where $
F(0) = 1 -\frac{\bl}{\mu} + \Big[\frac{\bl-\lambda}{\mu}\bF(T_2)
+ \frac{\lambda}{\mu}\bF(T_1)\Big].
$
\end{theorem}
\begin{proof}
The detailed proof is in ~\ref{app:ProofmgfGeneral}.
\end{proof}

\begin{corollary}
\label{cor:GeneralCase}
For an $N$ server system with \iid exponential service times of rate $\mu$ and Poisson arrivals of rate $N\lambda$, 
the single queue workload distribution under $\pi(d,T_1, T_2)$ policy is given by  
\eq{
&F(w) = F(0)\Big(1+\frac{\bl(1 - e^{-(\mu - \bl)w})}{\mu - \bl} \Big)  - \mu \bF(T_1) \Big(\frac{(1 - e^{-(\mu - \lambda)(w-T_1)_+})}{\mu - \lambda} - \frac{(1 - e^{-\mu(w-T_1)_+})}{\mu} \Big)\\
&+ ((\mu -\lambda)\bF(T_2) + \lambda\bF(T_1))\Big(\frac{(1 - e^{-(\mu - \lambda)(w-T_2)_+})}{\mu - \lambda} - \frac{(1 - e^{-(\mu - \bl)(w-T_2)_+})}{\mu - \bl} \Big).  
}
\end{corollary}

\begin{remark}
\label{Rem:IntExplanation}
We observe that when the workload in a cavity queue lies is less than the threshold $T_2$, 
the arrival rate to the queue is $\bl$. 
Therefore, we expect the cavity queue to behave like an $M/M/1$ queue with arrival rate $\bl$ and service rate $\mu$, 
when $w \in [0, T_2)$. 
Indeed, we observe from Corollary~\ref{cor:GeneralCase}, that the marginal workload distribution reduces to 
$
F(w) = F(0)\Big(\frac{\mu-\bl e^{-(\mu - \bl)w})}{\mu - \bl} \Big),~w < T_2. 
$
Similarly, when the workload in a cavity queue lies in the duration $[T_2, T_1)$, 
the arrival rate to the queue is $\lambda$. 
Accordingly, the behavior of cavity queue in this region is similar to an $M/M/1$ queue with arrival rate $\lambda$ and service rate $\mu$. 
As expected, the marginal workload distribution reduces to 
$
F(w) = F(T_2) + \frac{(\mu F(0)-(\mu-\bl)F(T_2))}{\mu-\lambda}
(1-e^{-(\mu - \lambda)(w-T_2)}),~w \in [T_2, T_1). 
$
Since there are no more arrivals to a cavity queue, when the workload is larger than the threshold $T_1$, 
the workload distribution is expected to decay exponentially with the service rate $\mu$. 
Unsurprisingly, the marginal workload distribution is 
$
F(w) = F(T_1) + \bF(T_1)(1- e^{-\mu (w-T_1)}),~w \ge T_1. 
$
\end{remark}

Next, we study some special cases of the $\pi(d,T_1,T_2)$ policy listed in Section~\ref{subsection:ThresholdCancel}.
\subsection{Replication with identical thresholds}
\label{subsec:RepIdenticalThreshold}

First, we study the system under the replication with identical thresholds policy, $\pi(d,T,T)$. 
Note that as the system allows loss, it is always stable.
Next result follows from Corollary~\ref{cor:GeneralCase} by substituting $T_1 = T_2$.
\begin{corollary}
\label{cor:F(w)Policy1}
For an $N$ server system with \iid exponential service times of rate $\mu$ and Poisson arrivals of rate $N\lambda$, 
the workload distribution at the cavity queue at stationarity under $\pi(d,T, T)$ policy, 
is given by
\EQ{
F(w) =\begin{cases}
F(0)(\frac{\mu}{\mu - \bl} - \frac{\bl}{\mu - \bl}e^{-(\mu - \bl)w}),& 0 < w \le T\\
F(T) + \frac{\bl}{\mu}e^{\bl T} F(0)(e^{-\mu T} - e^{-\mu w}),& w > T
\end{cases}
}
where $F(0) = \Big[\frac{(1 - \frac{\bl}{\mu})}{1 - (\frac{\bl}{\mu})^2 e^{-(\mu - \bl)T}}\Big]\SetIn{\mu \neq \bl} 
+ \frac{1}{\bl  T +  2}\SetIn{\mu =\bl}$ and $
F(T) = \frac{\mu}{\bl}(1-F(0))
$.  
\end{corollary}
Using the above corollary, we now compute the loss probability and conditional mean response time using Theorem~\ref{thm:MeanResponseTime}. 
\begin{corollary}
The loss probability of a job under discard threshold based dispatching policy $\pi(d,T,T)$ with equilibrium workload distribution $F$ and tail distribution of service time $\bG$, 
is given by
$
P_{L} = \left(1 - \frac{\mu}{\bl}(1-F(0))\right)^{d},
$
where probability of zero workload $F(0)$ is given in Corollary~\ref{cor:F(w)Policy1}.
\end{corollary}

\begin{remark}
Note that the effective arrival rate at each cavity queue under replication with identical thresholds policy is $\bl \SetIn{w \le T}$.
However, as the jobs are discarded as soon as the current workload exceeds the threshold $T$, the queues remain stable even when the arrival rate to the system exceeds the service rate.
In particular, the above results says that for $\bl = \mu$, we get the workload distribution
$
F(w) = \frac{\bl}{\bl T + 2}w \SetIn{0 < w \le T} + \frac{1-e^{-\bl(w-T)}}{\bl T + 2} \SetIn{w > T}
$
and the loss probability
$
P_L = \left(\frac{1}{\bl T + 2}\right)^d.
$
For the special case of $T = 0$, we get the expected workload and loss probability respectively as
$\E[W] = \frac{1}{2\mu} $ and 
$ P_L = \left(\frac{1}{2}\right)^d.
$
\end{remark}

From Theorem~\ref{thm:MeanResponseTime}, we know the conditional mean response time under $\pi(d,T,T)$ policy is
\eq{
\tau 
= &\frac{1}{1 - P_L} \int_x   \Big(( \bF(T) +k(x,T))^{d} 
- \bF(T)^{d}  \Big)dx.
}
Thus, we see that computing the term $k(x,T)$ shall allow us to evaluate the mean response time of the $N$ server system under the policy $\pi(d,T,T)$.
The next lemma provides us this result. 
\begin{lemma}
\label{lem:identical_th}
For an $N$ server system with \iid exponential service times of rate $\mu$ and Poisson arrivals of rate $N\lambda$, 
we can find the following constants under $\pi(d,T, T)$ policy, 
\begin{xalignat*}{1}
&\bF(T) = 1 - F(0)\Big[\frac{\mu}{\mu-\bl} - \frac{\bl}{\mu - \bl}e^{-(\mu - \bl)T}\Big],\\
&F(0)= \frac{(1 - \frac{\bl}{\mu})}{1 - (\frac{\bl}{\mu})^2 e^{-(\mu - \bl)T}}.
\end{xalignat*}
The function $k(x,T)$ is given by 
\EQ{  
\begin{cases}
F(0)e^{-\mu x} e^{\bl T}, & x \ge T\\
F(0)\Big(\frac{\mu}{\mu-\bl}e^{-(\mu - \bl)x} - \frac{\bl}{\mu - \bl}e^{-(\mu - \bl)T}\Big), & x < T.
\end{cases}
}
\end{lemma}

\begin{proof}
We know that the service time are exponential and
hence the tail service time distribution is $\bG(x) = e^{-\mu(x)_+}$, where $(x)_+ = \max\set{x,0}$.  
Therefore, we can write 
$
k(x,T) = \expect{\bG(x-W)\SetIn{W \le T}} 
= F(T) - F(T\wedge x) + e^{-\mu x}\int_0^{T\wedge x}e^{\mu W}dF(w).
$
Considering the two cases when $x \ge T$ and $x < T$, 
we get $k(x,T)$ as
\eq{
\begin{cases} 
e^{-\mu x}\int_0^{T\wedge x}e^{\mu W}dF(w), & x \ge T,\\
F(T) - F(x) + e^{-\mu x}\int_0^{x}e^{\mu W}dF(w), & x < T.
\end{cases} 
}
The result follows from the workload distribution $F$ given in Corollary~\ref{cor:F(w)Policy1}. 
\end{proof}

\begin{corollary}
For an $N$ server system with \iid exponential service times of rate $\mu$ and Poisson arrivals of rate $N\lambda$, the conditional mean response time under $\pi(d,0,0)$ policy is given by
\EQ{
\tau = \sum_{i = 0}^{d-1} \binom{d}{i} \frac{\bF(0)^i F(0)^{d-i}}{\mu(d-i)}
}
where $F(0) = \frac{1 - \frac{\bl}{\mu}}{1 - \left(\frac{\bl}{\mu}\right)^2}$ and $\bF(0) = 1 - F(0)$.
\end{corollary}

\begin{figure*}[h!]
\centering
\begin{subfigure}{.33\columnwidth}
		\centerline{\scalebox{0.6}{
\begin{tikzpicture}

\definecolor{color0}{rgb}{0.83921568627451,0.152941176470588,0.156862745098039}
\definecolor{color1}{rgb}{0.12156862745098,0.466666666666667,0.705882352941177}
\definecolor{color2}{rgb}{0.172549019607843,0.627450980392157,0.172549019607843}
\definecolor{color3}{rgb}{1,0.498039215686275,0.0549019607843137}
\definecolor{color4}{rgb}{0.833333333333333,0.866025403784439,0.5}
\definecolor{color5}{rgb}{1,0.494655843399779,0.255842777594436}
\definecolor{color6}{rgb}{1,1.22464679914735e-16,6.12323399573677e-17}

\begin{axis}[
legend cell align={left},
legend style={
  fill opacity=0.8,
  draw opacity=1,
  text opacity=1,
  at={(0.65,0.7)},
  anchor=north west,
  draw=white!80!black
},
tick align=outside,
tick pos=left,
x grid style={white!69.0196078431373!black},
xlabel={Number of replicas $d$},
xmajorgrids,
xmin=0.15, xmax=18.85,
xminorgrids,
xtick style={color=black},
y grid style={white!69.0196078431373!black},
ylabel={Conditional mean response time $\tau$},
ymajorgrids,
ymin=-0.0314867570763488, ymax=2.13679971722347,
yminorgrids,
ytick style={color=black}
]
\addplot [violet, mark=asterisk, mark size=1, mark options={solid}, line width=1pt, thick]
table {%
1 1.00442559663977
2 0.51057807547967
3 0.343604037258989
4 0.260296818549751
5 0.210370389943526
6 0.177118484485901
7 0.153392362416472
8 0.135619642706694
9 0.121816066586268
10 0.11079116679316
11 0.101787396079848
12 0.0942997312100914
13 0.0879785455164877
14 0.0825741095491379
15 0.0779032919405171
16 0.0738287475528317
17 0.0702454510055114
18 0.0670717190281885
};
\addlegendentry{$\lambda$=0.01}
\addplot [color0, mark=asterisk, mark size=1, mark options={solid}, line width=1pt, thick]
table {%
1 1.04906967849157
2 0.62427266257096
3 0.470822989676509
4 0.399755214346703
5 0.363533992125182
6 0.345600080769497
7 0.338842174703506
8 0.339736963759005
9 0.346344972945204
10 0.357501421344775
11 0.372446342410853
12 0.390638924993512
13 0.411657873791655
14 0.435145802732532
15 0.460778039819389
16 0.488245871994078
17 0.517248704534026
18 0.5474917879363
};
\addlegendentry{$\lambda$=0.11}
\addplot [color1, mark=asterisk, mark size=1, mark options={solid}, line width=1pt, thick]
table {%
1 1.09429265754655
2 0.748754227549012
3 0.63928378331482
4 0.60999231998377
5 0.618905216030749
6 0.649918608146881
7 0.69492335945375
8 0.748815226941593
9 0.80787434891696
10 0.869216281491494
11 0.930604091683257
12 0.990367208215281
13 1.04733609372936
14 1.10076794156612
15 1.1502643394999
16 1.19568905317237
17 1.23709340411587
18 1.27465361745456
};
\addlegendentry{$\lambda$=0.21}
\addplot [color2, mark=asterisk, mark size=1, mark options={solid}, line width=1pt, thick]
table {%
1 1.13990724491965
2 0.878528109115583
3 0.833308138571134
4 0.865462507100688
5 0.932999842392898
6 1.01615580546505
7 1.10337558740405
8 1.18768687647767
9 1.26524989705409
10 1.33438390802417
11 1.39478678868127
12 1.4469407990953
13 1.49170324347621
14 1.53005193239262
15 1.56294310399639
16 1.59124265716024
17 1.61570100885567
18 1.63695162293899
};
\addlegendentry{$\lambda$=0.31}
\addplot [color3, mark=asterisk, mark size=1, mark options={solid}, line width=1pt, thick]
table {%
1 1.18572104272948
2 1.0084907904641
3 1.03201148345267
4 1.121346039207
5 1.2300740566785
6 1.3365545498804
7 1.4312328081165
8 1.51127123863196
9 1.57721468911053
10 1.63096995579159
11 1.67473809806553
12 1.71054060737034
13 1.74006309766825
14 1.76464557767833
15 1.78532838654261
16 1.80291112232279
17 1.81800723476882
18 1.83108892189954
};
\addlegendentry{$\lambda$=0.41}
\addplot [color4, mark=asterisk, mark size=1, mark options={solid}, line width=1pt, thick]
table {%
1 1.23154017734865
2 1.13456343807676
3 1.21900079877369
4 1.34689067077006
5 1.47138073154306
6 1.57650405541016
7 1.66001414115673
8 1.72483616394121
9 1.7750195570438
10 1.81420002876889
11 1.8452188836892
12 1.8701724727992
13 1.89057348490924
14 1.90750904185594
15 1.92176508476078
16 1.93391632153307
17 1.94438896488063
18 1.95350391844553
};
\addlegendentry{$\lambda$=0.51}
\addplot [black, mark=asterisk, mark size=1, mark options={solid}, line width=1pt, thick]
table {%
1 1.27717294313461
2 1.25388186188684
3 1.38547760982267
4 1.53218795265149
5 1.65482587555272
6 1.74789545794877
7 1.81648712717941
8 1.86715270856652
9 1.90521399403162
10 1.93445490593263
11 1.95744639505763
12 1.97592033289829
13 1.99105407358179
14 2.00366223801869
15 2.01432049700204
16 2.02344463249331
17 2.03134132494093
18 2.03824124111893
};
\addlegendentry{$\lambda$=0.61}
\end{axis}

\end{tikzpicture}}}
		\caption{\small{
		$\tau$ vs $\lambda$.
		}}
		\label{Fig:STVsd-ArrRate-1.51.5}
\end{subfigure}
\begin{subfigure}{.33\columnwidth}
		\centerline{\scalebox{0.6}{
\begin{tikzpicture}

\definecolor{color0}{rgb}{0.83921568627451,0.152941176470588,0.156862745098039}
\definecolor{color1}{rgb}{0.12156862745098,0.466666666666667,0.705882352941177}
\definecolor{color2}{rgb}{0.172549019607843,0.627450980392157,0.172549019607843}
\definecolor{color3}{rgb}{1,0.498039215686275,0.0549019607843137}
\definecolor{color4}{rgb}{0.833333333333333,0.866025403784439,0.5}
\definecolor{color5}{rgb}{1,0.494655843399779,0.255842777594436}
\definecolor{color6}{rgb}{1,1.22464679914735e-16,6.12323399573677e-17}

\begin{axis}[
legend cell align={left},
legend style={
  fill opacity=0.8,
  draw opacity=1,
  text opacity=1,
  at={(0.65,0.9)},
  anchor=north west,
  draw=white!80!black
},
tick align=outside,
tick pos=left,
x grid style={white!69.0196078431373!black},
xlabel={Number of replicas $d$},
xmajorgrids,
xmin=0.15, xmax=18.85,
xminorgrids,
xtick style={color=black},
y grid style={white!69.0196078431373!black},
ylabel={Loss probability $P_L$},
ymajorgrids,
ymin=-0.00896350047626802, ymax=0.188233510001628,
yminorgrids,
ytick style={color=black},
ytick={-0.025,0,0.025,0.05,0.075,0.1,0.125,0.15,0.175,0.2},
yticklabels={−0.025,0.000,0.025,0.050,0.075,0.100,0.125,0.150,0.175,0.200}
]
\addplot [violet, mark=asterisk, mark size=1, mark options={solid}, line width=1pt, thick]
table {%
1 0.00224242396412633
2 2.03126344941573e-05
3 3.13514236245045e-07
4 6.86194675172449e-09
5 1.95175769807741e-10
6 6.85227152514231e-12
7 2.86995669099048e-13
8 1.39961959063396e-14
9 7.80442927535535e-16
10 4.90594718990319e-17
11 3.43740864798467e-18
12 2.65967482141275e-19
13 2.25497244298724e-20
14 2.08115975227757e-21
15 2.0790145278029e-22
16 2.23693831492596e-23
17 2.58113789244122e-24
18 3.18170132570863e-25
};
\addlegendentry{$\lambda$=0.01}
\addplot [color0, mark=asterisk, mark size=1, mark options={solid}, line width=1pt, thick]
table {%
1 0.0258454829237186
2 0.0029236959548983
3 0.000598918521003136
4 0.000181532138026398
5 7.33383595372444e-05
6 3.69284207605939e-05
7 2.20906685662021e-05
8 1.51387089010009e-05
9 1.15516926158547e-05
10 9.59489374053224e-06
11 8.518637221446e-06
12 7.96652852704703e-06
13 7.75550051318711e-06
14 7.78525847811776e-06
15 7.99764679840833e-06
16 8.3570339769452e-06
17 8.84024257833904e-06
18 9.43111404854603e-06
};
\addlegendentry{$\lambda$=0.11}
\addplot [color1, mark=asterisk, mark size=1, mark options={solid}, line width=1pt, thick]
table {%
1 0.051416561580273
2 0.0121442098928766
3 0.00520499440188158
4 0.00318719414179167
5 0.00244697569962497
6 0.00216900343334304
7 0.00210250745722334
8 0.00215038526601476
9 0.0022670657270281
10 0.00242719926648152
11 0.00261460957597005
12 0.0028179981948565
13 0.00302912246630329
14 0.00324192106739664
15 0.00345200866109404
16 0.00365632003712279
17 0.00385282515836061
18 0.00404029021937142
};
\addlegendentry{$\lambda$=0.21}
\addplot [color2, mark=asterisk, mark size=1, mark options={solid}, line width=1pt, thick]
table {%
1 0.0786684477576806
2 0.0289842344242039
3 0.0185755247057416
4 0.0157605880643788
5 0.0153982561030213
6 0.0160679548647522
7 0.0172112614480116
8 0.0185516118475909
9 0.0199363904474208
10 0.021280656394152
11 0.0225403447262259
12 0.0236964236112287
13 0.0247446703051495
14 0.025689051419016
15 0.0265375744787033
16 0.0272998088653866
17 0.0279854907359225
18 0.0286038014077916
};
\addlegendentry{$\lambda$=0.31}
\addplot [color3, mark=asterisk, mark size=1, mark options={solid}, line width=1pt, thick]
table {%
1 0.107278227173604
2 0.0535950311359052
3 0.0434540032940731
4 0.0428672346331436
5 0.0453799727066872
6 0.0488818982666204
7 0.0524967156607642
8 0.0558558145808325
9 0.0588319701753104
10 0.0614114511805962
11 0.063628341008109
12 0.0655319831661453
13 0.0671721464430299
14 0.0685931963489878
15 0.0698325284804563
16 0.0709208268667733
17 0.0718829711921727
18 0.0727390661893395
};
\addlegendentry{$\lambda$=0.41}
\addplot [color4, mark=asterisk, mark size=1, mark options={solid}, line width=1pt, thick]
table {%
1 0.136902851581796
2 0.0850412113511894
3 0.079065348036635
4 0.0831040022427031
5 0.0895075809854627
6 0.0958447273425772
7 0.101369924724132
8 0.105969641209299
9 0.109745459318449
10 0.112847788566822
11 0.115416899896349
12 0.117567354856613
13 0.119388054446808
14 0.120946738575568
15 0.122294896508625
16 0.123471884697796
17 0.124508075409139
18 0.12542717465281
};
\addlegendentry{$\lambda$=0.51}
\addplot [black, mark=asterisk, mark size=1, mark options={solid}, line width=1pt, thick]
table {%
1 0.167194818437206
2 0.12166749448819
3 0.122184177531756
4 0.13118105490577
5 0.14063031833213
6 0.148595081554465
7 0.154908342667289
8 0.159855730232578
9 0.163764630515115
10 0.166901012936492
11 0.169460974320852
12 0.17158502961464
13 0.173373757133154
14 0.174899945345758
15 0.176217121125174
16 0.177365341472871
17 0.178375104483924
18 0.17927000952536
};
\addlegendentry{$\lambda$=0.61}
\end{axis}

\end{tikzpicture}}}
		\caption{\small{
		$P_{L}$ vs $\lambda$.
		}}
		\label{Fig:PlossVsd-ArrRate-1.51.5}
\end{subfigure}
\begin{subfigure}{.33\columnwidth}
		\centerline{\scalebox{0.6}{
\begin{tikzpicture}

\definecolor{color0}{rgb}{0.83921568627451,0.152941176470588,0.156862745098039}
\definecolor{color1}{rgb}{0.12156862745098,0.466666666666667,0.705882352941177}
\definecolor{color2}{rgb}{0.172549019607843,0.627450980392157,0.172549019607843}
\definecolor{color3}{rgb}{1,0.498039215686275,0.0549019607843137}
\definecolor{color4}{rgb}{0.833333333333333,0.866025403784439,0.5}
\definecolor{color5}{rgb}{1,0.494655843399779,0.255842777594436}
\definecolor{color6}{rgb}{1,1.22464679914735e-16,6.12323399573677e-17}

\begin{axis}[
legend cell align={left},
legend style={fill opacity=0.8, draw opacity=1, text opacity=1, at={(0.7,0.58)}, anchor=north west, draw=white!80!black},
tick align=outside,
tick pos=left,
x grid style={white!69.01960784313725!black, densely dotted},
xlabel={Loss probability $P_{L}$},
xmajorgrids,
xmin=-0.05, xmax=0.2,
xminorgrids,
xtick style={color=black},
y grid style={white!69.01960784313725!black, densely dotted},
ylabel={Conditional mean response time $\tau$},
ymajorgrids,
ymin=-0.3, ymax=2.23555774770039,
yminorgrids,
ytick style={color=black}
]
\addplot [violet, mark=asterisk, mark size=1, mark options={solid}, line width=1pt, thick]
table {%
0.002242424	1.004425597
2.03E-05	0.510578075
3.13514E-07	0.343604037
6.86195E-09	0.260296819
1.95176E-10	0.21037039
6.85227E-12	0.177118484
2.86996E-13	0.153392362
1.39962E-14	0.135619643
7.80443E-16	0.121816067
4.90595E-17	0.110791167
3.43741E-18	0.101787396
2.65967E-19	0.094299731
2.25497E-20	0.087978546
2.08116E-21	0.08257411
2.07901E-22	0.077903292
2.23694E-23	0.073828748
2.58114E-24	0.070245451
3.1817E-25	0.067071719
};
\addlegendentry{$\lambda$=0.01}
\addplot [color0, mark=asterisk, mark size=1, mark options={solid}, line width=1pt, thick]
table {%
0.025845483	1.049069678
0.002923696	0.624272663
0.000598919	0.47082299
0.000181532	0.399755214
7.33384E-05	0.363533992
3.69284E-05	0.345600081
2.20907E-05	0.338842175
1.51387E-05	0.339736964
1.15517E-05	0.346344973
9.59489E-06	0.357501421
8.51864E-06	0.372446342
7.96653E-06	0.390638925
7.7555E-06	0.411657874
7.78526E-06	0.435145803
7.99765E-06	0.46077804
8.35703E-06	0.488245872
8.84024E-06	0.517248705
9.43111E-06	0.547491788
};
\addlegendentry{$\lambda$=0.11}
\addplot [color1, mark=asterisk, mark size=1, mark options={solid}, line width=1pt, thick]
table {%
0.051416562	1.094292658
0.01214421	0.748754228
0.005204994	0.639283783
0.003187194	0.60999232
0.002446976	0.618905216
0.002169003	0.649918608
0.002102507	0.694923359
0.002150385	0.748815227
0.002267066	0.807874349
0.002427199	0.869216281
0.00261461	0.930604092
0.002817998	0.990367208
0.003029122	1.047336094
0.003241921	1.100767942
0.003452009	1.150264339
0.00365632	1.195689053
0.003852825	1.237093404
0.00404029	1.274653617
};
\addlegendentry{$\lambda$=0.21}
\addplot [color2, mark=asterisk, mark size=1, mark options={solid}, line width=1pt, thick]
table {%
0.078668448	1.139907245
0.028984234	0.878528109
0.018575525	0.833308139
0.015760588	0.865462507
0.015398256	0.932999842
0.016067955	1.016155805
0.017211261	1.103375587
0.018551612	1.187686876
0.01993639	1.265249897
0.021280656	1.334383908
0.022540345	1.394786789
0.023696424	1.446940799
0.02474467	1.491703243
0.025689051	1.530051932
0.026537574	1.562943104
0.027299809	1.591242657
0.027985491	1.615701009
0.028603801	1.636951623
};
\addlegendentry{$\lambda$=0.31}
\addplot [color3, mark=asterisk, mark size=1, mark options={solid}, line width=1pt, thick]
table {%
0.107278227	1.185721043
0.053595031	1.00849079
0.043454003	1.032011483
0.042867235	1.121346039
0.045379973	1.230074057
0.048881898	1.33655455
0.052496716	1.431232808
0.055855815	1.511271239
0.05883197	1.577214689
0.061411451	1.630969956
0.063628341	1.674738098
0.065531983	1.710540607
0.067172146	1.740063098
0.068593196	1.764645578
0.069832528	1.785328387
0.070920827	1.802911122
0.071882971	1.818007235
0.072739066	1.831088922
};
\addlegendentry{$\lambda$=0.41}
\addplot [color4, mark=asterisk, mark size=1, mark options={solid}, line width=1pt, thick]
table {%
0.136902852	1.231540177
0.085041211	1.134563438
0.079065348	1.219000799
0.083104002	1.346890671
0.089507581	1.471380732
0.095844727	1.576504055
0.101369925	1.660014141
0.105969641	1.724836164
0.109745459	1.775019557
0.112847789	1.814200029
0.1154169	1.845218884
0.117567355	1.870172473
0.119388054	1.890573485
0.120946739	1.907509042
0.122294897	1.921765085
0.123471885	1.933916322
0.124508075	1.944388965
0.125427175	1.953503918
};
\addlegendentry{$\lambda$=0.51}
\addplot [black, mark=asterisk, mark size=1, mark options={solid}, line width=1pt, thick]
table {%
0.167194818	1.277172943
0.121667494	1.253881862
0.122184178	1.38547761
0.131181055	1.532187953
0.140630318	1.654825876
0.148595082	1.747895458
0.154908343	1.816487127
0.15985573	1.867152709
0.163764631	1.905213994
0.166901013	1.934454906
0.169460974	1.957446395
0.17158503	1.975920333
0.173373757	1.991054074
0.174899945	2.003662238
0.176217121	2.014320497
0.177365341	2.023444632
0.178375104	2.031341325
0.17927001	2.038241241
};
\addlegendentry{$\lambda$=0.61}
\end{axis}
\end{tikzpicture}}}
		\caption{\small{
		$\tau$ vs $P_{L}$.
		}}
		\label{Fig:STVsPloss-d-ArrRate-1.51.5}
\end{subfigure}
	\caption{
	We plot the conditional mean response time $\tau$ and the loss probability $P_L$ for the policy $\pi(d,T,T)$ as a function of number of replicas $d$, for a fixed discard threshold $T = 1.5$, the number of servers $N=20$, service rate $\mu=1$, and  different values of arrival rate $\lambda \in \set{0.01, 0.11, \dots, 0.61}$.  
}
\label{Fig:sameDeadLine-xVsd-lam}
\end{figure*}

In Fig.~\ref{Fig:sameDeadLine-xVsd-lam}, we plot the behavior of conditional mean response time $\tau$ and the loss probability $P_L$ for $\pi(d, T, T)$ as the number of replicas $d$ increases.  
We choose the number of servers $N=20$ and discard threshold $T=1.5$.  
Such a study is relevant for determining the ideal choice for the number of repicas $d$ for a given arrival rate.
Here are the main observations. 
\begin{compactenum}
\item  Fig.~\ref{Fig:STVsd-ArrRate-1.51.5} shows that there is an optimal number of replicas $d$ that minimizes the conditional mean response time for each arrival rate. 
In addition, the optimal number of replicas $d$ decreases with increase in arrival rate. This is expected as when the system load increases with a finite value for both the thresholds, the chances of replicas getting cancelled increases.
Even though, increasing the number of replicas ensures that more copies of the job are serviced in parallel, 
it results in an increase in load at individual servers. 
Thus, beyond a certain threshold, it can result in an increase in the conditional response time.
\item Fig.~\ref{Fig:PlossVsd-ArrRate-1.51.5} demonstrates that there is again an optimal number of replicas $d$ which minimizes the loss probability for each arrival rate. 
For small number of replicas, there is a high probability of the job getting canceled, 
since we are sampling less servers. 
However, larger number of replicas $d$ can cause increase in workload at the servers, 
which again results in an increase in the cancellation of replicas. 
Server workloads increase with arrival rate, and hence the loss probability increases with arrival rate. 
\item From the tradeoff presented in Fig.~\ref{Fig:STVsPloss-d-ArrRate-1.51.5}, it is clear that we can determine a suitable replication factor $d$ that minimizes both the conditional mean response time and the loss probability simultaneously for each value of arrival rate. 
Further, this optimal number of replicas decreases with increase in arrival rate.
\end{compactenum}

\begin{figure*}[h!]%
\centering
\begin{subfigure}{.33\columnwidth}
		\centerline{\scalebox{.6}{\input{Figures/STVsT-d-0.3}}}
		\caption{\small{$\tau$ vs threshold $T$.}}
		\label{Fig:STVsT-d-0.3}
\end{subfigure}
\begin{subfigure}{.33\columnwidth}
		\centerline{\scalebox{.6}{\input{Figures/PlossVsT-d-0.3}}}
		\caption{\small{$P_{L}$ vs threshold $T$.}
		\label{Fig:PlossVsT-d-0.3}}
\end{subfigure}
\begin{subfigure}{.33\columnwidth}
		\centerline{\scalebox{.6}{\input{Figures/STVsPloss-d-TT}}}
		\caption{\small{$\tau$ vs $P_{L}$ tradeoff.}
		\label{Fig:STVsPloss-d-TT}}
\end{subfigure}%
	\caption{\small{
	We plot the conditional mean response time $\tau$ and the loss probability $P_L$ for the policy $\pi(d,T,T)$ as a function of discard threshold $T$, for the number of servers $N=20$, arrival rate $\lambda=0.3$, service rate $\mu=1$, 
	and for the number of replicas $d \in \set{1,3,6,9}$.  
	}}
\label{Fig:sameDeadLine-xVsT-d-0}
\end{figure*}
\begin{figure*}[h!]%
\centering
\begin{subfigure}{.33\columnwidth}
		\centerline{\scalebox{.6}{\input{Figures/STVsPloss-d-TT-0.3}}}
		\caption{\small{$\tau$ vs $P_{L}$ tradeoff, $\lambda = 0.3$.}}
		\label{Fig:STVsPloss-d-TT-0.3}
\end{subfigure}
\begin{subfigure}{.33\columnwidth}
		\centerline{\scalebox{.6}{\input{Figures/STVsPloss-d-TT-0.4}}}
		\caption{\small{$\tau$ vs $P_{L}$ tradeoff, $\lambda = 0.4$.}
		\label{Fig:STVsPloss-d-TT-0.4}}
\end{subfigure}
\begin{subfigure}{.33\columnwidth}
		\centerline{\scalebox{.6}{\input{Figures/STVsPloss-d-TT-0.5}}}
		\caption{\small{$\tau$ vs $P_{L}$ tradeoff, $\lambda = 0.5$.}
		\label{Fig:STVsPloss-d-TT-0.5}}
\end{subfigure}%
	\caption{\small{
	For each arrival rate $\lambda \in \set{0.3, 0.4, 0.5}$, we plot the tradeoff between conditional mean response time $\tau$ and the loss probability $P_L$ for the policy $\pi(d,T,T)$ as a function of discard threshold $T$, for the number of servers $N=20$, service rate $\mu=1$, and for the number of replicas $d \in \set{1,3,6,9}$.
}}
\label{Fig:sameDeadLine-xVsT-d-1}
\end{figure*}

In Fig.~\ref{Fig:sameDeadLine-xVsT-d-0}, 
we plot the behavior of the conditional mean response time $\tau$ and the loss probability $P_L$ for the $\pi(d, T, T)$ policy as a function of the discard threshold $T$. 
We choose the number of servers $N=20$, the normalized arrival rate $\lambda = 0.3$, 
and the number of replicas $d \in \set{1,3,6,9}$. 
We list out our main observations below.
\begin{compactenum}
\item From Fig.~\ref{Fig:STVsT-d-0.3}, 
we see that the discard threshold $T$ that minimizes the conditional mean response time varies with the choice of replication factor $d$. 
Since less replicas will be discarded as the threshold $T$ increases, 
we expect the loss probability $P_L$ to decrease with discard threshold $T$. 
We verify this behavior in Fig.~\ref{Fig:PlossVsT-d-0.3}.  
\item When the discard threshold $T \in [0,1]$, we see significant reduction in conditional mean response time under $\pi(d,T,T)$ when compared to random routing. 
This gain comes at the cost of a nominal loss probability $P_L$ for $d \ge 3$. 
In fact, the maximum loss probability is observed to be around 0.095  for $d=3$.
\item The tradeoff curve in Fig.~\ref{Fig:STVsPloss-d-TT} helps in determining the best discard threshold $T$ for a fixed replication factor $d$.
It suggests that with an increase in the number of replicas, decreasing discard threshold could be beneficial as the corresponding increase in loss probabilities are nominal. 
It also provides a comparison with the conditional mean response time for the JSW($d$) policy. For the considered arrival rate, it shows that the proposed policy beats JSW($d$) policy if a loss is allowed and this loss to be admitted increases with $d$. 
To be specific, the loss percentage to be borne while using the proposed policy in order to provide a better performance than JSW($d$) policy are $0.65, 1.35$ and $1.8$ when the number of replicas are $3$, $6$ and $9$ respectively for a normalized arrival rate of $0.3$.
\end{compactenum}

Figure~\ref{Fig:sameDeadLine-xVsT-d-1} presents similar plots as Fig.~\ref{Fig:STVsPloss-d-TT} but for different normalized arrival rates. They show that the loss probability to be admitted by the proposed policy in order to provide a competitive performance to that of JSW($d$) policy increases with the increase in arrival rate.
\begin{figure*}[h!]%
\centering
\begin{subfigure}{.33\columnwidth}
		\centerline{\scalebox{.6}{\input{Figures/STVsAR-d-1.51.5}}}
		\caption{\small{$\tau$ vs $\lambda$.}}
		\label{Fig:STVsAR-d-1.51.5}
\end{subfigure}
\begin{subfigure}{.33\columnwidth}
		\centerline{\scalebox{.6}{
\begin{tikzpicture}

\definecolor{color0}{rgb}{0.83921568627451,0.152941176470588,0.156862745098039}
\definecolor{color1}{rgb}{0.12156862745098,0.466666666666667,0.705882352941177}
\definecolor{color2}{rgb}{0.172549019607843,0.627450980392157,0.172549019607843}
\definecolor{color3}{rgb}{1,0.498039215686275,0.0549019607843137}

\begin{semilogxaxis}[
legend cell align={left},
legend style={fill opacity=0.8, draw opacity=1, text opacity=1, at={(0.03,0.985)}, anchor=north west, draw=white!80!black},
tick align=outside,
tick pos=left,
x grid style={white!69.01960784313725!black, densely dotted},
xlabel={Arrival rate $\lambda$},
xmajorgrids,
xmin=0.01, xmax=1.2,
xminorgrids,
xtick style={color=black},
y grid style={white!69.0196078431373!black},
ylabel={Loss probability $P_{L}$},
ymajorgrids,
y grid style={white!69.01960784313725!black, densely dotted},
yminorgrids,
ytick style={color=black},
]
\addplot [violet, mark=asterisk, mark size=0.75, mark options={solid}, line width=0.8pt, thick]
table {%
0 0
0.2 0
0.4 0
0.6 0
0.8 0
1 0
};
\addlegendentry{Random Routing}
\addplot [color0, mark=asterisk, mark size=0.75, mark options={solid}, line width=0.8pt, thick]
table {%
0.01 0.00224242396412466
0.06 0.0137817942068013
0.11 0.025845482923718
0.16 0.0384019586385175
0.21 0.0514165615802725
0.26 0.064851927396871
0.31 0.0786684477576809
0.36 0.0928247561958179
0.41 0.107278227173603
0.46 0.121985476472752
0.51 0.136902851581797
0.56 0.151986901721247
0.61 0.167194818437206
0.66 0.182484839220041
0.71 0.197816608274524
0.76 0.213151490291723
0.81 0.228452834768132
0.86 0.243686190013257
0.91 0.258819467426622
0.96 0.273823057868724
};
\addlegendentry{$d=1$}
\addplot [color1, mark=asterisk, mark size=0.75, mark options={solid}, line width=0.8pt, thick]
table {%
0.01 3.1351423623871e-07
0.06 8.26247011144199e-05
0.11 0.000598918521003133
0.16 0.00209369625697631
0.21 0.0052049944018816
0.26 0.0105435140500709
0.31 0.0185755247057416
0.36 0.0295454806533088
0.41 0.043454003294073
0.46 0.0600853200749514
0.51 0.0790653480366349
0.56 0.0999294744528463
0.61 0.122184177531756
0.66 0.145354059738662
0.71 0.169012225149228
0.76 0.19279590259611
0.81 0.216410930153157
0.86 0.239628898710252
0.91 0.262280149644116
0.96 0.284244967124355
};
\addlegendentry{$d=3$}
\addplot [color2, mark=asterisk, mark size=0.75, mark options={solid}, line width=0.8pt, thick]
table {%
0.01 6.85229650798647e-12
0.06 6.39709651806264e-07
0.11 3.69284207606402e-05
0.16 0.000421521854555684
0.21 0.00216900343334303
0.26 0.00690613673255003
0.31 0.0160679548647522
0.36 0.0301957764333499
0.41 0.0488818982666204
0.46 0.071145470788247
0.51 0.0958447273425772
0.56 0.121938704285769
0.61 0.148595081554465
0.66 0.175202373744293
0.71 0.201339140927038
0.76 0.226731572247429
0.81 0.251213860597214
0.86 0.274696172029561
0.91 0.297140582878779
0.96 0.318543743269623
};
\addlegendentry{$d=6$}
\addplot [black, mark=asterisk, mark size=0.75, mark options={solid}, line width=0.8pt, thick]
table {%
0.01 7.7715611723761e-16
0.06 3.00236168149226e-08
0.11 1.15516926159076e-05
0.16 0.00030645731928669
0.21 0.00226706572702817
0.26 0.00827628325208785
0.31 0.0199363904474208
0.36 0.0371776348736795
0.41 0.0588319701753104
0.46 0.0834483632933112
0.51 0.109745459318449
0.56 0.136746644540173
0.61 0.163764630515114
0.66 0.190340662121109
0.71 0.21618256168129
0.76 0.241114770262449
0.81 0.265041529489983
0.86 0.287920823782659
0.91 0.309746283571899
0.96 0.330534728818797
};
\addlegendentry{$d=9$}
\end{semilogxaxis}

\end{tikzpicture}}}
		\caption{\small{Loss probability $P_{L}$ vs $\lambda$.}}
		\label{Fig:PlossVsT-d-1.51.5}
\end{subfigure}
\begin{subfigure}{.33\columnwidth}
		\centerline{\scalebox{.6}{
\begin{tikzpicture}

\definecolor{color0}{rgb}{0.83921568627451,0.152941176470588,0.156862745098039}
\definecolor{color1}{rgb}{0.12156862745098,0.466666666666667,0.705882352941177}
\definecolor{color2}{rgb}{0.172549019607843,0.627450980392157,0.172549019607843}
\definecolor{color3}{rgb}{1,0.498039215686275,0.0549019607843137}

\begin{semilogxaxis}[
legend cell align={left},
legend style={fill opacity=0.8, draw opacity=1, text opacity=1, at={(0.03,0.985)}, anchor=north west, draw=white!80!black},
tick align=outside,
tick pos=left,
x grid style={white!69.01960784313725!black, densely dotted},
xlabel={Loss probability $P_{L}$},
xmajorgrids,
xmin=-0.9, xmax=1.2,
xminorgrids,
xtick style={color=black},
y grid style={white!69.01960784313725!black, densely dotted},
ylabel={Conditional mean response time $\tau$},
ymajorgrids,
ymin=0, ymax=2.23555774770039,
yminorgrids,
ytick style={color=black}
]
\addplot [color0, mark=asterisk, mark size=0.75, mark options={solid}, line width=0.8pt, thick]
table {%
0.00224242396412466 1.00442559663977
0.0137817942068013 1.0266638661615
0.025845482923718 1.04906967849157
0.0384019586385175 1.07162034983284
0.0514165615802725 1.09429265754655
0.064851927396871 1.11706294747979
0.0786684477576809 1.13990724491965
0.0928247561958179 1.16280136804727
0.107278227173603 1.18572104272948
0.121985476472752 1.20864201747009
0.136902851581797 1.23154017734865
0.151986901721247 1.25439165580013
0.167194818437206 1.27717294313461
0.182484839220041 1.29986099075971
0.197816608274524 1.32243331014902
0.213151490291723 1.34486806569417
0.228452834768132 1.36714416068481
0.243686190013257 1.38924131577625
0.258819467426622 1.41114013942652
0.273823057868724 1.43282218990997
};
\addlegendentry{$d=1$}
\addplot [color1, mark=asterisk, mark size=0.75, mark options={solid}, line width=0.8pt, thick]
table {%
3.1351423623871e-07 0.343604037258989
8.26247011144199e-05 0.40159949168478
0.000598918521003133 0.470822989676509
0.00209369625697631 0.550612372174624
0.0052049944018816 0.63928378331482
0.0105435140500709 0.734425349167595
0.0185755247057416 0.833308138571134
0.0295454806533088 0.933274198178965
0.043454003294073 1.03201148345267
0.0600853200749514 1.12769257619302
0.0790653480366349 1.21900079877369
0.0999294744528463 1.30508456928308
0.122184177531756 1.38547760982267
0.145354059738662 1.4600112206975
0.169012225149228 1.5287333399449
0.19279590259611 1.59184071820704
0.216410930153157 1.64962551374547
0.239628898710252 1.7024351009276
0.262280149644116 1.75064292302574
0.284244967124355 1.7946280807084
};
\addlegendentry{$d=3$}
\addplot [color2, mark=asterisk, mark size=0.75, mark options={solid}, line width=0.8pt, thick]
table {%
6.85229650798647e-12 0.177118484485901
6.39709651806264e-07 0.245141776387569
3.69284207606402e-05 0.345600080769497
0.000421521854555684 0.48245241150661
0.00216900343334303 0.649918608146881
0.00690613673255003 0.833354917006263
0.0160679548647522 1.01615580546505
0.0301957764333499 1.18598454152783
0.0488818982666204 1.3365545498804
0.071145470788247 1.46632760162665
0.0958447273425772 1.57650405541016
0.121938704285769 1.66947193237102
0.148595081554465 1.74789545794877
0.175202373744293 1.81427293101845
0.201339140927038 1.8707679478457
0.226731572247429 1.91917850654603
0.251213860597214 1.96096642381999
0.274696172029561 1.99730732143574
0.297140582878779 2.02914270928132
0.318543743269623 2.05722665804052
};
\addlegendentry{$d=6$}
\addplot [black, mark=asterisk, mark size=0.75, mark options={solid}, line width=0.8pt, thick]
table {%
7.7715611723761e-16 0.121816066586268
3.00236168149226e-08 0.20260183748572
1.15516926159076e-05 0.346344972945204
0.00030645731928669 0.558168956679853
0.00226706572702817 0.80787434891696
0.00827628325208785 1.05254672532416
0.0199363904474208 1.26524989705409
0.0371776348736795 1.43890613183614
0.0588319701753104 1.57721468911053
0.0834483632933112 1.68705494710737
0.109745459318449 1.7750195570438
0.136746644540173 1.84640983632969
0.163764630515114 1.90521399403162
0.190340662121109 1.95436179079443
0.21618256168129 1.99599773001519
0.241114770262449 2.03170191585202
0.265041529489983 2.06265226339873
0.287920823782659 2.08973919750112
0.309746283571899 2.11364573214849
0.330534728818797 2.13490338193305
};
\addlegendentry{$d=9$}
\end{semilogxaxis}

\end{tikzpicture}}}
		\caption{\small{$\tau$ vs $P_{L}$ tradeoff.}}
		\label{Fig:STVsPloss-d-1.51.5}
\end{subfigure}%
	\caption{\small{
	We plot the conditional mean response time $\tau$ and the loss probability $P_L$ as the normalized arrival rate $\lambda $ increases under policy $\pi(d,T,T)$ for the number of servers $N=20$, discard threshold $T=1.5$, service rate $\mu=1$, and the number of replicas $d \in \set{1,3,6,9}$. 
	}}
\label{Fig:sameDeadLine-xVsT-ARate}
\end{figure*}
In Fig.~\ref{Fig:sameDeadLine-xVsT-ARate}, we study the behavior of conditional mean response time $\tau$ and the loss probability $P_L$ for $\pi(d, T, T)$ as the normalized arrival rate $\lambda$ increases.  
We choose the number of servers $N=20$, discard threshold $T=1.5$, 
and the number of replicas $d \in \set{1,3,6,9}$.  
We list our observations and inferences below which are similar for other discard thresholds. 
\begin{compactenum}
\item Fig.~\ref{Fig:STVsAR-d-1.51.5} shows that the conditional mean response time for $\pi(d,T,T)$ policy for $d > 1$ is uniformly smaller than random routing for all arrival rates. 
This performance improvements comes at the cost of some nominal loss probability for low arrival rates. 
\item Since the $\pi(d,T,T)$ policy admits loss, we observe from Fig.~\ref{Fig:STVsAR-d-1.51.5} that the conditional response time remains bounded even for higher arrival rates. 
However, this property results in a non trivial loss probability for higher arrival rates, as seen in Fig.~\ref{Fig:PlossVsT-d-1.51.5}. 
\item From the tradeoff in Fig~\ref{Fig:STVsPloss-d-1.51.5}, we again infer that as arrival rate increases, it is wiser to switch to lower number of replicas. 
\end{compactenum}

\begin{remark}
As mentioned earlier, this policy is to be adopted only in applications that can tolerate a certain amount of loss as in streaming applications.
In addition, the optimal value of policy parameters $d$ and $T$ depends on the application, especially on the minimum tolerable loss probability for the given application. We also note that a joint optimization of these parameters is difficult to perform analytically. However, in practice, one can always use the derived expressions to find the best operating point through grid search like approaches.
\end{remark}

\subsection{Replication with no loss}
\label{subsec:RepNoLoss}

We next study the $N$ server system under the replication with no loss policy. 
Specifically, we assume that the primary discard threshold $T_1 = \infty$, 
and the secondary discard threshold $T_2 < T_1$ is finite. 
In this case, the system is stable if and only if $\lambda < \mu$.  
First, we obtain the following result from Corollary~\ref{cor:GeneralCase} by substituting $T_1 = \infty$. 
\begin{corollary}
\label{cor:F(w)NoLoss}
For an $N$ server system with \iid exponential service times of rate $\mu$ 
and Poisson arrivals of rate $N\lambda$,
the stationary workload distribution at the cavity queue under $\pi(d,\infty, T_2)$ policy exists only for $\lambda < \mu$, 
and is given by
\EQ{
\begin{cases}
F(0)(\frac{\mu}{\mu-\bl} - \frac{\bl}{\mu - \bl}e^{-(\mu - \bl )w}),& w \le T_2\\
F(T_2) + \frac{\bl}{\mu - \lambda}F(0) e^{(\bl  - \lambda) T_2} (e^{-(\mu - \lambda)T_2} - e^{-(\mu - \lambda)w}), &w > T_2.
\end{cases}
}
where   
$
F(0) = \frac{(1 - \frac{\lambda}{\mu})(1 - \frac{\bl}{\mu})}{(1 - \frac{\lambda}{\mu}) + \frac{\bl}{\mu}(\frac{\lambda}{\mu} - \frac{\bl}{\mu})e^{-(\mu - \bl)T_2}}. 
$ 
\end{corollary}
\begin{remark}
Note that the loss probability is $0$ under this policy.
Then, from Theorem~\ref{thm:MeanResponseTime}, we have
\EQN{
\label{eq:MRTNoLoss}
\tau =\int_x   \Big[ k(x,\infty)( \bF(T_2) +k(x,T_2))^{d-1} 
 \Big] dx.
} 
\end{remark}
The next lemma provides us with the terms $k(x,T_2)$, $k(x,\infty)$ and $\bF(T_2)$ that enable us to compute the mean response time $\tau$ under the scheduling policy $\pi(d,\infty,T_2)$.
Note that, we provide the results only for the regime of arrival rates where the system is stable, 
that is, when $\lambda < \mu$.
\begin{lemma}
\label{lem:k(x,T)NoLoss}
For a stable $N$ server system with \iid exponential service times of rate $\mu$ and Poisson arrivals of rate $N\lambda$, 
the function $k(x,T_2)$ under the $\pi(d,\infty, T_2)$ policy is 
\EQ{
k(x,T_2) = 
\begin{cases}
F(0)e^{-\mu x}e^{\bl  T_2},& x \ge T_2\\
F(0)\Big[\frac{\mu}{\mu - \bl } e^{-(\mu - \bl )x} - \frac{\bl }{\mu - \bl } e^{-(\mu - \bl )T_2}\Big], & x < T_2. 
\end{cases}
}
We can also find the function $k(x,\infty)$ as 
\EQ{
 k(x,T_2) + 
\begin{cases}
F(0)\bl e^{(\bl -\lambda)T_2}e^{-\mu x} \Big[\frac{e^{\lambda x} - e^{\lambda T_2}}{\lambda} + \frac{e^{\lambda x}}{\mu-\lambda}\Big],& x \ge T_2\\
\frac{\bl}{\mu - \lambda}F(0)e^{-(\mu -\bl)T_2},& x < T_2.  
\end{cases}
}
 where
$F(0) = \Big[\bl \Big(\frac{1 - e^{-(\mu - \bl )T_2}}{\mu - \bl } + \frac{e^{-(\mu - \bl )T_2}}{\mu - \lambda} \Big)+1 \Big]^{-1}$.
\end{lemma}
\begin{proof}
Since the service time is exponentially distributed with rate $\mu$, 
we get $\bG(x) = e^{-\mu (x)_+}$. 
Therefore, we can write the function 
$
k(x,T) 
= \expect{\bG(x-W)\SetIn{W \le T}} = F(T) - F(T\wedge x) + e^{-\mu x}\int_0^{T\wedge x}e^{\mu w}dF(w). 
$
Setting $T = \infty$ in the above equation, 
we get $k(x,\infty) = F(\infty) - F(x)+e^{-\mu x}\int_0^xe^{\mu w}dF(w).$
Substituting the workload distribution $F$ from Corollary~\ref{cor:F(w)NoLoss}, 
we get the result.
\end{proof}

\begin{figure*}[h!]%
\centering
\begin{subfigure}{0.5\columnwidth}
		\centerline{\scalebox{0.85}{
\begin{tikzpicture}

\definecolor{color0}{rgb}{0.83921568627451,0.152941176470588,0.156862745098039}
\definecolor{color1}{rgb}{0.12156862745098,0.466666666666667,0.705882352941177}
\definecolor{color2}{rgb}{0.172549019607843,0.627450980392157,0.172549019607843}
\definecolor{color3}{rgb}{1,0.498039215686275,0.0549019607843137}

\begin{semilogyaxis}[
legend cell align={left},
legend style={fill opacity=0.8, draw opacity=1, text opacity=1, at={(0.03,0.97)}, anchor=north west, draw=white!80!black},
tick align=outside,
tick pos=left,
x grid style={white!69.01960784313725!black, densely dotted},
xmajorgrids,
xlabel={Arrival rate $\lambda$},
xmin=-0.039, xmax=1.039,
xminorgrids,
xtick style={color=black},
xtick={-0.2,0,0.2,0.4,0.6,0.8,1,1.2},
ymajorgrids,
yminorgrids,
y grid style={white!69.01960784313725!black, densely dotted},
ylabel={Conditional mean response time $\tau$},
ymin=0, ymax=140,
ytick style={color=black}
]
\addplot [color0, mark=asterisk, mark size=0.75, mark options={solid}, line width=0.8pt, thick]
table {%
0.01 1.01010101010101
0.03 1.03092783505155
0.05 1.05263157894737
0.07 1.0752688172043
0.09 1.0989010989011
0.11 1.12359550561798
0.13 1.14942528735632
0.15 1.17647058823529
0.17 1.20481927710843
0.19 1.23456790123457
0.21 1.26582278481013
0.23 1.29870129870071
0.25 1.33333333333333
0.27 1.36986301369863
0.29 1.40845070422535
0.31 1.44927536231884
0.33 1.49253731343284
0.35 1.53846153846154
0.37 1.5873015873016
0.39 1.63934426229508
0.41 1.69491525423729
0.43 1.75438596491228
0.45 1.8181818181818
0.47 1.8867924528301
0.49 1.96078431372549
0.51 2.04081632653061
0.53 2.12765957446809
0.55 2.22222222222222
0.57 2.32558139534884
0.59 2.4390243902439
0.61 2.56410256410256
0.63 2.7027027027027
0.65 2.85714285714285
0.67 3.03030303030303
0.69 3.2258064516129
0.71 3.44827586206897
0.73 3.70370370370357
0.75 4
0.77 4.34782608695652
0.79 4.76190476190476
0.81 5.26315789473684
0.83 5.88235294117646
0.85 6.66666666666667
0.87 7.69230769230769
0.89 9.09090909090909
0.91 11.1111111111111
0.93 14.2857142857143
0.95 20
0.97 33.3333333333333
0.99 100.000000000001
};
\addlegendentry{$d=1$}
\addplot [color1, mark=asterisk, mark size=0.75, mark options={solid}, line width=0.8pt, thick]
table {%
0.01 0.34360848276871
0.03 0.365858638892182
0.05 0.39060782932208
0.07 0.418157564753661
0.09 0.448833543309974
0.11 0.482982613009756
0.13 0.520967967403964
0.15 0.563162444282538
0.17 0.609939971864472
0.19 0.661665433333362
0.21 0.718683475331874
0.23 0.781307032457825
0.25 0.849806533952372
0.27 0.924400858583481
0.29 1.00525107971529
0.31 1.09245788663858
0.33 1.18606329794813
0.35 1.28605693886045
0.37 1.3923867926984
0.39 1.50497401637795
0.41 1.62373118107681
0.43 1.74858319563921
0.45 1.87949020448225
0.47 2.01647191832201
0.49 2.15963311758579
0.51 2.30919044449564
0.53 2.46550105623458
0.55 2.62909424889825
0.57 2.80070780375485
0.59 2.98133160967153
0.61 3.17226218005633
0.63 3.37517317896284
0.65 3.5922092770184
0.67 3.82611403369413
0.69 4.08040782868935
0.71 4.35964050694888
0.73 4.66975782863865
0.75 5.01864566294197
0.77 5.41696019668006
0.79 5.87943480047678
0.81 6.42701464311411
0.83 7.09050051614794
0.85 7.91711018182965
0.87 8.98309884717681
0.89 10.4201499746956
0.91 12.4769564844667
0.93 15.6863725773558
0.95 21.4337332275139
0.97 34.7984561299593
0.99 101.494877902713
};
\addlegendentry{$d=3$}
\addplot [color2, mark=asterisk, mark size=0.75, mark options={solid}, line width=0.8pt, thick]
table {%
0.01 0.177206617992375
0.03 0.201937214031062
0.05 0.232562081472575
0.07 0.270487522762788
0.09 0.317257060615679
0.11 0.374419712467351
0.13 0.443332681203365
0.15 0.52493076917285
0.17 0.619522468297502
0.19 0.726678183337217
0.21 0.845250817728254
0.23 0.973524404232039
0.25 1.10944619630067
0.27 1.2508800865743
0.29 1.39582701307083
0.31 1.54258082274616
0.33 1.68981252209747
0.35 1.83659322044094
0.37 1.98237405636072
0.39 2.12694195748283
0.41 2.27036655027569
0.43 2.41294873761375
0.45 2.55517709649085
0.47 2.69769501634606
0.49 2.84127947616811
0.51 2.98683133889361
0.53 3.13537675268002
0.55 3.28807946174396
0.57 3.44626439836878
0.59 3.61145379534669
0.61 3.78541825631903
0.63 3.97024687890704
0.65 4.16844289702415
0.67 4.38305484191054
0.69 4.61785869032234
0.71 4.87761523509394
0.73 5.16844144671057
0.75 5.49835953071277
0.77 5.8781317898393
0.79 6.32257182850646
0.81 6.85268312910731
0.83 7.49930643399641
0.85 8.30968424189267
0.87 9.36008403383373
0.89 10.7821914623618
0.91 12.8246935883475
0.93 16.0204324281774
0.95 21.7547258933613
0.97 35.1069709415528
0.99 101.791480769894
};
\addlegendentry{$d=6$}
\addplot [color3, mark=asterisk, mark size=0.75, mark options={solid}, line width=0.8pt, thick]
table {%
0.01 0.121938285548633
0.03 0.149556300954577
0.05 0.187812750421398
0.07 0.240650229366747
0.09 0.312399186062237
0.11 0.406706206208663
0.13 0.525055005982342
0.15 0.665684375023367
0.17 0.823710937990175
0.19 0.992461388334023
0.21 1.16521761567327
0.23 1.33652887565226
0.25 1.50275686700171
0.27 1.66199665151602
0.29 1.81368460071312
0.31 1.9581467995889
0.33 2.09622035918316
0.35 2.22898639055195
0.37 2.35760490494138
0.39 2.48322583054836
0.41 2.6069505541309
0.43 2.72982430060183
0.45 2.8528461231989
0.47 2.97698847568852
0.49 3.10322200994422
0.51 3.23254367491368
0.53 3.36600778206328
0.55 3.50476079549146
0.57 3.65008147419409
0.59 3.80342885633197
0.61 3.96650162836261
0.63 4.14131388735668
0.65 4.33029448539544
0.67 4.53642050961063
0.69 4.76340078119417
0.71 5.01593391339059
0.73 5.30007991731732
0.75 5.62380921708035
0.77 5.99783729279901
0.79 6.43693556468893
0.81 6.96206959869883
0.83 7.60404610117004
0.85 8.41007703265325
0.87 9.45640246898329
0.89 10.8746834505188
0.91 12.9135849073467
0.93 16.1059289240764
0.95 21.8370154256093
0.97 35.1862251012733
0.99 101.867856396122
};
\addlegendentry{$d=9$}
\addplot [black, mark=asterisk, mark size=0.75, mark options={solid}, line width=0.8pt, thick]
table {%
0.01 0.094460047634223
0.03 0.125387692578125
0.05 0.173395538361517
0.07 0.246997430083601
0.09 0.354874600566214
0.11 0.501198017769898
0.13 0.681252195152926
0.15 0.882116169400202
0.17 1.08833380783126
0.19 1.28760695370047
0.21 1.47307647523553
0.23 1.64262955630031
0.25 1.79712158376446
0.27 1.93878748862272
0.29 2.07022912833445
0.31 2.19390519119985
0.33 2.31193924909213
0.35 2.42609373709711
0.37 2.53781565008717
0.39 2.64830455893525
0.41 2.75858060225762
0.43 2.86954434288354
0.45 2.98202709150369
0.47 3.09683309733043
0.49 3.2147759572353
0.51 3.3367117845197
0.53 3.46357164137071
0.55 3.59639573332036
0.57 3.73637202367096
0.59 3.88488234486512
0.61 4.04355986743277
0.63 4.21436309337565
0.65 4.39967363020284
0.67 4.60242831636695
0.69 4.82630157276927
0.71 5.0759624998224
0.73 5.35744568429988
0.75 5.67869955283962
0.77 6.05042046811216
0.79 6.48736316098939
0.81 7.01047856233497
0.83 7.65056048777956
0.85 8.4548094939657
0.87 9.49945552482023
0.89 10.9161505809795
0.91 12.9535514938421
0.93 16.1444730627343
0.95 21.8742086324019
0.97 35.2221329248172
0.99 101.902538951336
};
\addlegendentry{$d=12$}
\end{semilogyaxis}

\end{tikzpicture}}}
		\caption{\small{$\tau$ vs arrival rate $\lambda$ for discard threshold $T_2 = 2$.}}
		\label{Fig:STVsAR-d-Inf2}
\end{subfigure}
\begin{subfigure}{0.5\columnwidth}
		\centerline{\scalebox{0.85}{
\begin{tikzpicture}

\definecolor{color0}{rgb}{0.83921568627451,0.152941176470588,0.156862745098039}
\definecolor{color1}{rgb}{0.12156862745098,0.466666666666667,0.705882352941177}
\definecolor{color2}{rgb}{0.172549019607843,0.627450980392157,0.172549019607843}
\definecolor{color3}{rgb}{1,0.498039215686275,0.0549019607843137}

\begin{semilogyaxis}[
legend cell align={left},
legend style={fill opacity=0.8, draw opacity=1, text opacity=1, at={(0.03,0.97)}, anchor=north west, draw=white!80!black},
tick align=outside,
tick pos=left,
x grid style={white!69.01960784313725!black, densely dotted},
xlabel={Threshold $T_2$},
xmajorgrids,
xmin=-0.195, xmax=4.095,
xminorgrids,
xtick style={color=black},
y grid style={white!69.01960784313725!black, densely dotted},
ylabel={Conditional mean response time $\tau$},
ymajorgrids,
ymin=0.68, ymax=4.2,
yminorgrids,
ytick style={color=black}
]
\addplot [color0, mark=asterisk, mark size=0.75, mark options={solid}, line width=0.8pt, thick]
table {%
0 1.42857142857142
0.1 1.42857142857142
0.2 1.42857142857142
0.3 1.42857142857142
0.4 1.42857142857142
0.5 1.42857142857142
0.6 1.42857142857142
0.7 1.42857142857142
0.8 1.42857142857142
0.9 1.42857142857142
1 1.42857142857142
1.1 1.42857142857142
1.2 1.42857142857143
1.3 1.42857142857143
1.4 1.42857142857143
1.5 1.42857142857143
1.6 1.42857142857143
1.7 1.42857142857143
1.8 1.42857142857143
1.9 1.42857142857143
2 1.42857142857143
2.1 1.42857142857143
2.2 1.42857142857143
2.3 1.42857142857143
2.4 1.42857142857143
2.5 1.42857142857143
2.6 1.42857142857143
2.7 1.42857142857143
2.8 1.42857142857143
2.9 1.42857142857143
3 1.42857142857143
3.1 1.42857142857143
3.2 1.42857142857143
3.3 1.42857142857143
3.4 1.42857142857143
3.5 1.42857142857143
3.6 1.42857142857143
3.7 1.42857142857143
3.8 1.42857142857143
3.9 1.42857142857143
};
\addlegendentry{$d=1$}
\addplot [color1, mark=asterisk, mark size=0.75, mark options={solid}, line width=0.8pt, thick]
table {%
0 0.901991925789156
0.1 0.874792731912408
0.2 0.857390021973573
0.3 0.847775304649869
0.4 0.844429430717959
0.5 0.846189568385017
0.6 0.852155803340448
0.7 0.861624607357231
0.8 0.874040810541946
0.9 0.888962500561461
1 0.906035073744746
1.1 0.924971846001472
1.2 0.945539420137901
1.3 0.967546539286
1.4 0.990835521316032
1.5 1.01527562228442
1.6 1.04075785454088
1.7 1.06719091101311
1.8 1.09449793735494
1.9 1.12261395884638
2 1.15148381651215
2.1 1.18106050194252
2.2 1.21130380628298
2.3 1.24217921828804
2.4 1.27365702096883
2.5 1.30571154746504
2.6 1.33832056524848
2.7 1.37146476427903
2.8 1.40512732977015
2.9 1.43929358413682
3 1.47395068576127
3.1 1.50908737461881
3.2 1.54469375670821
3.3 1.58076112074096
3.4 1.61728178174859
3.5 1.65424894723252
3.6 1.69165660225758
3.7 1.72949941051855
3.8 1.76777262891819
3.9 1.80647203361044
};
\addlegendentry{$d=4$}
\addplot [color2, mark=asterisk, mark size=0.75, mark options={solid}, line width=0.8pt, thick]
table {%
0 0.823267831107235
0.1 0.794590110664533
0.2 0.781708443632892
0.3 0.780723554293803
0.4 0.788960343766503
0.5 0.804541554407419
0.6 0.826122503738298
0.7 0.852721675299704
0.8 0.883609954887633
0.9 0.918236654632471
1 0.956179144692377
1.1 0.997107952588583
1.2 1.04076219193773
1.3 1.08693201137528
1.4 1.13544589261705
1.5 1.18616134869781
1.6 1.23895803983499
1.7 1.29373263067729
1.8 1.35039491700426
1.9 1.40886488819691
2 1.46907048663062
2.1 1.5309458910216
2.2 1.59443019707891
2.3 1.65946640175413
2.4 1.72600062106202
2.5 1.79398148864593
2.6 1.86335969488074
2.7 1.9340876356531
2.8 2.0061191469458
2.9 2.07940930662465
3 2.1539142888374
3.1 2.22959125951303
3.2 2.3063983038327
3.3 2.38429437840237
3.4 2.46323928231859
3.5 2.54319364247651
3.6 2.62411890939207
3.7 2.70597736055121
3.8 2.78873210889705
3.9 2.87234711455115
};
\addlegendentry{$d=6$}
\addplot [color3, mark=asterisk, mark size=0.75, mark options={solid}, line width=0.8pt, thick]
table {%
0 0.767352107142155
0.1 0.74042036013092
0.2 0.737211752498912
0.3 0.750746332648036
0.4 0.776730938468384
0.5 0.812398216375513
0.6 0.855883663584049
0.7 0.905876314547193
0.8 0.961415426415426
0.9 1.02176836660028
1 1.08635532080777
1.1 1.15470184652587
1.2 1.22640842621258
1.3 1.30113062492724
1.4 1.37856597681526
1.5 1.45844519582372
1.6 1.54052618914564
1.7 1.62458989566519
1.8 1.7104373140811
1.9 1.79788730516349
2 1.88677489588003
2.1 1.97694990762659
2.2 2.06827579357481
2.3 2.16062861195788
2.4 2.25389608988264
2.5 2.34797675053489
2.6 2.44277908848462
2.7 2.53822078526637
2.8 2.63422796193002
2.9 2.73073446780878
3 2.82768120601721
3.1 2.92501549664116
3.2 3.02269047853807
3.3 3.12066455034945
3.4 3.21890085088182
3.5 3.31736677853306
3.6 3.41603354898419
3.7 3.5148757899754
3.8 3.61387117165514
3.9 3.71300007073607
};
\addlegendentry{$d=9$}
\addplot [black, mark=asterisk, mark size=0.75, mark options={solid}, line width=0.8pt, thick]
table {%
0 0.738255881650325
0.1 0.714629187110025
0.2 0.721524575606218
0.3 0.74870925075698
0.4 0.7903529248829
0.5 0.842863764480081
0.6 0.903864753302593
0.7 0.971679116284161
0.8 1.04505839378813
0.9 1.12303241840696
1 1.2048231167588
1.1 1.28979282156216
1.2 1.37741173060435
1.3 1.46723623596637
1.4 1.55889358128129
1.5 1.65207032638621
1.6 1.74650321193187
1.7 1.84197163411639
1.8 1.93829128181311
1.9 2.03530867598504
2 2.13289645241584
2.1 2.23094928210164
2.2 2.32938035116279
2.3 2.42811833627985
2.4 2.52710481930634
2.5 2.62629208952035
2.6 2.72564128578037
2.7 2.82512083446934
2.8 2.92470514283092
2.9 3.02437351114991
3 3.12410923112275
3.1 3.22389884159176
3.2 3.32373151647582
3.3 3.42359856314475
3.4 3.5234930126028
3.5 3.62340928564496
3.6 3.72334292162336
3.7 3.82329035861874
3.8 3.9232487556741
3.9 4.02321584934001
};
\addlegendentry{$d=12$}
\end{semilogyaxis}

\end{tikzpicture}}}
		\caption{\small{$\tau$ vs discard threshold $T_2$ for arrival rate $\lambda = 0.3$}
		\label{Fig:STVsT-d-InfT2-0.3}}
\end{subfigure}
	\caption{\small{
	We plot the mean response time $\tau$ for the policy $\pi(d,\infty,T_2)$ for the number of servers $N=20$, service rate $\mu=1$, and the number of replicas $d$. 
	}}
\label{Fig:xVsST-d}
\end{figure*}

%
%

We compare the mean response time $\tau$ for jobs under policy $\pi(d,\infty,T_2)$ in Fig.~\ref{Fig:xVsST-d} for different number of replicas $d$, 
when the number of servers $N=20$ and the exponential service rates of jobs is $\mu =1$.
We plot $\tau$ as a function of normalized arrival rate $\lambda$ in Fig.~\ref{Fig:STVsAR-d-Inf2}, 
where we select a secondary discard threshold $T_2 = 2$ which is twice the mean service time of a job.  
We plot $\tau$ as a function of secondary discard threshold $T_2$ in Fig.~\ref{Fig:STVsT-d-InfT2-0.3}, 
where we choose the normalized arrival rate $\lambda = 0.3$. 
We list out the observations in the following.
\begin{compactenum}
\item Fig.~\ref{Fig:STVsAR-d-Inf2} shows that the lower replication factor $d$ is preferable for larger arrival rates $\lambda$. 
This is due to the fact that system load increases due to larger number of redundant replicas, 
adversely impacting the mean response time performance at high arrival rates. 
\item Fig.~\ref{Fig:STVsT-d-InfT2-0.3} shows the existence of an optimal discard threshold $T_2$ for a fixed number of replicas $d$, and this optimal threshold decreases with increase in the number of replicas.
\end{compactenum}
To conclude, as the normalized arrival rate increases, it is preferable to decrease the number of replicas while choosing an appropriate value for the secondary discard threshold. 

%
\begin{remark}
\label{rem:Tinfinity}
Let us consider the $\pi(d, \infty, \infty)$ policy, 
which is a special case of $\pi(d, T,T)$ for $T = \infty$ as well as of $\pi(d,\infty,T_2)$ for $T_2 = \infty$.
We note that no jobs are lost in such a system and therefore, the loss probability is zero.
This is a r.w.c policy and has been studied in ~\cite{Vulimiri2015Thesis}. 
Under this policy, the arrival rate to any queue is $\bl$, and hence the system is stable if only if $\bl < \mu$. 
Using Lemma~\ref{lem:identical_th} it can be shown that $k(x,\infty) = e^{-(\mu - \bl)x}$ for this policy under stability.  
Using this, the mean response time for exponential service time distribution can be found to be
$
\tau = \frac{1}{(\mu - \bl)d}.
$
\end{remark}

We plot the mean response time $\tau$ for policy $\pi(d,\infty,\infty)$ as a function of arrival rate $\lambda$ in Fig.~\ref{Fig:STVsAR-d-InfInf}, 
for the number of servers $N=20$, service rate $\mu=1$, and different number of replicas $d$.  
The figure is indicative of the stability condition $\lambda < \frac{1}{d}$ for this policy. 
The performance gain from using larger values of $d$ is also evident, but this comes at a cost of requiring a stricter stability condition. 
Of course, the clear advantage of this policy over random routing ($d=1$) is limited to lower arrival rates. 
At higher arrival rates $\lambda$, the fact that the redundant replicas cannot be canceled adversely impacts the system performance. 
For better clarity, we also provide the percentage improvement of mean response time of the policy $\pi(d, \infty, \infty)$ over random routing policy \emph{across stable regions} in Table~\ref{Tab:NoDeadLineComaprison-d}.

\begin{figure*}[h!]%
\centering
\begin{subfigure}{0.5\columnwidth}
		\centerline{\scalebox{0.85}{\input{Figures/STVsAR-d-InfInf}}}
		\caption{\small{$\tau$ vs arrival rate $\lambda$ for discard threshold $T_2 = \infty$.}}
		\label{Fig:STVsAR-d-InfInf}
\end{subfigure}
\begin{subfigure}{0.5\columnwidth}
		\centerline{\scalebox{0.85}{
\begin{tikzpicture}

\definecolor{color0}{rgb}{0.83921568627451,0.152941176470588,0.156862745098039}
\definecolor{color1}{rgb}{0.12156862745098,0.466666666666667,0.705882352941177}
\definecolor{color2}{rgb}{0.172549019607843,0.627450980392157,0.172549019607843}
\definecolor{color3}{rgb}{1,0.498039215686275,0.0549019607843137}

\begin{semilogyaxis}[
legend cell align={left},
legend style={fill opacity=0.8, draw opacity=1, text opacity=1, at={(0.66,0.55)}, anchor=north west, draw=white!80!black},
tick align=outside,
tick pos=left,
x grid style={white!69.01960784313725!black, densely dotted},
xlabel={Arrival rate $\lambda$},
xmajorgrids,
xmin=-0.0375, xmax=1.0075,
xminorgrids,
xtick style={color=black},
y grid style={white!69.01960784313725!black, densely dotted},
ylabel={Conditional mean response time $\tau$},
ymajorgrids,
ymin=-1.16786106081546, ymax=30,
yminorgrids,
ytick style={color=black}
]
\addplot [violet, mark=asterisk, mark size=0.75, mark options={solid}, line width=0.8pt, thick]
table {%
0.01 1.01010101010101
0.06 1.06382978723404
0.11 1.12359550561798
0.16 1.19047619047619
0.21 1.26582278481013
0.26 1.35135135135135
0.31 1.44927536231884
0.36 1.5625
0.41 1.69491525423729
0.46 1.85185185185185
0.51 2.04081632653061
0.56 2.27272727272727
0.61 2.56410256410256
0.66 2.94117647058823
0.71 3.44827586206896
0.76 4.16666666666667
0.81 5.26315789473684
0.86 7.14285714285714
0.91 11.1111111111111
0.96 25
};
\addlegendentry{d = 1}
\addplot [color0, mark=asterisk, mark size=0.75, mark options={solid}, line width=0.8pt, thick]
table {%
0.01 0.346962082685152
0.06 0.423940024532127
0.11 0.51496678448918
0.16 0.618881594973866
0.21 0.734958533143902
0.26 0.8632465170462
0.31 1.00470046943843
0.36 1.1613058375742
0.41 1.33629408017224
0.46 1.53453149755909
0.51 1.76320117040921
0.56 2.0329986978752
0.61 2.3602889921808
0.66 2.77120060574421
0.71 3.30998473033329
0.76 4.05784842403545
0.81 5.18154812826118
0.86 7.08613640980514
0.91 11.076894662358
0.96 24.9858199344086
};
\addlegendentry{$d=3$}
\addplot [color1, mark=asterisk, mark size=0.75, mark options={solid}, line width=0.8pt, thick]
table {%
0.01 0.178578838485481
0.06 0.246749378661575
0.11 0.332537167031598
0.16 0.437524126005587
0.21 0.560605191055723
0.26 0.700213129055748
0.31 0.85562275742264
0.36 1.02747664334659
0.41 1.21804994692255
0.46 1.43157427633869
0.51 1.67483565250329
0.56 1.95829786369462
0.61 2.2982044786396
0.66 2.72063408438777
0.71 3.26983222741965
0.76 4.02702621079597
0.81 5.1590061470693
0.86 7.07086325722933
0.91 11.0679167694427
0.96 24.9821966086651
};
\addlegendentry{$d=6$}
\addplot [color2, mark=asterisk, mark size=0.75, mark options={solid}, line width=0.8pt, thick]
table {%
0.01 0.122452320007261
0.06 0.187221602987039
0.11 0.269546624697853
0.16 0.373030422208774
0.21 0.497329408542122
0.26 0.640420632409122
0.31 0.80077632846684
0.36 0.978326644241142
0.41 1.1748330849259
0.46 1.39419494220174
0.51 1.64300033985204
0.56 1.93160601752454
0.61 2.27620659663718
0.66 2.70286723366979
0.71 3.25584064694189
0.76 4.0163723078935
0.81 5.15127540415951
0.86 7.06566531393293
0.91 11.0648841039762
0.96 24.9809816832427
};
\addlegendentry{$d=9$}
\addplot [color3, mark=asterisk, mark size=0.75, mark options={solid}, line width=0.8pt, thick]
table {%
0.01 0.0943897016632806
0.06 0.157450201252118
0.11 0.237770591577156
0.16 0.340052391915594
0.21 0.464637671149719
0.26 0.609357985708144
0.31 0.772233129044431
0.36 0.952765123112946
0.41 1.15240471510812
0.46 1.37485266465476
0.51 1.62658159610062
0.56 1.91788796455492
0.61 2.26494062554737
0.66 2.69379949700709
0.71 3.24872356674797
0.76 4.01097042908184
0.81 5.14736776934469
0.86 7.06304572845491
0.91 11.0633601313193
0.96 24.9803728695882
};
\addlegendentry{$d=12$}
\addplot [black, mark=asterisk, mark size=0.75, mark options={solid}, line width=0.8pt, thick]
table {%
0.01 0.0775523199094928
0.06 0.139592178239329
0.11 0.218638748189571
0.16 0.320042692237552
0.21 0.444675981366488
0.26 0.590326134620871
0.31 0.754725407536833
0.36 0.937091804107705
0.41 1.13866905643604
0.46 1.36302666554077
0.51 1.61656193271175
0.56 1.90953281107438
0.61 2.258092362086
0.66 2.68829799478762
0.71 3.24441348589728
0.76 4.00770482128559
0.81 5.14500944678052
0.86 7.06146730517483
0.91 11.0624432855176
0.96 24.9800071477656
};
\addlegendentry{$d=15$}
\end{semilogyaxis}

\end{tikzpicture}}}
		\caption{\small{$\tau$ vs arrival rate $\lambda$ for discard threshold $T_2=0$.}
		\label{Fig:STVsAR-d-Inf0}}
\end{subfigure}
	\caption{\small{
	We plot the mean response time $\tau$ as a function of arrival rate $\lambda$ for $\pi(d,\infty, T_2)$ policy for fixed number of servers $N=20$, service rate $\mu=1$, and different number of replicas $d$. 
	}}
\label{Fig:ArrivalRateVsST-d}
\end{figure*}

%

\begin{table}[hhh]
\centering
\caption{{\small{
Percentage improvement in the mean response time of the policy $\pi(d,\infty,\infty)$ over random routing with the number of servers $N=20$ and service rate $\mu = 1$. 
See Fig.~\ref{Fig:STVsAR-d-InfInf}.}}}
 \begin{tabular}{||c | c | c | c | c ||} 
 \hline
 Replicas & $\lambda=0.1$ &  $\lambda=0.15$ & $\lambda=0.2$ & $\lambda=0.25$  \\ [0.5ex] 
 \hline\hline
 d=2 & 43.6\% & 39.18 \% & 33.19\% & 24.79\%\\
 \hline
 d=3 & 57\% & 48.26\%& 32.91\%& -1\%\\
 \hline
 d=4 & 62.29\% &  46.4\% & -1.91\%& NA\\
 \hline
\end{tabular}
\label{Tab:NoDeadLineComaprison-d}
\end{table}
From the above studies, we observe that introduction of secondary replicas add to the system load and deteriorates the system performance for high arrival rates.
Therefore, in the following section, we study a policy where secondary replications occur only on idle servers.
\subsection{Replication on idle secondary servers }
\label{subsec:RISS}
As mentioned above, we next study the special case of $\pi(d,\infty,T_2)$ policy where the secondary discard threshold $T_2 = 0$.  
In this case, the secondary replicas are added only if the sampled secondary servers are idle. 
Here, we would like to point out a seemingly similar policy which is the Redundant to idle queue (RIQ($d$))~\cite{Gardner2017TON}.
We note that, unlike our policy that utilizes absolutely no feedback information, the RIQ($d$) policy utilizes information about availability of idle servers.
If there are no more than a single idle server, RIQ($d$) policy is identical to the Join Threshold Queue (JTQ($d,T$))~\cite[Section 6.6]{Hellemans19} with threshold $T$ set to zero which we discuss in Section~\ref{sec:NumericalComparison}.
Although the RIQ($d$) policy is studied under a more general service model, the analysis is only approximate and no closed form expressions are provided for the performance metrics under this general model.
More importantly, the analysis for RIQ($d$) policy is only valid in the regime where the number of replicas is much lower than the number of servers.
On the other hand, we provide closed form expressions for mean workload under our proposed policy with \iid exponential service times. 
In addition, we have also provided implicit expressions for general \iid service times and our analysis is not restricted to any regime of any of the system or policy parameters.

The replication on idle secondary servers policy that we discuss here is a special case of replication with no loss policy and 
we can obtain the mean response time directly from the previously obtained result. 
\begin{lemma}
\label{lem:E[R]Policy3}
The mean response time of any job under the dispatching policy $\pi(d,\infty,0)$ when service times of each job is \iid exponential with rate $\mu$ and arrivals are Poisson with rate $N\lambda$, 
is given by 
\eqn{
\label{eq:RISSMRT}
\tau &= \sum_{n=0}^{d-1} {d-1 \choose n}\bF(0)^{d-1 - n}F(0)^{n + 1}\Big[\Big(\frac{d\mu\lambda }{(\mu - \lambda)(\mu(n+1) - \lambda)\lambda} -\frac{\lambda(d-1)}{\lambda \mu (n+1)}\Big) \Big],
}
for $\lambda < \mu$ and $F(0) = \frac{\mu - \lambda}{\mu + \lambda(d-1)}$ and $\bF(0) = 1 - F(0)$.
\end{lemma}
\begin{proof}
Substituting $T_2 = 0$ in Lemma~\ref{lem:k(x,T)NoLoss} and substituting the terms in~\eqref{eq:MRTNoLoss} gives the result.
\end{proof}

\begin{remark}
\label{rem:TRTSimplified}
To better understand the behavior of $\pi(d,\infty,0)$ policy, we can simplify the expression for tail response time distribution for the cavity queue under this policy 
with $N\lambda$ Poisson arrivals and \iid exponential service rate $1$ as 
$
\bH_d(x) = k(x,\infty)( \bF(0) +k(x,0))^{d-1} =  e^{-x}\Big(1+\frac{d(e^{\lambda x}-1)}{(d-1)\lambda+1}\Big)\Big(1-\frac{(1-\lambda)(1-e^{-x})}{(d-1)\lambda+1}\Big)^{d-1}.  
$
\end{remark}
The next Lemma shows that the response time under $\pi(d,\infty,0)$ policy is stochastically decreasing in $d$.
As random routing corresponds to $\pi(d,\infty,0)$ policy for $d=1$, this result shows that the performance of the $\pi(d,\infty,0)$ policy can never be worse than that of random routing.

\begin{lemma}
\label{lem:RRProposedComparison}
The response time under the policy $\pi(d,\infty,0)$ with $N\lambda$ Poisson arrivals and \iid exponential service rate $1$ is stochastically decreasing in $d$.
\end{lemma}
\begin{proof}
In order to show the stochastic ordering, it suffices to show that the tail response time follows $\bH_{d+1}(x) \le \bH_d(x)$ for all $x \in \R_+$ and $d \in \N$.
To this end, we first observe from Remark~\ref{rem:TRTSimplified} that $\bH_d(x) = e^{-x}e^{f(d,x)}$ where the function $f:[1,\infty)\times\R_+\to \R$ can be defined for each $y \ge 1$ and $x \in \R_+$ as 
\begin{equation}
\label{eqn:f(y,x)}
f(y,x) \triangleq 
(y-1)\ln(y\lambda+(1-\lambda)e^{-x})+\ln(ye^{\lambda x}-(y-1)(1-\lambda)) - y\ln(y\lambda+1-\lambda).
\end{equation}
We will show that $f(y,x)$ is nonincreasing in $y \in [1,\infty)$ for all $x \in \R_+$ and hence the result follows.
It suffices to show that the first partial derivative of $f$ with respect to $y$ is upper bounded by zero. 
To this end, we write 
\EQ{
\frac{\partial f(y,x)}{\partial y} 
= \frac{e^{\lambda x}-1+ \lambda}{y(e^{\lambda x}-1+\lambda)+1-\lambda}
+ \frac{(y-1)\lambda}{y\lambda+(1-\lambda)e^{-x}}
-\frac{y\lambda}{y\lambda+1-\lambda}+ \ln\Big(1 -\frac{(1-\lambda)(1-e^{-x})}{y\lambda+1-\lambda}\Big).
}
We recall that $(e^{\lambda x}-1)\le \lambda(e^x-1)$ for all $\lambda \in [0,1]$ and $x \in \R_+$, 
$\frac{a}{ya+1-\lambda} \le \frac{a_0}{ya_0+1-\lambda}$ for $0 < a \le a_0$, and $\ln(1-x) \le -x$ for all $x \in [0,1]$, to upper bound the partial derivative of $f$ with respect to $y$ as 
\begin{align*}
\frac{\partial f(y,x)}{\partial y} &\le
\frac{y\lambda}{y\lambda+(1-\lambda)e^{-x}}
-\frac{y\lambda+(1-\lambda)(1-e^{-x})}{y\lambda+1-\lambda}
= -\frac{(1-\lambda)^2e^{-x}(1-e^{-x})}{(y\lambda+1-\lambda)(y\lambda+(1-\lambda)e^{-x})} \le 0.
\end{align*}
\end{proof}

When the system is lightly loaded, we expect that the replicas of an arriving  job will find most servers idle. 
Thus, all $d$ replicas get served by $d$ parallel servers, leading to improvement in the mean response time performance.
However, under heavy traffic regimes, this policy behaves similar to the random routing policy where only the primary replica gets served, while all secondary replicas are likely to get canceled. 
In this regime, the policies with queue state information can perform better, 
although this improvement in performance comes at the cost of procurement of information.  
\begin{table}[hhh]
\centering
\caption{{\small{
Percentage improvement in mean response time of the policy $\pi(d,\infty,0)$ over random routing with the number of servers $N=20$ and service rate $\mu = 1$. See Fig.~\ref{Fig:STVsAR-d-Inf0}.}}}
 \begin{tabular}{||c | c | c | c | c ||} 
 \hline
 Replicas $d$ & $\lambda=0.2$ &  $\lambda=0.4$ & $\lambda=0.6$ & $\lambda=0.8$  \\ [0.5ex] 
 \hline\hline
 3 & 43.14\% & 22.02\%& 7.9\%& 1.74\%\\
 \hline
 6 & 57.23\% & 29.30\%& 10.37\%& 2.22\%\\
 \hline
  9 & 62.33\% & 31.97\%& 11.22\%& 2.39\%\\
 \hline
 12 & 64.96\% & 33.35\%& 11.66\%& 2.47\%\\
 \hline
\end{tabular}
\label{Tab:ZeroDeadLineComaprison-d}
\end{table}
We plot the mean response time $\tau$ for policy $\pi(d, \infty, 0)$ as a function of arrival rate $\lambda$ in Fig.~\ref{Fig:STVsAR-d-Inf0},
for the number of servers $N = 20$, service rate $\mu = 1$, and different number of replicas $d$. 
Here, are the main observations.
\begin{compactenum}
\item The mean response time for $\pi(d,\infty,0)$ is uniformly better for larger number of replicas $d$, 
and the gains are highest  for lower arrival rates. 
\item Here, the additional replicas are executed only if the server is idle in this policy. 
Therefore, a higher choice of replication factor $d$ does not increase the system load significantly.
\item For moderate to higher values of arrival rates, 
all the different choices of number of replicas have a similar performance under the stability region of $\lambda < \mu$, 
independent of number of replicas $d$. 
\end{compactenum}
We also provide the percentage improvement of conditional mean response time of the policy $\pi(d, \infty, 0)$ over random routing policy for various values of normalized arrival rate in Table~\ref{Tab:ZeroDeadLineComaprison-d}.

We note from the numerical comparisons that $\pi(d, \infty, 0)$ policy offers a superior performance 
among all $\pi(d,\infty, T_2)$ policies. 
It is also clear that $\pi(d, \infty, 0)$ policy performs better than random routing for any value of $d$. 
We now proceed to study the performance of this policy with respect to some of the best known load balancing policies in the literature. 

\section{Comparison with feedback based policies}
\label{sec:NumericalComparison}

\begin{figure*}[h!]
\begin{subfigure}{0.33 \columnwidth}
		\centerline{\scalebox{0.6}{
\begin{tikzpicture}

\definecolor{color0}{rgb}{0.83921568627451,0.152941176470588,0.156862745098039}
\definecolor{color1}{rgb}{0.12156862745098,0.466666666666667,0.705882352941177}
\definecolor{color2}{rgb}{0.172549019607843,0.627450980392157,0.172549019607843}
\definecolor{color3}{rgb}{1,0.498039215686275,0.0549019607843137}

\begin{axis}[
legend cell align={left},
legend style={fill opacity=0.8, draw opacity=1, text opacity=1, at={(0.45,0.73)}, anchor=north west, draw=white!80!black},
tick align=outside,
tick pos=left,
x grid style={white!69.01960784313725!black,   dotted},
xlabel={Arrival rate $\lambda$},
xmajorgrids,
xmin=-0.0375, xmax=1.0075,
xminorgrids,
xtick style={color=black},
y grid style={white!69.01960784313725!black,   dotted},
ylabel={Mean response time loss $\%$, $\gamma$},
ymajorgrids,
ymin= -100.16786106081546, ymax=354.4,
yminorgrids,
ytick style={color=black}
]
\addplot [color0, mark=asterisk, mark size=1, mark options={solid}, line width=1.4cm, very thick]
table {%
0.01	279.3843959
0.06	188.275783
0.11	127.6136596
0.16	80.5850224
0.21	49.0810336
0.26	23.3385367
0.31	4.5831626
0.36	-10.3278152
0.41	-21.7272218
0.46	-30.6254459
0.51	-38.2850711
0.56	-45.1167896
0.61	-51.7945671
0.66	-55.802407
0.71	-59.4402843
0.76	-62.2214482
0.81	-66.8668825
0.86	-68.0903284
0.91	-71.0617017
0.96	-73.116573
};

\addlegendentry{$\gamma_{JIQ(d)-\pi(d,\infty,0)}$}
\addplot [color1, mark=+, mark size=2.4, mark options={solid}, line width=1.4cm, very thick]
table {%
0.01	281.6289479
0.06	188.1749724
0.11	125.8803392
0.16	82.1130552
0.21	48.8572076
0.26	22.7511371
0.31	3.8330042
0.36	-11.2626078
0.41	-23.0713607
0.46	-32.0595623
0.51	-40.4400779
0.56	-47.8606793
0.61	-55.8521165
0.66	-60.5267231
0.71	-66.5415365
0.76	-70.1714484
0.81	-76.3457685
0.86	-80.7602035
0.91	-85.6049031
0.96	-89.1581398
};

\addlegendentry{$\gamma_{JSW(d)-\pi(d,\infty,0)}$}
\addplot [black, mark=asterisk, mark size=1, mark options={solid}, line width=1.4cm, very thick]
table {%
0.01	281.8103232
0.06	189.0433857
0.11	125.785302
0.16	79.5010236
0.21	49.9821617
0.26	22.873011
0.31	4.4928671
0.36	-9.9766645
0.41	-21.279059
0.46	-29.3657021
0.51	-36.7306871
0.56	-44.2931556
0.61	-50.5181648
0.66	-55.0530757
0.71	-60.332459
0.76	-63.9447533
0.81	-71.22972
0.86	-76.0287868
0.91	-81.8583055
0.96	-87.3376017
};

\addlegendentry{$\gamma_{JSQ(d)-\pi(d,\infty,0)}$}
\addplot [color3, mark=asterisk, mark size=1, mark options={solid}, line width=1.4cm, very thick]
table {%
0.01	-4.2202917
0.06	-25.0177681
0.11	-39.1377388
0.16	-49.6789259
0.21	-56.8841037
0.26	-63.3490318
0.31	-67.683635
0.36	-71.1357213
0.41	-74.0444073
0.46	-75.9541132
0.51	-77.5621864
0.56	-79.4373921
0.61	-81.6861514
0.66	-82.5472035
0.71	-84.3761658
0.76	-85.0748524
0.81	-87.55219
0.86	-88.7593173
0.91	-91.3461721
0.96	-90.9694969
};

\addlegendentry{$\gamma_{c.o.c.(d)-\pi(d,\infty,0)}$}

\end{axis}

\end{tikzpicture}}}
				\caption{\small{
		$\gamma$ vs $\lambda$
		}}
		\label{Fig:Compare-GammaSTVsAR-d}
\end{subfigure}
\begin{subfigure}{0.33 \columnwidth}
		\centerline{\scalebox{0.6}{
\begin{tikzpicture}

\definecolor{color0}{rgb}{0.83921568627451,0.152941176470588,0.156862745098039}
\definecolor{color1}{rgb}{0.12156862745098,0.466666666666667,0.705882352941177}
\definecolor{color2}{rgb}{0.172549019607843,0.627450980392157,0.172549019607843}
\definecolor{color3}{rgb}{1,0.498039215686275,0.0549019607843137}

\begin{semilogyaxis}[
legend cell align={left},
legend style={fill opacity=0.8, draw opacity=1, text opacity=1, at={(0.05,0.9)}, anchor=north west, draw=white!80!black},
tick align=outside,
tick pos=left,
x grid style={white!69.01960784313725!black,   dotted},
xlabel={Arrival rate $\lambda$},
xmajorgrids,
xmin=-0.0375, xmax=1.0075,
xminorgrids,
xtick style={color=black},
y grid style={white!69.01960784313725!black,   dotted},
ylabel={Mean response time $\tau$},
ymajorgrids,
ymin=-1.16786106081546, ymax=24,
yminorgrids,
ytick style={color=black}
]

\addplot [color2, mark=asterisk, mark size=1, mark options={solid}, line width=1.4cm, very thick]
table {%
0.01	0.262271175
0.06	0.346378008
0.11	0.441448453
0.16	0.55288107
0.21	0.669483506
0.26	0.814311311
0.31	0.964806039
0.36	1.132527637
0.41	1.312845536
0.46	1.494458711
0.51	1.719121612
0.56	1.991228658
0.61	2.374475061
0.66	2.722801563
0.71	3.283676822
0.76	3.890415462
0.81	5.193553245
0.86	6.966877796
0.91	11.4523176
0.96	22.15137221
};
\addlegendentry{$\pi(d,\infty,0)$}

\addplot [color0, mark=square, mark size=1, mark options={solid}, line width=1.4cm, very thick]
table {%
0.01	0.995015913
0.06	0.998523916
0.11	1.004796978
0.16	0.998420404
0.21	0.99807293
0.26	1.004359655
0.31	1.009024669
0.36	1.015562276
0.41	1.027600674
0.46	1.036774068
0.51	1.060954681
0.56	1.092850215
0.61	1.144625982
0.66	1.203412733
0.71	1.331849984
0.76	1.469742621
0.81	1.720786101
0.86	2.223107823
0.91	3.314105836
0.96	5.955047975
};
\addlegendentry{${JIQ(d)}$}
\addplot [color1, mark=+, mark size=2.4, mark options={solid}, line width=1.4cm, very thick]
table {%
0.01	1.000902726
0.06	0.998174731
0.11	0.997145262
0.16	1.006868608
0.21	0.996574452
0.26	0.999576394
0.31	1.001787096
0.36	1.004975491
0.41	1.009954206
0.46	1.01534179
0.51	1.023907493
0.56	1.038213096
0.61	1.048280485
0.66	1.074778999
0.71	1.098667811
0.76	1.160454583
0.81	1.228495108
0.86	1.340413109
0.91	1.648572217
0.96	2.401620801
};
\addlegendentry{${JSW(d)}$}
\addplot [mark=asterisk, mark size=1, mark options={solid}, line width=1.4cm, very thick]
table {%
0.01	1.001378421
0.06	1.001182723
0.11	0.996725722
0.16	0.99242718
0.21	1.004105834
0.26	1.000568827
0.31	1.008153493
0.36	1.019539154
0.41	1.03348436
0.46	1.055600419
0.51	1.087676432
0.56	1.10925065
0.61	1.174933836
0.66	1.223815558
0.71	1.302553851
0.76	1.402698892
0.81	1.494199808
0.86	1.670045132
0.91	2.077644476
0.96	2.80489498
};
\addlegendentry{${JSQ(d)}$}
\addplot [color3, mark=asterisk, mark size=1, mark options={solid}, line width=1.4cm, very thick]
table {%
0.01	0.251202566
0.06	0.259721962
0.11	0.26867551
0.16	0.278215693
0.21	0.288653814
0.26	0.29845298
0.31	0.311790241
0.36	0.326895933
0.41	0.340756841
0.46	0.35935585
0.51	0.385733303
0.56	0.409448542
0.61	0.434857767
0.66	0.475205016
0.71	0.513036224
0.76	0.580650249
0.81	0.646483639
0.86	0.783124629
0.91	0.991063855
0.96	2.000380363
};
\addlegendentry{$COC(d)$}

\end{semilogyaxis}

\end{tikzpicture}}}
		\caption{\small{
		$\tau$ vs $\lambda$
		}}
		\label{Fig:Compare-STVsAR-d}
\end{subfigure}
\begin{subfigure}{0.33\columnwidth}
		\centerline{\scalebox{0.6}{
\begin{tikzpicture}

\definecolor{color0}{rgb}{0.83921568627451,0.152941176470588,0.156862745098039}
\definecolor{color1}{rgb}{0.12156862745098,0.466666666666667,0.705882352941177}
\definecolor{color2}{rgb}{0.172549019607843,0.627450980392157,0.172549019607843}
\definecolor{color3}{rgb}{1,0.498039215686275,0.0549019607843137}
\definecolor{color4}{rgb}{0.498039215686275,0.0549019607843137,1}

\begin{axis}[
legend cell align={left},
legend style={fill opacity=0.8, draw opacity=1, text opacity=1, at={(0.05,0.85)}, anchor=north west, draw=white!80!black},
tick align=outside,
tick pos=left,
x grid style={white!69.01960784313725!black,   dotted},
xlabel={Arrival rate $\lambda$},
xmajorgrids,
xmin=-0.0375, xmax=1.0075,
xminorgrids,
xtick style={color=black},
y grid style={white!69.01960784313725!black,   dotted},
ylabel={Mean workload $\mathbb{E}[W]$},
ymajorgrids,
ymin=-1.16786106081546, ymax=15,
yminorgrids,
ytick style={color=black}
]
\addplot [color0, mark=asterisk, mark size=1, mark options={solid}, line width=1.4cm, very thick]
table {%
0.01	0.0104274816
0.06	0.0595566333
0.11	0.1134435188
0.16	0.1573295831
0.21	0.20515928
0.26	0.2599556446
0.31	0.3024927871
0.36	0.3552677875
0.41	0.4454992764
0.46	0.5078428716
0.51	0.5746113859
0.56	0.5970475871
0.61	0.72035991
0.66	0.7663197794
0.71	0.9815749302
0.76	1.1406066471
0.81	1.2935750888
0.86	2.0217772733
0.91	2.9004769046
0.96	6.0506079524
};

\addlegendentry{${JIQ(d)}$}
\addplot [color1, mark=+, mark size=2.4, mark options={solid}, line width=1.4cm, very thick]
table {%
0.01	0.0094808076
0.06	0.0615061252
0.11	0.1116497805
0.16	0.1597440179
0.21	0.2204865773
0.26	0.2610003184
0.31	0.3176535661
0.36	0.3573419357
0.41	0.4003176622
0.46	0.4877747446
0.51	0.5163854114
0.56	0.5805758744
0.61	0.6782124252
0.66	0.7098383879
0.71	0.7836716635
0.76	0.8617865254
0.81	0.9819676687
0.86	1.1422342177
0.91	1.4973570911
0.96	2.1397751282
};
\addlegendentry{$JSW(d)$}
\addplot [black, mark=asterisk, mark size=1, mark options={solid}, line width=1.4cm, very thick]
table {%
0.01	0.0106263417
0.06	0.0585364481
0.11	0.1146843904
0.16	0.1594622644
0.21	0.2117381845
0.26	0.255037594
0.31	0.3254099164
0.36	0.375037105
0.41	0.4177914148
0.46	0.4870132169
0.51	0.5589661901
0.56	0.6427144734
0.61	0.7224455586
0.66	0.8010975926
0.71	0.8530933475
0.76	1.0411721695
0.81	1.1949797859
0.86	1.5265731877
0.91	1.8702730634
0.96	3.6425319422
};
\addlegendentry{$JSQ(d)$}
\addplot [violet, densely dotted, mark=asterisk, mark size=1, mark options={solid}, line width=1.4cm, very thick]
table {%
0.01	0.0384258238
0.06	0.217351861
0.11	0.3805172393
0.16	0.5105306685
0.21	0.6483144153
0.26	0.7896890494
0.31	0.9016980562
0.36	1.0752503033
0.41	1.3170355769
0.46	1.4495316827
0.51	1.6128251436
0.56	1.9219681272
0.61	2.3552689797
0.66	2.8889691484
0.71	3.2744773021
0.76	3.6945754622
0.81	4.6906573377
0.86	5.9318985997
0.91	7.6467974635
0.96	15.341939508
};
\addlegendentry{$\pi(d,\infty,0)$}
\addplot [color3, mark=asterisk, mark size=1, mark options={solid}, line width=1.4cm, very thick]
table {%
0.01	0.0144274315
0.06	0.0840937838
0.11	0.1555808336
0.16	0.2299202806
0.21	0.3013130009
0.26	0.3965373647
0.31	0.4768655725
0.36	0.5755715397
0.41	0.6609472891
0.46	0.7533693786
0.51	0.8384078583
0.56	0.9784833253
0.61	1.1050659881
0.66	1.243906164
0.71	1.3412444776
0.76	1.507432352
0.81	1.7391846252
0.86	1.899797257
0.91	2.0963230909
0.96	2.2937278645
};
\addlegendentry{$c.o.c.(d)$}

\end{axis}

\end{tikzpicture}}}
		\caption{\small{
		$\E[W]$ vs $\lambda$
		}}
		\label{Fig:Compare-WorkloadVsAR-d-Exp}
\end{subfigure}
\caption{
	\small{The mean response time loss percentage $\gamma$ with respect to $\pi(d,\infty,0)$ policy, mean response time $\tau$ and the mean workload $\E[W]$ at the cavity queue as a function of normalized arrival rate $\lambda$ for the policies $\pi(d,\infty,0)$, JIQ($d$), JSW($d$), JSQ($d$), c.o.c.($d$) for exponential service distribution of rate $\mu=1$, the number of servers $N=20$ and the number of replicas $d=4$. } 
	} 
\end{figure*}
As a benchmark, 
we compare the performance of the proposed $\pi(d,\infty, 0)$ policy 
to some popular policies like  redundancy-$d$ cancel on start (c.o.s.), redundancy-$d$ cancel on complete (c.o.c.), JSQ($d$), 
and Join Threshold Queue (JTQ($d,T$))~\cite[Section 6.6]{Hellemans19}
that have information feedback and/or synchronized cancellation of replicas. 
We choose to set threshold $T$ to $0$ in JTQ($d,T$) policy where it is identical to JIQ($d$) policy.
It is easy to see that the c.o.s.($d$) policy is identical to the JSW($d$) policy. 
For fairness of comparison, we are comparing only no loss policies, in which case, 
the conditional mean response time is the mean response time for any job. 
We first present analytical comparison between c.o.s.($d$) and c.o.c.($d$) followed by the comparison between c.o.c.($d$) and the proposed $\pi(d,\infty,0)$ policy. 
Then, we proceed to present the comparison through simulation studies.
The unavailability of closed form expressions for some of the existing policies under comparison and the complicated expressions for the mean response time distributions restricts us from providing a complete analytical comparison between all the policies considered.

\begin{remark}
\label{rem:coc,cos}
For an $N$ server system with a Poisson arrival rate of $N\lambda$, \iid exponential service times of mean $1$ and under the assumption of asymptotic independence, 
the tail response time distribution for all $x\in\R_+$, under replication-$d$ for $d\ge 2$ with c.o.s.~\cite[Theorem 5.3]{Hellemans18} and c.o.c.~\cite[Theorem 6]{Gardner17}, is given by
\begin{xalignat*}{2}
&\bH_{\textrm{c.o.s.}(d)}(x) = \frac{1}{(\lambda^d+(1-\lambda^d)e^{(d-1)x})^{\frac{1}{d-1}}},&
&\bH_{\textrm{c.o.c.}(d)}(x) 
= \frac{1}{(\lambda+ e^{(d-1)x}(1-\lambda))^{\frac{d}{d-1}}}.
\end{xalignat*}
\end{remark}
\begin{proposition}
\label{pro:coscocComparison}
For an $N$ server system with a Poisson arrival rate of $N\lambda$, \iid exponential service times of mean $1$, the response time of c.o.c.($d$) policy is always stochastically dominated by that of c.o.s.($d$) policy. 
That is, $\bH_{\textrm{c.o.c.}(d)}(x) \le\bH_{\textrm{c.o.s.}(d)}(x)$ for all $x \in \R_+$ and $d \ge 2$.
\end{proposition}
\begin{proof}
Let $x\in\R_+$ and $d\ge 2$. 
From Remark~\ref{rem:coc,cos}, we observe that we only need to show $\lambda^d+(1-\lambda^d)e^{(d-1)x} \le (\lambda+ e^{(d-1)x}(1-\lambda))^d.$ 
However, it follows from the observation $
(\lambda+ e^{(d-1)x}(1-\lambda))^d 
\ge \lambda^d + e^{d(d-1)x}(1-\lambda)^d 
\ge \lambda^d + e^{(d-1)x}(1-\lambda)^d.
$
\end{proof}

\begin{proposition}
\label{pro:cocProposedComparison}
For an $N$ server system with a Poisson arrival rate of $N\lambda$, \iid exponential service times of mean $1$, the response time of c.o.c.($d$) policy is stochastically dominated by that of $\pi(d, \infty, 0)$ policy. 
That is, $\bH_{\textrm{c.o.c.}(d)}(x) \le \bH_d(x)$ for all $x \in \R_+$ and $d \ge 2$.
\end{proposition}
\begin{proof} 
From Remark~\ref{rem:TRTSimplified} and Remark~\ref{rem:coc,cos}, it suffices to show that for all $x \in \R_+$
\begin{align*}
\frac{1}{(\lambda+ e^{(d-1)x}(1-\lambda))^\frac{d}{d-1}} \le e^{-x}\Big(1+\frac{d(e^{\lambda x}-1)}{(d-1)\lambda+1}\Big)\Big(1-\frac{(1-\lambda)(1-e^{-x})}{(d-1)\lambda+1}\Big)^{d-1}.
\end{align*}
In order to prove this, we consider a function $g:[2,\infty)\times\R_+\to\R$ defined as 
\begin{align*}
g(y,x) &\triangleq 
-x+ \ln(y(e^{\lambda x} -1)+y\lambda+1-\lambda)+(y-1)\ln(y\lambda+(1-\lambda)e^{-x})-y\ln(y\lambda+1-\lambda)+\frac{y}{y-1}\ln(\lambda+e^{(y-1)x}(1-\lambda)).
\end{align*}
We observe that $g(y,0) = 0$. 
Then, we obtain the required result by showing that $g(y,x)$ is increasing in $x$ for all $x \in \R_+$. 
To this end, we compute the first partial derivative of $g(y,x)$ with respect to $x$, and write 
\begin{align*}
\frac{\partial g(y,x)}{\partial x}
&= -(1-\lambda)+\frac{\lambda(1-\lambda)(y-1)}{y(e^{\lambda x}-1)+y\lambda+1-\lambda}-\frac{(y-1)(1-\lambda)e^{-x}}{y\lambda+(1-\lambda)e^{-x}}+\frac{y(1-\lambda)e^{(y-1)x}}{\lambda+(1-\lambda)e^{(y-1)x}}.
\end{align*}
We use the fact that $e^{\lambda x}-1 \le \lambda(e^x-1)$ for all $\lambda \in [0,1]$ and $x\in\R_+$, to observe that $g(y,x)$ is increasing in $x$ as
\begin{align*}
\frac{1}{(1-\lambda)}\frac{\partial g(y,x)}{\partial x}
&\ge (y-1)-\frac{(y-1)(1-\lambda)e^{-x}}{y\lambda+(1-\lambda)e^{-x}}
+\frac{y\lambda(e^{(y-1)x}-1)}{\lambda+(1-\lambda)e^{(y-1)x}}= \frac{y(y-1)\lambda}{y\lambda+(1-\lambda)e^{-x}}
+\frac{y\lambda(e^{(y-1)x}-1)}{\lambda+(1-\lambda)e^{(y-1)x}} \ge 0.
\end{align*}
\end{proof}


Although closed form expressions for mean response time for c.o.s.($d$) and c.o.c.($d$) policies can be found in ~\cite[Theorem 5.4]{Hellemans18} and ~\cite[Theorem 6]{Gardner17} respectively, we do not have closed form expressions for mean response time of JIQ($d$)  and JSQ($d$) policies under the given settings. 
The comparison of these policies against the proposed policy via numerical simulations is presented next.
All the experiments reported in this section have been run for $10^5$ iterations with the number of servers $N = 20$ and the number of replicas $d = 4$.
If the mean response time of the policies $x$ and $y$ are denoted by $\tau_x$ and $\tau_y$ respectively, 
then the  \emph{mean response time loss percentage} of policy $x$ against policy $y$ is defined as 
$
\gamma_{x-y} \triangleq \frac{\tau_x - \tau_y}{\tau_y} \times 100 \%. 
$
\begin{remark}
\label{rem:MRTLossValue}
Note that whenever $\tau_y \le \tau_x$,  the value of the mean response time loss percentage $\gamma_{x-y}$ is non-negative.
Thus, $\gamma_{x-y}$ value being non-negative for any value of arrival rate indicates that the policy $y$ is superior to the policy $x$ for the given load conditions.
\end{remark}

In Fig.~\ref{Fig:Compare-GammaSTVsAR-d}, we plot the mean response time loss percentage for JIQ($d$), JSW($d$), JSQ($d$) and c.o.c.($d$) policies against the replicate on idle secondary servers ($\pi(d,\infty,0)$) policy when the service times are \iid exponentially distributed with rate $1$. 
From the figure, we observe that the mean response time loss percentage with respect to the $\pi(d,\infty,0)$ policy is non-negative for low arrival rates for JIQ($d$), JSW($d$), and JSQ($d$) policies.
As mentioned in Remark~\ref{rem:MRTLossValue}, this  shows that the $\pi(d,\infty,0)$ policy performs better than JIQ($d$), JSW($d$), and JSQ($d$) policies in the regime of low arrival rates.
For a better clarity, we plot the mean response times of these policies in Fig.~\ref{Fig:Compare-STVsAR-d} which clearly indicates the  performance improvement of $\pi(d,\infty,0)$ policy.
Although at higher arrival rates, the JIQ($d$), JSW($d$), and JSQ($d$) policies performs better than our policy, this performance improvement comes at the price of information exchange between the servers and dispatcher.
In addition, we observe that the cancel on complete policy performs the best for all arrival rates. 
This is due to the fact that cancel on complete is equivalent to water filling at the $d$ sampled servers for \iid exponential service~\cite{Shneer2021QS}, 
and the water filling policy has an additional degree of freedom to divide jobs arbitrarily on different servers.
It should also be kept in mind that c.o.c policy requires strict coordination and communication among the servers to achieve this performance. 
Moreover, we will see in the next section that the performance improvement of c.o.c. policy degrades for non-exponential service distributions like Weibull and Pareto and it further suffers from stability issues.

We now provide a comparison of the expected workloads at the cavity queues in each of these policies in Fig.~\ref{Fig:Compare-WorkloadVsAR-d-Exp}.
Compared to other policies, the expected workload at the individual queues is higher in our policy.
This is expected as a larger number of replicas are processed per job under our policy unlike the other policies that performs coordinated cancellation of additional replicas. 
However, the extra workload is not huge in the low arrival rate regime. Also, to be noted is that our policy provides a performance improvement in this regime in spite of the increment in the average workload.
In fact, the increment of the workload is caused by the additional redundant replicas and it is this additional redundancy that helps in bringing down the overall response time of the job.

\begin{figure}[h!]
\begin{subfigure}{0.33\columnwidth}
		\centerline{\scalebox{0.6}{
\begin{tikzpicture}

\definecolor{color0}{rgb}{0.83921568627451,0.152941176470588,0.156862745098039}
\definecolor{color1}{rgb}{0.12156862745098,0.466666666666667,0.705882352941177}
\definecolor{color2}{rgb}{0.172549019607843,0.627450980392157,0.172549019607843}
\definecolor{color3}{rgb}{1,0.498039215686275,0.0549019607843137}

\begin{axis}[
legend cell align={left},
legend style={fill opacity=0.8, draw opacity=1, text opacity=1, at={(0.05,0.4)}, anchor=north west, draw=white!80!black},
tick align=outside,
tick pos=left,
x grid style={white!69.01960784313725!black,   dotted},
xlabel={Arrival rate $\lambda$},
xmajorgrids,
xmin=-0.0375, xmax=1.0075,
xminorgrids,
xtick style={color=black},
y grid style={white!69.01960784313725!black,   dotted},
ylabel={Mean response time loss $\%$, $\gamma$},
ymajorgrids,
ymin=-100.16786106081546, ymax=100,
yminorgrids,
ytick style={color=black}
]
\addplot [color0, mark=asterisk, mark size=1, mark options={solid}, line width=1.4cm, very thick]
table {%
0.01	30.6106332
0.06	24.7382842
0.11	18.8501486
0.16	12.8386573
0.21	06.8151956
0.26	1.6178527
0.31	-4.1262456
0.36	-9.5269894
0.41	-14.5582998
0.46	-19.3807592
0.51	-24.0600071
0.56	-28.8633835
0.61	-33.9420355
0.66	-37.9745134
0.71	-42.6089019
0.76	-46.8395106
0.81	-50.798813
0.86	-55.0622005
0.91	-58.7929349
0.96	-61.0399531
};
\addlegendentry{$\gamma_{JIQ(d)-\pi(d,\infty,0)}$}
\addplot [color1, mark=+, mark size=2.4, mark options={solid}, line width=1.4cm, very thick]
table {%
0.01	30.6010895
0.06	24.5714578
0.11	18.9335383
0.16	13.0266388
0.21	7.0947926
0.26	1.5115072
0.31	-4.2593514
0.36	-9.7150842
0.41	-14.7209922
0.46	-19.7404827
0.51	-24.6583649
0.56	-29.8533951
0.61	-35.5231901
0.66	-40.22812
0.71	-45.3487407
0.76	-50.3341434
0.81	-55.4282643
0.86	-60.9810857
0.91	-67.0323313
0.96	-71.5517755
};
\addlegendentry{$\gamma_{JSW(d)-\pi(d,\infty,0)}$}
\addplot [black, mark=asterisk, mark size=1, mark options={solid}, line width=1.4cm, very thick]
table {%
0.01	30.8629392
0.06	24.5955396
0.11	18.9356218
0.16	12.8277741
0.21	6.9222637
0.26	1.4920869
0.31	-4.0595756
0.36	-9.1531522
0.41	-13.9179756
0.46	-18.8620289
0.51	-23.321314
0.56	-28.0879708
0.61	-33.413377
0.66	-37.3747232
0.71	-42.1993946
0.76	-46.6488813
0.81	-51.4612653
0.86	-56.9088885
0.91	-63.214688
0.96	-67.4399787
};
\addlegendentry{$\gamma_{JSQ(d)-\pi(d,\infty,0)}$}
\addplot [color3, mark=asterisk, mark size=1, mark options={solid}, line width=1.4cm, very thick]
table {%
0.01	-0.3244324
0.06	-1.9541261
0.11	-3.2534714
0.16	-4.5127478
0.21	-5.4859439
0.26	-6.0517746
0.31	-6.430771
0.36	-5.3825841
0.41	-3.1695729
0.46	1.6143458
0.51	14.8971253
0.56	43.3687461
0.61	8249.961375
};
\addlegendentry{$\gamma_{c.o.c(d)-\pi(d,\infty,0)}$}

\end{axis}

\end{tikzpicture}}}
		\caption{\small{
		Weibull distribution with scale $1$ and shape $5$.
		}}
		\label{Fig:Compare-GammaSTVsAR-d-Weibull}
\end{subfigure}
\begin{subfigure}{0.33 \columnwidth}
		\centerline{\scalebox{0.6}{
\begin{tikzpicture}

\definecolor{color0}{rgb}{0.83921568627451,0.152941176470588,0.156862745098039}
\definecolor{color1}{rgb}{0.12156862745098,0.466666666666667,0.705882352941177}
\definecolor{color2}{rgb}{0.172549019607843,0.627450980392157,0.172549019607843}
\definecolor{color3}{rgb}{1,0.498039215686275,0.0549019607843137}

\begin{axis}[
legend cell align={left},
legend style={fill opacity=0.8, draw opacity=1, text opacity=1, at={(0.05,0.4)}, anchor=north west, draw=white!80!black},
tick align=outside,
tick pos=left,
x grid style={white!69.01960784313725!black,   dotted},
xlabel={Arrival rate $\lambda$},
xmajorgrids,
xmin=-0.0375, xmax=1.0075,
xminorgrids,
xtick style={color=black},
y grid style={white!69.01960784313725!black,   dotted},
ylabel={Mean response time loss $\%$, $\gamma$},
ymajorgrids,
ymin=-100.16786106081546, ymax=100,
yminorgrids,
ytick style={color=black}
]
\addplot [color0, mark=asterisk, mark size=1, mark options={solid}, line width=1.4cm, very thick]
table {%
0.01	16.015984
0.06	13.5342085
0.11	9.8891946
0.16	5.8862129
0.21	1.0091184
0.26	-3.5386508
0.31	-8.8995349
0.36	-14.1289273
0.41	-19.1043945
0.46	-24.0332229
0.51	-29.0407521
0.56	-33.8790961
0.61	-38.2869231
0.66	-42.6399129
0.71	-48.5537911
0.76	-52.7136274
0.81	-57.1938849
0.86	-61.0390344
0.91	-64.8512514
0.96	-67.6973742
};
\addlegendentry{$\gamma_{JIQ(d)-\pi(d,\infty,0)}$}
\addplot [color1, mark=+, mark size=2.4, mark options={solid}, line width=1.4cm, very thick]
table {%
0.01	16.8535673
0.06	13.3270147
0.11	9.7683929
0.16	6.0819631
0.21	1.2299029
0.26	-3.3776302
0.31	-9.3241003
0.36	-14.3633567
0.41	-19.3492312
0.46	-24.8141571
0.51	-30.2116361
0.56	-35.4895467
0.61	-40.8032129
0.66	-44.9055369
0.71	-52.5477702
0.76	-58.4432272
0.81	-63.7262471
0.86	-70.6307151
0.91	-79.599328
0.96	-86.4216344
};
\addlegendentry{$\gamma_{JSW(d)-\pi(d,\infty,0)}$}
\addplot [black, mark=asterisk, mark size=1, mark options={solid}, line width=1.4cm, very thick]
table {%
0.01	15.9291868
0.06	13.8008116
0.11	10.3253948
0.16	6.3111147
0.21	0.8170503
0.26	-3.1572089
0.31	-8.4722728
0.36	-13.43269
0.41	-18.350936
0.46	-23.0239304
0.51	-28.3048054
0.56	-33.3514027
0.61	-37.2762074
0.66	-42.5779613
0.71	-49.4742063
0.76	-53.5366607
0.81	-59.9321486
0.86	-66.4019977
0.91	-75.5848084
0.96	-82.7036653
};
\addlegendentry{$\gamma_{JSQ(d)-\pi(d,\infty,0)}$}
\addplot [color3, mark=asterisk, mark size=1, mark options={solid}, line width=1.4cm, very thick]
table {%
0.01	-0.1124771
0.06	-0.5867062
0.11	-1.6264307
0.16	-2.3244684
0.21	-2.7244697
0.26	-3.1826673
0.31	-2.4454146
0.36	-1.1027202
0.41	3.115823
0.46	15.3131653
0.51	60.2722669
0.56	774.903511
};
\addlegendentry{$\gamma_{c.o.c.(d)-\pi(d,\infty,0)}$}

\end{axis}

\end{tikzpicture}}}
		\caption{\small{
		Pareto distribution with scale $0.83$ and shape $5.5$.
		}}
		\label{Fig:Compare-GammaSTVsAR-d-Pareto}
\end{subfigure}\hfill%
\begin{subfigure}{0.33\columnwidth}
		\centerline{\scalebox{0.6}{
\begin{tikzpicture}

\definecolor{color0}{rgb}{0.83921568627451,0.152941176470588,0.156862745098039}
\definecolor{color1}{rgb}{0.12156862745098,0.466666666666667,0.705882352941177}
\definecolor{color2}{rgb}{0.172549019607843,0.627450980392157,0.172549019607843}
\definecolor{color3}{rgb}{1,0.498039215686275,0.0549019607843137}

\begin{axis}[
legend cell align={left},
legend style={fill opacity=0.8, draw opacity=1, text opacity=1, at={(0.05,0.4)}, anchor=north west, draw=white!80!black},
tick align=outside,
tick pos=left,
x grid style={white!69.01960784313725!black,   dotted},
xlabel={Arrival rate $\lambda$},
xmajorgrids,
xmin=-0.0375, xmax=1.0075,
xminorgrids,
xtick style={color=black},
y grid style={white!69.01960784313725!black,   dotted},
ylabel={Mean response time loss $\%$, $\gamma$},
ymajorgrids,
ymin=-100.16786106081546, ymax=100,
yminorgrids,
ytick style={color=black}
]
\addplot [color0, mark=asterisk, mark size=1, mark options={solid}, line width=1.4cm, very thick]
table {%
0.01	13.1195103
0.06	10.6460125
0.11	6.9374136
0.16	3.2915792
0.21	-0.4520389
0.26	-5.9375222
0.31	-9.3525766
0.36	-14.7499231
0.41	-18.6969644
0.46	-23.3911167
0.51	-28.6910403
0.56	-32.9034808
0.61	-37.8476519
0.66	-41.7363669
0.71	-45.7467709
0.76	-51.5376173
0.81	-56.7119657
0.86	-62.0926473
0.91	-68.531052
0.96	-71.5650119
};
\addlegendentry{$\gamma_{JIQ(d)-\pi(d,\infty,0)}$}
\addplot [color1, mark=+, mark size=2.4, mark options={solid}, line width=1.4cm, very thick]
table {%
0.01	13.4465919
0.06	10.6385855
0.11	6.8954891
0.16	3.300029
0.21	-0.7518317
0.26	-5.763115
0.31	-9.3768878
0.36	-15.1229599
0.41	-19.0817449
0.46	-23.9146317
0.51	-29.7240893
0.56	-34.4372474
0.61	-39.8715557
0.66	-46.1925909
0.71	-49.8167855
0.76	-57.4391965
0.81	-62.8132392
0.86	-70.1638851
0.91	-77.8075242
0.96	-83.382348
};
\addlegendentry{$\gamma_{JSW(d)-\pi(d,\infty,0)}$}
\addplot [black, mark=asterisk, mark size=1, mark options={solid}, line width=1.4cm, very thick]
table {%
0.01	12.9688923
0.06	10.5566821
0.11	6.7330709
0.16	3.2145475
0.21	-0.4653921
0.26	-5.7925495
0.31	-9.0817822
0.36	-14.2393053
0.41	-18.2158182
0.46	-22.5997541
0.51	-27.7068356
0.56	-32.5822266
0.61	-37.4524866
0.66	-43.5202859
0.71	-46.0030123
0.76	-53.6035702
0.81	-58.5848106
0.86	-67.5683247
0.91	-73.6522196
0.96	-82.0360445
};
\addlegendentry{$\gamma_{JSQ(d)-\pi(d,\infty,0)}$}
\addplot [color3, mark=asterisk, mark size=1, mark options={solid}, line width=1.4cm, very thick]
table {%
0.01	-29.4698331
0.06	-26.7172274
0.11	-24.2349675
0.16	-20.1910045
0.21	-16.5740969
0.26	-10.5795266
0.31	-5.8040338
0.36	2.7182695
0.41	18.028695
0.46	36.6046562
0.51	79.5792952
0.56	175.6305034
};
\addlegendentry{$\gamma_{c.o.c.(d)-\pi(d,\infty,0)}$}

\end{axis}

\end{tikzpicture}}}
		\caption{\small{
		Uniform distribution in $[0.5, 1.5]$.
		}}
		\label{Fig:Compare-GammaSTVsAR-d-Uniform}
\end{subfigure}
	\caption{
	\small{The mean response time loss percentage $\gamma$ as a function of normalized arrival rate $\lambda$ for the policies JIQ($d$), JSW($d$), JSQ($d$), c.o.c.($d$) with respect to the $\pi(d,\infty,0)$ policy, for the number of servers $N=20$ and the number of replicas $d=4$. } 
	} 
	\label{Fig:MRL-ParetoAndUniform}
\end{figure}

\subsection{General service time distribution}
\label{subsec:GeneralService}

We see from our previous analysis that obtaining closed form expressions for the mean response time for our policy can be difficult when service times are not exponentially distributed.
In this section, we provide observations on numerical studies conducted on our policy under non-exponential service time distributions.
In ~\cite{Hellemans19}, the authors discuss the analysis for several workload dependent load balancing policies when job sizes follow a general distribution. 
However, closed form expressions are lacking and the solution is determined numerically. 
Further, the authors of~\cite{Liu2006QS} provide a method to derive the expected workload at cavity queue of the $N$ server system when the jobs are serviced only when the workload at arrival is less than a threshold and when service times are \iid and follow a general distribution. 
They provide expressions, implicit in some cases, for the expected workload when the service times are deterministic or follow phase type, Erlang, or exponential distribution.
This setting matches our special case of $\pi(d,T,T)$  policy and their expressions hold valid for this special case.
However, the computation of mean response time requires numerical evaluations.
Therefore, we do not adopt this methodology in our work and we provide only simulation results for the comparison of our policy under non-exponential service time distributions.

The mean response time loss percentage of JIQ($d$), JSW($d$), JSQ($d$) and c.o.c.($d$) policies against $\pi(d,\infty,0)$ policy is plotted as a function of normalized arrival rate $\lambda$ in Fig.~\ref{Fig:Compare-GammaSTVsAR-d-Weibull}, Fig.~\ref{Fig:Compare-GammaSTVsAR-d-Pareto} and Fig.~\ref{Fig:Compare-GammaSTVsAR-d-Uniform} when the service times follow  Weibull distribution with scale parameter $1$ and shape parameter $5$, Pareto distribution with scale parameter $0.83$ and shape parameter $5.5$, and uniform distribution in the range $[0.5,1.5]$, respectively. 
We observe that except the c.o.c.($d$) policy, the behavior of remaining policies remains similar to that in Fig.~\ref{Fig:Compare-GammaSTVsAR-d} for the exponential distribution case. 
We observe that the response time performance of c.o.c.($d$) policy degrades and the stability region shrinks with the change in service time distribution.
The $\pi(d,\infty,0)$ policy achieves almost the same performance as c.o.c. without any coordination or communication requirements in the low to medium arrival rate regimes under Weibull and Pareto distributions.
From moderate to high arrival rate regime, the c.o.c. policy tends to get unstable and the proposed policy is superior to c.o.c.($d$) in this regime for non-exponential service distributions. 
This plot also demonstrates that the performance improvement of the proposed policy against feedback based policies is not an artifact of choosing exponential service times.



\section{Discussion \& Future work}
\label{sec:discussion}

In this work, we consider load balancing policies without feedback and propose a policy based on timed replicas. 
For every replica that is created, 
the policy sends cancellation instructions to servers along with the replica. 
This instruction specifies an expiry time for the replica and thereby prevents potentially wasteful replicas from being executed.
In this work, we have shown that this policy and several of its special cases, 
offer a marked improvement over the random routing policy for suitable  choice of parameters such as normalized arrival rate $\lambda$ and number of replicas $d$. 
We also observed that under certain parameter regimes, 
the proposed dispatch policy has better performance when compared to feedback based policies. 
We analyze this policy using the cavity queue approach and the assumption on asymptotic independence of queues.
Using the MGF approach, we characterize the mean conditional response time of a job and the loss probability for the policy as part of our key result.

A key assumption in most of our analysis has been the exponential service requirements for jobs, 
and that the job replicas require \iid service time.
We believe that relaxing these assumptions and analyzing the proposed $\pi(d,T_1,T_2)$ policy for more general service time distributions and for the case of identical replicas is an interesting open direction.
One can think of more nuanced policies such as replicating only short jobs if the service requirement of a job is known at arrival. 
This would require no feedback from the server. 
One can also think of incorporating feedback in our proposed policy, and consider replicating only if the primary copy is discarded, 
or decide the number of replicas based on queue state. 
Analyzing such policies is also part of our agenda. 
Further, while the performance of $\pi(d,T_1,T_2)$ seems to be good for lower values of normalized arrival rates $\lambda$, 
it would be interesting to investigate if there exist other no feedback policies that are better than random routing or feedback based policies even for higher values of normalized arrival rates $\lambda$. 
Another interesting direction is to find mean response time optimal load balancing policies, and policies that can utilized server feedback in a more efficient way. 
Finally, we plan to  study the use case for such no feedback policies in an $(n,k)$ fork-join system, 
where a parallelizable job is distributed across $n$ servers and is considered completed when a certain fraction of jobs are executed.

\section*{References}
\bibliography{peva-2022} 
\appendix
\section{Proof sketch for Proposition~\ref{pro:AsymptoticInd}}
\label{app:AsymptoticInd}
This proof for asymptotic independence of server workload during any finite time horizon, 
is based on the results provided in ~\cite[Section 7]{Bramson12}.
Let the $N$ dimensional workload process in the $N$ server system be denoted by $W^N(t) \in \R^N$ for $t \ge 0$ under the given load balancing policy. 
The workload at queue $k$ at time $t$ is denoted by $W^{N,(k)}(t)$.
We assume that all queues start with zero workload, and hence are mutually independent. 
Let $M^N$ be the measure of the workloads over the Borel sets $\cB(\R^N)$ over $\R^N$.
Also, let $M^{N,N'}$ denote the projection of $M^N$ on to the first $N'$ queues.

Recall that the replica service times are assumed to be independent across the queues and the arrival at each queue follows a Poisson distribution with rate $\lambda d$.
At any arrival instant, the set of $d$ servers selected by the servers is referred to as the selection set corresponding to the arrival.
In the load balancing policy studied in~\cite{Bramson12}, whether a newly arrived job is accepted at a certain queue or not depend on the workloads of the other servers in the selection set.
However, in our policy whether a job gets accepted at a queue or not depends only on the current workload at that queue..
That is, suppose that the latest arrival to queue $n$ before time $t$ happened at time $t'$.

Under the policy of threshold based cancellation, whether the job joins the queue $k$ at time $t'$ or not depends only on whether $W^{N,(k)}(t')$ is less than the preset threshold or not.
That is, $W^{N,(k)}(t)$ depends only on $W^{N,(k)}(t')$ and service time of the arriving job.
Recall that we assumed the workloads at all queues to be independent initially.
However, if the job gets accepted at more than one queue in the corresponding selection set, then the correlated arrival of jobs in these queues will make the workloads dependent.

Let us introduce a measure $M^{T,\infty, N'}$ over $\R^{N'}$ which is the $N'$ fold product of $M^{T,\infty, 1}$.  
We need to prove that over a finite time horizon, the joint workload measure of any $N'$ queues converges to the \iid measure $M^{T,\infty, N'}$ asymptotically in the number of queues.
We consider the convergence of the measures in total variation distance.
More precisely, we need to prove that as $N \to \infty$,
$
\lim_{N \to \infty} \sup_{A \in \cB(\R^{N'})} \abs{M^{T,N,N'}(A) - M^{T,\infty, N'}(A)} = 0.
$
Next, we outline the main steps in the proof.

\textbf{Step 1: Construction of an influence process and the number of influencing servers.}

Consider a reversed time process $I^{N,N'}(T-t)$ for $t \in [0,T]$ constructed as given next.
We define $I^{N,N'}(T) \triangleq \set{1,2,\cdots, N'}$.
Now, if there is a potential arrival at time $T-t$ at a queue $n \in I^{N,N'}((T-t)^-)$, then
$
I_{N,N'}(T-t) \triangleq I_{N,N'}((T-t)^-) \cup S
$
where $S$ is the selection set of $d$ servers selected by the new arrival.
Note that the knowledge of service times and intersecting selection sets at each arrival instant can completely describe the workload process $W^{N}(t)$.
We define the number of influencing servers at time $T-t$ as $C^{N,N'}(T-t) \triangleq \abs{I^{N,N'}(T-t)}$ for $t \in [0,T]$.

\textbf{Step 2: Coupling the number of influencing servers with a $d$-ary branching process.}

Couple the number of influencing servers $C^{N,N'}(T-t)$ with a process $C^{\infty,N,N'}(T-t)$ constructed as follows.
We first let $C^{\infty,N,N'}(T) = C^{N,N'}(T) = N'$ and recursively define the process at each arrival instant $t$ as 
$
C^{\infty,N,N'}(T-t) \triangleq C^{\infty,N,N'}((T-t)^-) + d-1 
$   
if there is an intersection between the influence process $I^{N,N'}((T-t)^-)$ and the set of $d$ servers selected by the arrival at time $t$.
Note that $C^{\infty,N,N'}(T-t) \ge C^{N,N'}((T-t))$ always.
We also observe that 
$C^{N,N'}(T-t) = C^{\infty,N,N'}(T-t)$ for all $t \in [0,T]$ 
if the selection set of servers at all arrival instants intersect with either zero or one server 
in $I^{N,N'}(T)$ within the time $[0,T]$. 
We may therefore think of $C^{\infty,N,N'}(T-t)$ as the number of influencing servers in an alternate system with 
the maximum intersection of one server for selections sets with $I^{N,N'}(T-t)$
over the time horizon $T$.
 
As the arrival to each queue occurs according to a Poisson process with rate $\lambda d$, the process $C^{\infty,N,N'}((T-t))$ is a $\lambda d$ branching process. 
It follows from~\cite[Proposition 7.2]{Bramson12} that the process $C^{N,N'}(T)$ converges to $C^{\infty,N,N'}(T)$ in probability for large $N$.
Note that, $C^{N,N'}(T)$ being equal to $C^{\infty,N,N'}(T)$ guarantees that the number of influencing servers $C^{N,N'}(T-t)$ will be equal to the process $C^{\infty,N,N'}(T-t)$ for all $t \in [0,T]$.

\textbf{Step 3: Extension of the influence process to an infinite server system.}

We extend the influence process $I^{N,N'}(T-t)$ to the process  $I^{\infty,N,N'}(T-t)$ satisfying $C^{\infty,N,N'}(T) = \abs{I^{\infty,N,N'}(T)}$.
As hinted earlier, we will be constructing an influence process $I^{\infty,N,N'}(T-t)$ in an alternate system comprising of infinitely many servers where there is an intersection of atmost one server for $I^{\infty,N,N'}(T-t)$ with selection sets over the time horizon $T$.
Therefore, the workloads at the $N'$ servers will stay independent over the time horizon $[0,T]$ in this alternate system. 
This extended influence process is constructed the same way as we construct $I^{N,N'}(T-t)$ except for the following.
If an arrival happens at time $t$ and if $C^{\infty,N,N'}(T-t) \neq C^{N,N'}(T-t)$, include $(C^{\infty,N,N'}(T-t) - C^{N,N'}(T-t))$ new servers to the set $I^{\infty,N,N'}((T-t)^-)$ besides the ones in the new selection set. 
Inclusion of new servers are always possible as we suppose the system has infinitely many servers.
Observe that the branches of this extended influence process starting from each of the $N'$ servers will not intersect each other at any point and remain independent of each other.  
That is, the construction does not allow any correlated arrivals to occur for this workload process, 
preserving the independence of the $N'$ servers under consideration.
That is, the measure corresponding to the workload process at the $N'$ servers of this infinite server system with the given influence process denoted as $M^{T,\infty, N'}$ will be the $N$ fold product of $M^{T,\infty, 1}$ due to the independence of the queues.
From \cite[Lemma 7.2]{Bramson12}, it can be seen that $M^{T,\infty, N'}$ is always independent of $N$.
Further, whenever $C^{N,N'}(T) =C^{\infty,N,N'}(T)$, the processes $I^{N,N'}(T-t)$  and $I^{\infty,N,N'}(T-t)$ are also equal.
As seen in Step 2 that $C^{N,N'}(T)$ converges to $C^{\infty,N,N'}(T)$ in probability for large value of $N$, 
the result follows.
\begin{remark}
We remark that the above proof does not make any assumption on the distribution of the service time except that they are \iid across jobs.
\end{remark}
The extension of this independence across the servers to infinite time interval requires the following monotonicity conditions to be satisfied and do not follow directly.
Let us first define the fraction of servers with workload greater than $w$ at time $t \ge 0$ as
$
x^N_w(t) \triangleq  \frac{1}{N} \sum_{i=1}^N \SetIn{W^{N,(i)}(t) > w}.
$
Then, \cite[Lemma 3]{Shneer2021QS} shows that the workloads under JSW($d$) and water filling($d$) dispatch policies, satisfy the following monotonicity property.

\begin{proposition}
Consider two versions of the process, $x^N(.)$ and $\hat{x}^N(.)$, such that $x^N (0) \le \hat{x}^N(0)$
Then these processes can be coupled so that, with probability $1$,
$x^N (t) \le \hat{x}^N(t)$ for all $t \ge 0$.
\end{proposition}
This monotonicity property holds for our proposed dispatch and cancellation policy only for the special case where the thresholds $T_1$ and $T_2$ are infinity.
The following simplified example shows that the proposed $\pi(d,T_1,T_2)$ policy need not always offer this monotonicity property.
\begin{exmp}{2}
\label{example:MonotonicityViolation}
Consider two single server systems with processes  $x^1_w(t) = \SetIn{W^1(t) > w}$ and $\hat{x}^1_w(t) = \SetIn{\hat{W}^1(t)> w}$ indicating the events that the workloads $W^1(t)$ and $\hat{W}^1(t)$ at the respective servers in these two systems exceed the value $w$ time $t$.
The arrivals to the system follows Poisson distribution with rate $\lambda$ and an arrival is accepted at the server only if the current workload at the server is less than a threshold $T$.
We further suppose that the server services any job at a unit rate.
Assume $W^1(0) = 0$ and $\hat{W}^1(0) = T+ t'$ where $t'$ is a positive constant.
That is, 
$x^1_w(0) = 0 < \hat{x}^1_w(0) = 1 \text{ for all } w < T+t'.
$ 
Suppose the first arrival happens at time $t_1 = t'-h$ which brings in a job of size $c > T+t'$.
As the workload in the second system exceeds the threshold $t$, the new job is accepted only in the first system 
and  the workloads in the coupled systems will be respectively $W^1(t_1) = c$ and $\hat{W}^1(t_1) = T+ h$.
That is,
$ 
x^1_w(t_1) = 1 > \hat{x}^1_w(t_1) = 0 \text{ for } T_1+h < w \le c.
$
This shows that the monotonicity property need not hold under threshold based policies. 
\end{exmp}
We remark that the lack of monotonicity does not necessarily imply that the asymptotic independence of limiting marginal server workloads does not hold, and our simulation studies suggest that the assumption of asymptotic independence remains valid for our selected choice of system parameters.

\section{Model Validation}
\label{app:ModelValidation}
In this section, we discuss the accuracy of our theoretical results and compare them with simulation experiments. 
We obtained the conditional mean sojourn time $\tau$ for undiscarded jobs and the probability of discard $P_L$ under proposed probabilistic redundancy policy $\pi(d,T_1,T_2)$, 
based on the conjecture of the asymptotic independence of the queues. 
The workload distribution for the cavity queue under policy $\pi(d,T_1,T_2)$ has a closed form expression for exponentially distributed service time, 
and is provided in Corollary~\ref{cor:GeneralCase}. 
The expression for the conditional mean sojourn time under policy $\pi(d,T_1,T_2)$ is complex, 
and hence we have omitted it.  
Instead, we restrict our validation results for three special cases: 
(a) deterministic $d$ replicas with identical finite discard threshold $\pi(d,T, T)$,  (b) deterministic $d$ replicas with no discard $\pi(d,\infty, \infty)$, and (c) deterministic $d$ replicas with secondary replicas only at idle servers $\pi(d,\infty, 0)$. 

Findings of the simulation experiments under the policy $\pi(d,T,T)$ are reported in Fig.~\ref{Fig:STVsAR-N-TT}. 
We note that this is a lossy system, where some jobs can be discarded if none of the sampled servers have workload smaller than the threshold $T$.  
We plot the conditional response time for $\pi(d,T,T)$ as a function of normalized arrival rate $\lambda$, 
when the jobs have \iid exponential service times with unit mean. 
The identical discard threshold for primary and secondary replicas is taken as $T_1 = T_2 = 5$ and total number of replicas is selected as $d=3$. 
Each experiment is run over \num{e5} iterations and we repeat this experiment for increasing number of servers $N$. 
We empirically compute the average response time of undiscarded jobs, 
as a function of normalized arrival rate $\lambda$. 
We observe that the empirical curve approaches our analytical computation under asymptotic independence conjecture, 
as the number of servers $N$ increases. 
This provides an empirical validation of the asymptotic independence conjecture, 
and hence our theoretical results.  
In particular, it indicates that even for the most general of our policies, 
the asymptotic independence of queues is indeed true. 

\begin{figure}[h!]
\centerline{\scalebox{0.65}{
\begin{tikzpicture}

\definecolor{color0}{rgb}{0.83921568627451,0.152941176470588,0.156862745098039}
\definecolor{color1}{rgb}{0.12156862745098,0.466666666666667,0.705882352941177}
\definecolor{color2}{rgb}{0.172549019607843,0.627450980392157,0.172549019607843}
\definecolor{color3}{rgb}{1,0.498039215686275,0.0549019607843137}

\begin{axis}[
legend cell align={left},
legend style={fill opacity=0.8, draw opacity=1, text opacity=1, at={(0.47,0.42)}, anchor=north west, draw=white!80!black},
tick align=outside,
tick pos=left,
x grid style={white!69.01960784313725!black, densely dotted},
xlabel={Arrival rate $\lambda$},
xmajorgrids,
xmin=-0.0475, xmax=1.2175,
xminorgrids,
xtick style={color=black},
xtick={-0.2,0,0.2,0.4,0.6,0.8,1,1.2,1.4},
xticklabels={ˆ'0.2,0.0,0.2,0.4,0.6,0.8,1.0,1.2,1.4},
y grid style={white!69.01960784313725!black, densely dotted},
ylabel={Conditional mean response time $\tau$},
ymajorgrids,
ymin=0.0901446454580589, ymax=5.66702006201173,
yminorgrids,
ytick style={color=black}
]
\addplot [semithick, color0, mark=asterisk, mark size=0.75, mark options={solid},line width=0.8pt]
table {%
0.01 0.347009289607324
0.06 0.434409706165839
0.11 0.567519197284339
0.16 0.741601930482381
0.21 1.02059058068424
0.26 1.37971887345767
0.31 1.85685275731543
0.36 2.37812158217408
0.41 2.90086376055552
0.46 3.41167419757119
0.51 3.81142978034758
0.56 4.15184079416535
0.61 4.39626259127593
0.66 4.60375214369891
0.71 4.78534232071293
0.76 4.90772535177907
0.81 5.02961925780632
0.86 5.10185745185504
0.91 5.16814632873345
0.96 5.22809830881278
1.01 5.28609411455868
1.06 5.32749821675266
1.11 5.36555426110933
1.16 5.3981499643883
};
\addlegendentry{Sim-$N=3$}
\addplot [semithick, color1, mark=asterisk, mark size=0.75, mark options={solid},line width=0.8pt]
table {%
0.01 0.344931324608243
0.06 0.421724233210289
0.11 0.522144927354556
0.16 0.682269807063593
0.21 0.898234313277983
0.26 1.24482533562724
0.31 1.67889383250749
0.36 2.20286270541311
0.41 2.70746507612927
0.46 3.25583559533167
0.51 3.67596142504814
0.56 4.03687567545695
0.61 4.31812792319943
0.66 4.579999492159
0.71 4.7304745418793
0.76 4.89073464909779
0.81 4.98603417378011
0.86 5.08441525735044
0.91 5.15573643034084
0.96 5.22880650593677
1.01 5.28262885549231
1.06 5.32716730446062
1.11 5.36071140967278
1.16 5.40648281235059
};
\addlegendentry{Sim-$N=5$}
\addplot [semithick, color2, mark=asterisk, mark size=0.75, mark options={solid},line width=0.8pt]
table {%
0.01 0.344155911036865
0.06 0.412101641462321
0.11 0.510152753631578
0.16 0.651678516083823
0.21 0.875691281560583
0.26 1.17360949837264
0.31 1.59988409408178
0.36 2.10730251175116
0.41 2.65375741506152
0.46 3.11647838526674
0.51 3.63729919325254
0.56 4.00522119828657
0.61 4.30774358650838
0.66 4.52531184086838
0.71 4.69783300618747
0.76 4.86472098084627
0.81 4.97777496211291
0.86 5.07685127806157
0.91 5.14960044216328
0.96 5.22117761094853
1.01 5.27567087058282
1.06 5.33159803171641
1.11 5.3615545735171
1.16 5.40656971173241
};
\addlegendentry{Sim-$N=8$}
\addplot [semithick, color3, mark=asterisk, mark size=0.75, mark options={solid},line width=0.8pt]
table {%
0.01 0.345843831292788
0.06 0.413986474135688
0.11 0.503046226188076
0.16 0.64489511203026
0.21 0.856561759348594
0.26 1.14699010914951
0.31 1.58109595005699
0.36 2.10114050628034
0.41 2.67285954269605
0.46 3.12482774927301
0.51 3.63765593021285
0.56 3.99451791864629
0.61 4.3069818197258
0.66 4.50670811930641
0.71 4.70971964415851
0.76 4.85630988686377
0.81 4.98607080429798
0.86 5.07752770602302
0.91 5.15707207291595
0.96 5.21750413721227
1.01 5.28360133654187
1.06 5.33213708081461
1.11 5.37350320774974
1.16 5.40910599011778
};
\addlegendentry{Sim-$N=10$}
\addplot [semithick, black, mark=diamond*, mark size=0.75, mark options={solid},line width=0.8pt]
table {%
0.01 0.343638982574135
0.06 0.406176922329729
0.11 0.494671716826265
0.16 0.624814379617827
0.21 0.820569884450149
0.26 1.11128390481113
0.31 1.51665262390415
0.36 2.02325953194564
0.41 2.57818212697554
0.46 3.11490615922136
0.51 3.58589567409122
0.56 3.97337591234843
0.61 4.28103504450832
0.66 4.5219762124029
0.71 4.7108002056787
0.76 4.86016470773571
0.81 4.97994688248431
0.86 5.07749106070225
0.91 5.15815246822889
0.96 5.22582211484143
1.01 5.28334315810097
1.06 5.33281566240513
1.11 5.37581180233453
1.16 5.41352572489565
};
\addlegendentry{Th}
\end{axis}

\end{tikzpicture}}}
\caption{\small{For the policy $\pi(d,T,T)$ with fixed thresholds $T_1=T_2=5$, number of replicas $d=3$,service rate $\mu=1$, conditional mean response time $\tau$ as function of arrival rate $\lambda$ for different values of servers $N \in \{3,5,8,10\}$.}}
\label{Fig:STVsAR-N-TT}
\end{figure}

When the primary discard threshold is infinite, 
then all jobs get served.  
We illustrate a similar validation for two special cases where the primary replica is never discarded. 
The results for deterministic $d$ replicas with no discard ($\pi(d,\infty, \infty)$ policy) is presented in Fig.~\ref{Fig:STVsAR-N-InfInf},   
and for deterministic $d$ replicas with secondary on idle servers ($\pi(d,\infty, 0)$ policy) in Fig.~\ref{Fig:STVsAR-N-Inf0}.  
The closed form theoretical expressions of the conditional mean response time of these policies are provided in Remark~\ref{rem:Tinfinity} and Lemma~\ref{lem:E[R]Policy3} respectively. 
As in the case of $\pi(d,T,T)$, 
we see that the empirically computed mean response time of undiscarded job converge to the corresponding theoretical expression with increase in number of servers $N$. 
This indicates that as the number of servers $N$ increases the workload across queues tend to be independent, 
validating our conjecture on the asymptotic independence of queues. 
It is remarkable to note that the theoretical values and those obtained empirically from the simulation, 
coincide even when the number of servers $N$ is as low as 10. 

\begin{figure}[h!]
\centerline{\scalebox{0.665}{
\begin{tikzpicture}

\definecolor{color0}{rgb}{0.83921568627451,0.152941176470588,0.156862745098039}
\definecolor{color1}{rgb}{0.12156862745098,0.466666666666667,0.705882352941177}
\definecolor{color2}{rgb}{0.172549019607843,0.627450980392157,0.172549019607843}
\definecolor{color3}{rgb}{1,0.498039215686275,0.0549019607843137}

\begin{semilogyaxis}[
legend cell align={left},
legend style={fill opacity=0.8, draw opacity=1, text opacity=1, at={(0.03,0.97)}, anchor=north west, draw=white!80!black},
tick align=outside,
tick pos=left,
x grid style={white!69.01960784313725!black, densely dotted},
xlabel={Arrival rate $\lambda$},
xmajorgrids,
xmin=-0.006, xmax=0.34,
xminorgrids,
xtick style={color=black},
xtick={-0.05,0,0.05,0.1,0.15,0.2,0.25,0.3,0.35},
xticklabels={0.05,0.00,0.05,0.10,0.15,0.20,0.25,0.30,0.35},
y grid style={white!69.01960784313725!black, densely dotted},
ylabel={Conditional mean response time $\tau$},
ymajorgrids,
ymin=0, ymax=10,
yminorgrids,
ytick style={color=black}
]
\addplot [thick, color0, mark=asterisk, mark size=0.75, mark options={solid},line width=1pt]
table {%
0.01 0.349241879950632
0.03 0.378707489706853
0.05 0.416050728796163
0.07 0.458331550216724
0.09 0.508497942164563
0.11 0.565333708525054
0.13 0.637263820793518
0.15 0.717793540786269
0.17 0.837392166134741
0.19 0.962536627302006
0.21 1.14468573404966
0.23 1.37552012110609
0.25 1.77130051276225
0.27 2.51449807762828
0.29 3.66846201661279
0.31 7.21770612911541
};
\addlegendentry{Sim-$N=3$}
\addplot [thick, color1, mark=asterisk, mark size=0.75, mark options={solid},line width=1pt]
table {%
0.01 0.34363445996354
0.03 0.373949192676979
0.05 0.40493294355595
0.07 0.43648800045713
0.09 0.478451962817448
0.11 0.523987499266581
0.13 0.578334006304094
0.15 0.653899586409193
0.17 0.732336148826539
0.19 0.854605906517816
0.21 1.01682813670843
0.23 1.20594426732495
0.25 1.4998058602392
0.27 2.02297052569715
0.29 3.00033305027128
0.31 5.73693245829251
};
\addlegendentry{Sim-$N=5$}
\addplot [thick, color2, mark=asterisk, mark size=0.75, mark options={solid},line width=1pt]
table {%
0.01 0.345717389094007
0.03 0.369997212126001
0.05 0.398129251120963
0.07 0.428955590017917
0.09 0.469143563328467
0.11 0.514917565676957
0.13 0.565701926451354
0.15 0.637899473647716
0.17 0.718128149344169
0.19 0.817765449513664
0.21 0.942256911127097
0.23 1.15909817066189
0.25 1.42051133889116
0.27 1.90200848817428
0.29 2.88297043732052
0.31 4.8832282455336
};
\addlegendentry{Sim-$N=8$}
\addplot [thick, color3, mark=asterisk, mark size=0.75, mark options={solid},line width=1pt]
table {%
0.01 0.342769976835491
0.03 0.368799380341911
0.05 0.397499091796345
0.07 0.427990170357506
0.09 0.463279928009664
0.11 0.507487881662547
0.13 0.565597216642268
0.15 0.626713124332357
0.17 0.713723003771239
0.19 0.811468907271856
0.21 0.937025236812803
0.23 1.15332901132011
0.25 1.3903814776239
0.27 1.89232225023049
0.29 2.66197030319242
0.31 5.26393016364552
};
\addlegendentry{Sim-$N=10$}
\addplot [thick, black, mark=diamond*, mark size=0.75, mark options={solid},line width=1pt]
table {%
0.01 0.343642611683849
0.03 0.366300366300366
0.05 0.392156862745098
0.07 0.421940928270042
0.09 0.45662100456621
0.11 0.497512437810945
0.13 0.546448087431694
0.15 0.606060606060606
0.17 0.680272108843537
0.19 0.775193798449613
0.21 0.900900900900901
0.23 1.0752688172043
0.25 1.33333333333333
0.27 1.75438596491228
0.29 2.56410256410256
0.31 4.76190476190476
};
\addlegendentry{Th}
\end{semilogyaxis}

\end{tikzpicture}}}
\caption{\small{For the policy $\pi(d,\infty,\infty)$ with number of replicas $d=3$, service rate $\mu=1$, conditional mean response time $\tau$ as function of arrival rate $\lambda$ for different values of servers $N \in \{3,5,8,10\}$.}}
\label{Fig:STVsAR-N-InfInf}
\end{figure}

\begin{figure}[h!]
\centerline{\scalebox{0.65}{
\begin{tikzpicture}

\definecolor{color0}{rgb}{0.83921568627451,0.152941176470588,0.156862745098039}
\definecolor{color1}{rgb}{0.12156862745098,0.466666666666667,0.705882352941177}
\definecolor{color2}{rgb}{0.172549019607843,0.627450980392157,0.172549019607843}
\definecolor{color3}{rgb}{1,0.498039215686275,0.0549019607843137}

\begin{semilogyaxis}[
legend cell align={left},
legend style={fill opacity=0.8, draw opacity=1, text opacity=1, at={(0.03,0.97)}, anchor=north west, draw=white!80!black},
tick align=outside,
tick pos=left,
x grid style={white!69.01960784313725!black, densely dotted},
xlabel={Arrival rate $\lambda$},
xmajorgrids,
xmin=-0.006, xmax=0.346,
xminorgrids,
xtick style={color=black},
xtick={-0.05,0,0.05,0.1,0.15,0.2,0.25,0.3,0.35},
xticklabels={0.05,0.00,0.05,0.10,0.15,0.20,0.25,0.30,0.35},
y grid style={white!69.01960784313725!black, densely dotted},
ylabel={Conditional mean response time $\tau$},
ymajorgrids,
ymin=0.303684499674208, ymax=1.29808458081528,
yminorgrids,
ytick style={color=black},
ytick={0.2,0.4,0.6,0.8,1,1.2,1.4},
yticklabels={0.2,0.4,0.6,0.8,1.0,1.2,1.4}
]
\addplot [thick, color0, mark=asterisk, mark size=0.75, mark options={solid},line width=1pt]
table {%
0.01 0.362325747934106
0.03 0.416562408000323
0.05 0.472034553737977
0.07 0.525126811573398
0.09 0.580435934287591
0.11 0.631445847799273
0.13 0.683868709149909
0.15 0.735809298554199
0.17 0.784496784184296
0.19 0.851620453294554
0.21 0.903926197634315
0.23 0.957330769209906
0.25 0.990713419962578
0.27 1.05833000587738
0.29 1.10871658606827
0.31 1.16991014427906
0.33 1.25288457712705
};
\addlegendentry{Sim-$N=3$}
\addplot [thick, color1, mark=asterisk, mark size=0.75, mark options={solid},line width=1pt]
table {%
0.01 0.352441828612103
0.03 0.392038354229535
0.05 0.431797190774525
0.07 0.480142723895274
0.09 0.517309269053038
0.11 0.558162091585514
0.13 0.606190831777223
0.15 0.65712419311844
0.17 0.704998123090926
0.19 0.758397400328677
0.21 0.806587984826848
0.23 0.865034126357889
0.25 0.912651218198097
0.27 0.970613812408621
0.29 1.00899945338457
0.31 1.08043024390118
0.33 1.12852727248397
};
\addlegendentry{Sim-$N=5$}
\addplot [thick, color2, mark=asterisk, mark size=0.75, mark options={solid},line width=1pt]
table {%
0.01 0.348884503362439
0.03 0.385230422486953
0.05 0.420942898159203
0.07 0.46297481414006
0.09 0.499999025118603
0.11 0.543592702240949
0.13 0.584296820047099
0.15 0.631277191313202
0.17 0.680744372030957
0.19 0.729153619334683
0.21 0.769004537644601
0.23 0.816699575717084
0.25 0.874227699623874
0.27 0.937661295283558
0.29 0.995389718656944
0.31 1.04638922014491
0.33 1.09882652248872
};
\addlegendentry{Sim-$N=8$}
\addplot [thick, color3, mark=asterisk, mark size=0.75, mark options={solid},line width=1pt]
table {%
0.01 0.349684779709869
0.03 0.381623366640269
0.05 0.419772536658788
0.07 0.452612062681607
0.09 0.495345075838337
0.11 0.539232073695502
0.13 0.578614379894868
0.15 0.618273282212782
0.17 0.67542132239494
0.19 0.70983994451452
0.21 0.765160005886813
0.23 0.818267464077013
0.25 0.862275300644053
0.27 0.922376461505449
0.29 0.975918511956584
0.31 1.04566903348875
0.33 1.09196485136214
};
\addlegendentry{Sim-$N=10$}
\addplot [violet, black, mark=asterisk, mark size=0.75, mark options={solid},line width=1pt]
table {%
0.01 0.346962082685152
0.03 0.375997673416833
0.05 0.407383506293497
0.07 0.441059775150267
0.09 0.476947986772122
0.11 0.51496678448918
0.13 0.555041916726913
0.15 0.597112458814879
0.17 0.641134666899586
0.19 0.68708437377452
0.21 0.734958533143902
0.23 0.784776322324343
0.25 0.836580086580087
0.27 0.890436327682623
0.29 0.946436889943445
0.31 1.00470046943843
0.33 1.06537456057651
};
\addlegendentry{Th}
\end{semilogyaxis}

\end{tikzpicture}}}
\caption{\small{For the policy $\pi(d,\infty,0)$ with number of replicas $d=3$, service rate $\mu=1$, conditional mean response time $\tau$ as function of arrival rate $\lambda$ for different values of servers $N \in \{3,5,8,10\}$.}}
\label{Fig:STVsAR-N-Inf0}
\end{figure}

Even though, 
we have performed extensive validations for different values of $T_1$ and $T_2$ (for which closed form results are available) and have observed a similar behavior with increase in number of servers $N$, 
we have presented only a select few of the plots validating our models. 
\section{Proof of Theorem~\ref{thm:MeanResponseTime}}
\label{App:ProofMeanResponseTime}
\begin{proof}
The tail distribution of an undiscarded job in the system is denoted by $\bH$ and defined in Definition~\ref{def:MeanResponseTime}. 
Therefore, the mean response time for an undiscarded job can be written as
$\E[R] = \int_{x}  \bH(x) dx.$ 
Next, we derive an expression for the tail distribution $\bH$ of the response time for each undiscarded job under $\pi(d,T_1,T_2)$ policy. 
Recall that $I_1,I_2$ denote the disjoint random sets of servers where primary and secondary replicas are dispatched.   
Then, the indicator that the job replica at server~$j \in I_1\cup I_2$ with workload $W_j$ is not discarded is defined in~\eqref{eqn:IndicatorJobDiscardServer}. 
Recall that the set of servers, 
where the job replicas are not discarded is denoted by  
$I = \set{j \in I_1: \xi_j=1}\cup\set{j \in I_2:  \xi_j = 1}$, 
and the indicator of an undiscarded job is $\xi = \SetIn{I \neq \emptyset}$. 
Therefore, we can write the indicator of response time of an undiscarded job being larger than a threshold $x \in R_+$ as 
\EQ{
\SetIn{R > x} = \xi\prod_{j \in I}\SetIn{W_j+X_j > x}
= \xi\prod_{j \in I_1}(\xi_j\SetIn{W_j+X_j > x}+ \bxi_j)\prod_{j \in I_2}[(\xi_j\SetIn{W_j+X_j > x}+ \bxi_j)] .
}
Substituting~\eqref{eqn:IndicatorJobDiscard} for the indicator $\xi$ in the above equation,  
using the fact that $\xi_j\bxi_j= 0$, 
and re-arranging the terms, 
we can write 
\EQ{
\SetIn{R > x}  
= \prod_{j \in I_1\cup I_2}(\xi_j\SetIn{W_j+X_j > x}+ \bxi_j) - \prod_{j \in I_1\cup I_2}\bxi_j.
}
 
Taking expectation on both sides of the above equations,  
using the independence of indicators $(\xi_j: j \in I_1\cup I_2)$ 
with the mean $\E[\xi_j\SetIn{j \in I_1\cup I_2}\given I_1,I_2] = F(T_1)\gamma^1_j + F(T_2)\gamma^2_j$, 
and the definition of $k(x,T) = \expect{\SetIn{W_j \le T}\SetIn{X_j+W_j > x}}$, 
we obtain the tail distribution of the response time for an undiscarded job as 
$
\E[\SetIn{R > x}|I_1,I_2] 
= (k(x,T_1)+\bF(T_1))(k(x,T_2)+\bF(T_2))^{d-1}
 -\bF(T_1)\bF(T_2)^{d-1}.
$
Since the right hand side of the preceding equation doesn't depend on $I_1, I_2$, 
 we have $\bar{H}(x) = \E[\SetIn{R > x}|I_1,I_2]$.    
The result follows from Eq.~\eqref{eq:mrtDefinition} for conditional mean of response time. 
\end{proof}

\section{Proof of Theorem~\ref{thm:mgfGeneral}}
\label{app:ProofmgfGeneral}
This section provides the moment generating function based approach for deriving the stationary workload distribution in a single queue in an $N$ server system with \iid service times and Poisson arrivals with thtreshold based dispatching policy, $\pi(d,T_1,T_2)$ .
Although, the proof is provided only for the case where the service times are exponentially distributed with rate $\mu$, the same approach can be used when the service times follow a shifted exponential distribution.
We omit the details due to space constraints.
Let us now begin the proof by providing two simple results.
\begin{lemma}
For the interarrival time sequence $(T_n: n \in \N)$, 
we have
\EQN{
\label{eqn:TailMGFInterArrival}
\expect{e^{\theta T_{n+1}}\SetIn{W_n+X_n > T_{n+1}}|W_n,X_n} 
 = \frac{N\lambda}{N\lambda-\theta}(1-e^{-(N\lambda-\theta)(W_n+X_n)}). 
}
\end{lemma}
\begin{proof}
Recall that interarrival times $(T_n: n\in \N)$ are \iid exppnential with rate $N\lambda$,  
and duration $T_{n+1}$ is independent of past workloads $(W_1, \dots, W_n)$ and past and present service times $(X_1, \dots, X_n)$ for all $n \in \Z_+$.  
Hence, the result follows. 
\end{proof}

\begin{lemma}
For \iid exponential service time sequence $(X_n: n \in \N)$ with rate $\mu$, 
we have 
\EQN{
\label{eqn:TailMGFService}
\expect{e^{-\theta X_n}\SetIn{X_n < T-W_n}|W_n} 
= \Phi_X(\theta)(1-e^{-(\mu+\theta)(T-W_n)_+}). 
}
In addition, we have the following identity
\EQN{
\label{eqn:MGFEquality}
\frac{\Phi_X(\theta)-1}{\theta} = -\frac{1}{\mu}\Phi_X(\theta).
}
\end{lemma}
\begin{proof}
The $n$th service time $X_n$ is independent of workloads $(W_1, \dots, W_n)$ seen by first $n$ incoming arrivals.  
The first equality follows from this observation.
The second equality follows from the fact that $\Phi_X(\theta) = \frac{\mu}{\mu+\theta}$. 
\end{proof}
\begin{proposition}
\label{pro:mgfGeneral}
For an $N$ server system with i.i.d. exponential service times of rate $\mu$, Poisson arrivals of rate $N\lambda$ under $\pi(d,T_1,T_2)$ policy and the moment genrating functions of the limiting workload $W$ in a single queue defined in definition~\ref{def:mgf},
\eqn{
\label{eqn:MGFWorkload}
\Phi_W(\theta) &= F(0)(1 + \frac{\bl}{\theta + \mu - \bl}) + \big((\mu - \lambda)\bF(T_2) + \lambda \bF(T_1)\big)e^{-\theta T_2}\big[\frac{1}{\theta + \mu - \lambda} - \frac{1}{\theta + \mu - \bl}\big]  \\ \nonumber
&- \bF(T_1)\mu e^{-\theta T_1}\big[\frac{1}{\theta + \mu - \lambda} - \frac{1}{\theta + \mu} \big]. 
}
%
%
\end{proposition}
\begin{proof}
From~\eqref{eq:LindleysRecursion}, we can write the restricted moment generating function for $W_{n+1}$ in terms of $W_n$. 
We assume that there exists a limiting workload distribution $\lim_{n\in \N}P\set{W_n \le w}$ seen by an arriving customer, 
which equals the limiting distribution of workload in the system by the PASTA property.    
At stationarity, we will take the distribution of both $W_{n+1}$ and $W_n$ as the limiting distribution $F$.  

Now let us compute $\Phi_W(\theta)$.
From the definition of moment generating function for workload at $(n+1)$th arrival is given by 
$
\Phi_W(\theta) 
= \expect{e^{-\theta W_{n+1}}\SetIn{W_n > T_1}} 
+ \expect{e^{-\theta W_{n+1}}\SetIn{T_2 < W_n \le T_1}} 
+ \expect{e^{-\theta W_{n+1}}\SetIn{W_n \le T_2}}. 
$
We will derive the three terms separately. 
We first observe that in the region $W_n > T_1$, 
we have $W_{n+1} = (W_n - T_{n+1})\SetIn{W_n > T_{n+1}}$ from~\eqref{eq:WorkloadRecursionGeneral}.
Using the identity in~\eqref{eqn:TailMGFInterArrival}, we can write 
\eq{
\expect{e^{-\theta W_{n+1}}\SetIn{W_n > T_1}} 
= \expect{\SetIn{W_n > T_1}e^{-N\lambda W_n}} +\frac{N\lambda}{N\lambda-\theta}\expect{\SetIn{W_n > T_1}(e^{-\theta W_n}-e^{-N\lambda W_n})}= \frac{N\lambda \Phi_1(\theta)-\theta \Phi_1(N\lambda)}{N\lambda-\theta}. 
}
We next observe that in the region $W_n \in (T_2, T_1]$, 
we have $W_{n+1} = (W_n-T_{n+1})\SetIn{W_n  > T_{n+1}}$ with probability $1- \frac{1}{N}$, 
and $W_{n+1} = (W_n+X_n-T_{n+1})\SetIn{W_n+ X_n > T_{n+1}}$ with probability $\frac{1}{N}$. 
We can write 
\eq{
\expect{e^{-\theta(X_n+W_n-T_{n+1})_+}\SetIn{W_n \in (T_2,T_1]}}
= \frac{1}{N\lambda-\theta} \Big( N\lambda(\Phi_2(\theta)-\Phi_1(\theta))\Phi_X(\theta)-\theta(\Phi_2(N\lambda)-\Phi_1(N\lambda))\Phi_X(N\lambda) \Big).
}  
For $X_n=0$ the moment generating function $\Phi_X(\theta) = 1$, 
and hence combining the above two equations, we get 
$
\expect{e^{-\theta W_{n+1}}\SetIn{W_n > T_2}} 
= \Big[-\frac{\theta}{N\lambda-\theta}\Phi_2(N\lambda) + \frac{N\lambda}{N\lambda-\theta}\Phi_2(\theta)\Big]+ \frac{\lambda\theta}{N\lambda-\theta}\Big[-(\Phi_2(N\lambda)-\Phi_1(N\lambda))\frac{(\Phi_X(N\lambda)-1)}{N\lambda} 
+ (\Phi_2(\theta)-\Phi_1(\theta))\frac{(\Phi_X(\theta)-1)}{\theta}\Big].
$
We next observe that in the region $W_n \le T_2$, 
we have $W_{n+1} = (W_n-T_{n+1})\SetIn{W_n  > T_{n+1}}$ with probability $1- \frac{\bl}{N\lambda}$, 
and $W_{n+1} = (W_n+X_n-T_{n+1})\SetIn{W_n+ X_n > T_{n+1}}$ with probability $\frac{\bl}{N\lambda}$. 
Repeating the steps followed above for the region $W_n > T_1$ and rearranging, we get 
$
\Phi_W(N\lambda)
+ \Big[\lambda(\Phi_2(N\lambda)-\Phi_1(N\lambda))
+\bl(\Phi_W(N\lambda)-\Phi_2(N\lambda))\Big]
\frac{(\Phi_X(N\lambda)-1)}{N\lambda}
= \Phi_W(\theta)
+ \Big[\lambda(\Phi_2(\theta)-\Phi_1(\theta))
+ \bl(\Phi_W(\theta)-\Phi_2(\theta))\Big]
\frac{(\Phi_X(\theta)-1)}{\theta}.
$
We observe that LHS and RHS have the form $f(\theta) = f(N\lambda)$ for an arbitrary function $f$ and variables $\theta$ and $\lambda$. 
Therefore, we conclude that $f(\theta) = f(0)$.
Further, note that  $\Phi_i(0) = \bF_{T_i}$ for $i \in [2]$.
Then, using equation~\eqref{eqn:MGFEquality}, we can write for exponential service times, 
\eq{
\Phi_W(\theta)\Big(1-\frac{\bl}{\mu}\Phi_X(\theta)\Big)
+ \Big[\frac{\bl-\lambda}{\mu}\Phi_2(\theta)
+ \frac{\lambda}{\mu}\Phi_1(\theta)\Big]\Phi_X(\theta)
= 1 -\frac{\bl}{\mu} + \Big[\frac{\bl-\lambda}{\mu}\bF(T_2)
+ \frac{\lambda}{\mu}\bF(T_1)\Big].
}  
Now, we substitute $\Phi_1(\theta)$ and $\Phi_2(\theta)$ from equations~\eqref{eqn:MGFTailPri} and \eqref{eqn:MGFTailSec} respectively in the above equation. 
Further incorporating equations ~\eqref{eq:bFT1} and \eqref{eq:bFT2} and rearranging the terms will yield equation~\eqref{eqn:MGFWorkload}.
\end{proof}
\begin{remark}
Upon inverting the moment generating function in equation~\eqref{eqn:MGFWorkload}, we see that the complementary workload distribution function for $w \ge 0$ is given by
$
\bF(w) = 1 - F(0)\Big(1+\frac{\bl(1 - e^{-(\mu - \bl)w})}{\mu - \bl} \Big)  + \mu \bF(T_1) \Big(\frac{(1 - e^{-(\mu - \lambda)(w-T_1)_+})}{\mu - \lambda} - \frac{(1 - e^{-\mu(w-T_1)_+})}{\mu} \Big)
- ((\mu -\lambda)\bF(T_2) + \lambda\bF(T_1))\Big(\frac{(1 - e^{-(\mu - \lambda)(w-T_2)_+})}{\mu - \lambda} - \frac{(1 - e^{-(\mu - \bl)(w-T_2)_+})}{\mu - \bl} \Big).  
$
In addition, we can find the constant,
$
F(0) = 1 -\frac{\bl}{\mu} + \Big[\frac{\bl-\lambda}{\mu}\bF(T_2)
+ \frac{\lambda}{\mu}\bF(T_1)\Big].
$  
\end{remark}

\begin{proposition}
For an $N$ server system with i.i.d. exponential service times of rate $\mu$, Poisson arrivals of rate $N\lambda$ under $\pi(d,T_1,T_2)$ policy and the moment generating functions of the limiting workload $W$ in a single queue defined in definition~\ref{def:mgf},
\eqn{
\label{eqn:MGFTailPri}
\Phi_1(\theta) = e^{-\mu T_1}\Big[\frac{\lambda}{\mu} 
(\Phi_2(-\mu)-\Phi_1(-\mu))
+  \frac{\bl}{\mu}(\Phi(-\mu)-\Phi_2(-\mu))\Big]e^{-\theta T_1}\Phi_X(\theta).
}
This implies that for $w > T_1$,
$
\bF(w) = \bF(T_1)e^{-\mu(w-T_1)_+},
$,
where 
\eqn{
\label{eq:bFT1}
\bF(T_1) =  e^{-\mu T_1}\Big[\frac{\lambda}{\mu} 
(\Phi_2(-\mu)-\Phi_1(-\mu))
+ \frac{\bl}{\mu}(\Phi(-\mu)-\Phi_2(-\mu))\Big].
}
\end{proposition}
\begin{proof}
The computation remains similar to the previous step, with an additional restriction of $W_{n+1} > T_1$. 
Therefore, we can write
\eqn{
\label{eq:mgf1}
\Phi_1(\theta) = \E[e^{-\theta W_{n+1}}\SetIn{W_{n+1}>T_1}
\Big(\SetIn{W_n > T_1}  
+ \SetIn{T_2 < W_n \le T_1} + \SetIn{W_n \le T_2}\Big)].
}
We sequentially compute the first term, 
the summation of first two terms, 
and the summation of all three terms as before. 
In the region $W_n > T_1$, 
we have $e^{-\theta W_{n+1}}\SetIn{W_{n+1}>T_1}\SetIn{W_n > T_1} = e^{-\theta(W_n-T_{n+1})}\SetIn{T_{n+1}<W_n-T_1}\SetIn{W_n > T_1}$.  
Then, it follows that 
\eq{
&\expect{e^{-\theta W_{n+1}}\SetIn{W_{n+1}>T_1}\SetIn{W_n > T_1}} = \frac{N\lambda}{N\lambda-\theta}\expect{e^{-\theta W_n}\SetIn{W_n > T_1}(1-e^{-(N\lambda-\theta)(W_n-T_1)})}= \frac{N\lambda}{N\lambda-\theta}\Big(\Phi_1(\theta)- e^{(N\lambda-\theta)T_1}\Phi_1(N\lambda)\Big).
}
Note that, in the region $W_n \le T_1$, 
it is not possible for $W_{n+1} > T_1$, unless the $n$th arrival with service time $X_n$ is admitted at the cavity queue. 
This occurs with probability $\frac{1}{N}$ in region $T_2 < W_n \le T_1$, 
and with probability $\frac{\bl}{N\lambda}$ in region $W_n \le T_2$. 
Therefore, for the region $T_2 < W_n \le T_1$, 
we can write  
\eq{
\expect{e^{-\theta W_{n+1}}\SetIn{W_{n+1}>T_1}\SetIn{T_2 < W_n \le T_1}}
= \frac{\lambda e^{-(\mu+\theta)T_1}}{N\lambda-\theta}
(\Phi_2(-\mu)-\Phi_1(-\mu))(\Phi_X(\theta)- \Phi_X(N\lambda)).
} 

Similarly, for the region $W_n \le T_2$, 
we can write  
\eq{
\expect{e^{-\theta W_{n+1}}\SetIn{W_{n+1}>T_1}\SetIn{W_n \le T_2}} 
= \frac{\bl e^{-(\mu+\theta)T_1}}{N\lambda-\theta}(\Phi(-\mu)-\Phi_2(-\mu))(\Phi_X(\theta)- \Phi_X(N\lambda)).
} 
Substituting the above three equations in equation~\eqref{eq:mgf1} and simplifying as in the previous proof, we get 
$
\Phi_1(\theta) =  \Big[\lambda 
(\Phi_2(-\mu)-\Phi_1(-\mu))
+ \bl (\Phi(-\mu)-\Phi_2(-\mu))\Big] \frac{e^{-(\mu + \theta) T_1}}{\mu}\Phi_X(\theta).
$
The result follows by inverting the moment generating function and noting that $\Phi_1(0) = \bF(T_1)$.
\end{proof}
\begin{proposition}
For an $N$ server system with i.i.d. exponential service times of rate $\mu$, Poisson arrivals of rate $N\lambda$ under $\pi(d,T_1,T_2)$ policy and the moment generating functions of the limiting workload $W$ in a single queue defined in definition~\ref{def:mgf},
\eqn{
\label{eqn:MGFTailSec}
\Phi_2(\theta) 
= \frac{\lambda}{\mu}(\Phi_2(\theta)-\Phi_1(\theta))\Phi_X(\theta) 
+ \frac{\bl}{\mu}e^{-\mu T_2}\big(\Phi(-\mu) 
-\Phi_2(-\mu)\big)e^{-\theta T_2}\Phi_X(\theta). 
}
This implies that for $w > T_2$,
$
\bF(w) = \bF(T_1)\Big(e^{-\mu(w-T_1)_+}-\frac{\mu}{\mu-\lambda}e^{-(\mu-\lambda)(w-T_1)_+}\Big)+\Big[e^{-\mu T_2}(\Phi(-\mu)-\Phi_2(-\mu))\Big]\frac{\bl}{\mu-\lambda}e^{-(\mu-\lambda)(w-T_2)_+}.
$
In addition,
\EQN{
\label{eq:bFT2}
\bF(T_2) = \frac{\bl}{\mu-\lambda}e^{-\mu T_2}(\Phi(-\mu)-\Phi_2(-\mu)) - \frac{\lambda}{\mu-\lambda}\bF(T_1).
}
\end{proposition}
\begin{proof}
The computation remains similar to the previous case but here we have the restriction of $W_{n+1} > T_2$. 
Then, we can write
$
\Phi_2(\theta) = \E[e^{-\theta W_{n+1}}\SetIn{W_{n+1}>T_2}
\Big(\SetIn{W_n > T_2} + \SetIn{T_2 < W_n \le T_1}
  + \SetIn{W_n \le T_2}\Big)].
$
We sequentially compute the first term, 
the summation of first two terms, 
and the summation of all three terms as before. 
The indicator $W_{n+1} > T_2$ implies that $W_{n+1}$ can't be zero. 
In the region $W_n > T_1$, 
we have $e^{-\theta W_{n+1}}\SetIn{W_{n+1}>T_2}\SetIn{W_n > T_1} = e^{-\theta(W_n-T_{n+1})}\SetIn{T_{n+1}<W_n-T_2}\SetIn{W_n > T_1}$.  
Therefore, it follows that 
$
\expect{e^{-\theta W_{n+1}}\SetIn{W_{n+1}>T_2}\SetIn{W_n > T_1}}=  
\frac{N\lambda \Big(\Phi_1(\theta)- e^{(N\lambda-\theta)T_2}\Phi_1(N\lambda)\Big)}{N\lambda-\theta}.
$
Similarly, for the region $T_2 < W_n \le T_1$, 
an external arrival is admitted with probability $\frac{1}{N}$. 
When there is no arrival $W_{n+1} = W_n - T_{n+1}$, and we have  
$
\expect{e^{-\theta(W_n-T_{n+1})}\SetIn{W_n - X_{n+1}>T_2}\SetIn{T_2 < W_n \le T_1}} = \frac{N\lambda\Big(\Phi_2(\theta)-\Phi_1(\theta) - e^{(N\lambda-\theta)T_2}(\Phi_2(N\lambda)-\Phi_1(N\lambda))\Big)}{N\lambda-\theta}.
$
In the region $T_2 < W_n \le T_1$, the $n$th arrival with service time $X_n$ is admitted at the cavity queue with probability $\frac{1}{N}$.  
In this case, $W_{n+1} = W_n + X_n - T_{n+1}$, and we can write  
$
\expect{e^{-\theta(W_n+X_n-T_{n+1})}\SetIn{W_n+X_n-T_{n+1}>T_2}\SetIn{T_2 < W_n \le T_1}}  
= \frac{N\lambda}{N\lambda-\theta}\Big[(\Phi_2(\theta)-\Phi_1(\theta))\Phi_X(\theta)- e^{(N\lambda-\theta)T_2}(\Phi_2(N\lambda)-\Phi_1(N\lambda))\Phi_X(N\lambda)\Big].
$
Combining these results in the region $W_n > T_2$, we can write 
\eq{
\expect{e^{-\theta(W_{n+1})}\SetIn{W_{n+1}>T_2}\SetIn{W_n> T_2}} =& \frac{N\lambda \Big[\Phi_2(\theta)-e^{(N\lambda-\theta)T_2}\Phi_2(N\lambda)\Big]}{N\lambda-\theta} \\
&+ \frac{\lambda}{N\lambda-\theta}\Big[(\Phi_2(\theta)-\Phi_1(\theta))(\Phi_X(\theta)-1)-e^{(N\lambda-\theta)T_2}(\Phi_2(N\lambda)-\Phi_1(N\lambda))(\Phi_X(N\lambda)-1)\Big].  
} 
In the region $W_n \le T_2$, 
it's not possible for $W_{n+1} > T_2$, 
unless the $n$ arrival with service time $X_n$ is admitted at the cavity queue. 
This occurs with probability $\frac{\bl}{N\lambda}$, 
and we can write  
\eq{
 \expect{e^{-\theta W_{n+1}}\SetIn{W_{n+1}>T_1}\SetIn{W_n \le T_2}}= \frac{\bl e^{-(\mu+\theta)T_2}}{N\lambda-\theta}  
(\Phi(-\mu)-\Phi_2(-\mu))(\Phi_X(\theta)- \Phi_X(N\lambda)).
} 
Combining the above equations and simplifying as in the previous proof, we obtain
\eqn{
\Phi_2(\theta) 
= \frac{\lambda}{\mu}(\Phi_2(\theta)-\Phi_1(\theta))\Phi_X(\theta) 
+\frac{\bl}{\mu}e^{-\mu T_2}(\Phi(-\mu)-\Phi_2(-\mu))e^{-\theta T_2}\Phi_X(\theta). 
}
To prove the second statement, note that $\Phi_1(\theta) = \bF(T_1)e^{-\theta T_1}\Phi_X(\theta)$ from equation~\eqref{eqn:MGFTailPri}. 
Substitution and simplification tells us that 
$
\Phi_2(\theta) = \Big(\frac{1}{\mu+\theta}-\frac{1}{\mu-\lambda+\theta}\Big) \mu\bF(T_1)e^{-\theta T_1}
+ \frac{\bl e^{-\theta T_2}}{(\mu-\lambda+\theta)} e^{-\mu T_2}(\Phi(-\mu)-\Phi_2(-\mu))
$
when service times are exponentially distributed with rate $\mu$.
The result follows by inverting this moment generating function and the fact that $\Phi_2(0) = \bF(T_2)$.
\end{proof}

\section{Mean response time under identical replicas}
\label{App:DeterministicSlowdown}

In this appendix, we analyze performance of our proposed policy when the service time distribution follows a special case of the S\&X model.
We assume that the slowdown factor takes a deterministic value $S_i = s$ for all servers $i \in [N]$ and some finite $s \ge 1$ and the job service time $X_n$ is \iid exponential with rate $\mu$. 
Therefore, the service time of the $n$th job at all the servers at which it get accepted for processing will be identical and is a realization of the scaled exponential random variable  $sX_n$. 
For this service model, we will derive the mean response time of a job for the $\pi(d,T_1,T_2)$ policy under the assumption of asymptotic independence among the workloads at different queues.
\begin{remark}
We observe that the mean workload at the cavity queue under this model remains identical to the case when the job sizes are \iid exponential. 
Therefore, the loss probability for this model will remain identical to the case when the job sizes are \iid exponential, and is given by Lemma~\ref{lem:LossProb}.
\end{remark}

\begin{lemma}
The conditional mean response time of a job under the $\pi(d,T_1,T_2)$ policy and identical job replica size of mean $s/\mu$
is given by
$
\tau = \frac{1}{1-P_L} \Bigg[ \int_0^{T_2}\bF(w)^{d}  - \bF(T_1)\bF(T_2)^{d-1}dw 
+ \int_{T_2}^{T_1}(\bF(w) - \bF(T_1))\bF(T_2)^{d-1} dw \Bigg]+\frac{s}{\mu} 
$
where $P_L$ is the loss probability and $\bF(w)$ is the marginal complementary workload distribution at equilibrium.
\end{lemma}
\begin{proof}
Consider a job arriving at the set of $d$ randomly selected set of primary and secondary servers, $I_1$ and $I_2$.
Note that the job gets admitted only at a set of servers $I \subseteq I_1 \cup I_2$.
Suppose that the current workload at server $j$ is denoted by $W_j$ and the indicator of a job being undiscarded by $\xi = \SetIn{I \neq \emptyset}$ as defined previously.
Then, the response time of an undiscarded job is given by
$
R = Z + \xi sX,
$
where we define $Z \triangleq \xi(\min\set{W_j: j \in I}$ and the mean response time is
\EQN{
\label{eq:IndependentCondMRT}
\E[R] = \E[Z] +\frac{s(1-P_L)}{\mu}.
}
Next, we observe that
$
\SetIn{Z>z} = \xi\SetIn{\min\{W_j: j \in I\} > z} = \xi\left(\prod_{I_1 \cup I_2} (\xi_j\SetIn{W_j > z} + \bar{\xi}_j) \right) .
$
From the independence of the workloads and therefore of the indicators $\xi_j$ across queues and by substituting for $\xi$ from Equation~\ref{eqn:IndicatorJobDiscard}, we get
\eq{
\E[\SetIn{Z>z}] &= \E[\prod_{I_1 \cup I_2} (\xi_j\SetIn{W_j > z} + \bar{\xi}_j)] - \E[\prod_{I_1 \cup I_2}\bar{\xi}_j] = \prod_{I_1 \cup I_2}\E[ (\xi_j\SetIn{W_j > z} + \bar{\xi}_j)] - \prod_{I_1 \cup I_2}\E[\bar{\xi}_j] \\
& = (\bF(z)^{d}  - \bF(T_1)\bF(T_2)^{d-1})\SetIn{z \le T_2} + (\bF(z) - \bF(T_1))\bF(T_2)^{d-1}\SetIn{T_2 < z \le T_1}.
}
Since $\E[Z] = \int_0^\infty P\set{Z > z}dz$, we obtain 
$
\E[Z] = \int_0^{T_2}\bF(z)^{d}  - \bF(T_1)\bF(T_2)^{d-1}dz 
+ \int_{T_2}^{T_1}(\bF(z) - \bF(T_1))\bF(T_2)^{d-1} dz.
$
The result follows from Equation~\eqref{eq:IndependentCondMRT} and Equation~\eqref{eq:mrtDefinition}. 
\end{proof}
\begin{corollary}
For the special case of replication on idle secondary servers in $N$ server system with Poisson arrivals of rate $\lambda N$, the conditional mean response time under deterministic slowdown and exponential job sizes with rate $\mu$ is given by
$
\tau = \bF(0)^{d} + \int_{0^+}^{\infty}\bF(z)\bF(0)^{d-1}dz + \frac{s}{\mu}
$
where $
F(0) = \frac{(1 - \frac{\lambda}{\mu})(1 - \frac{\bl}{\mu})}{(1 - \frac{\lambda}{\mu}) + \frac{\bl}{\mu}(\frac{\lambda}{\mu} - \frac{\bl}{\mu})} 
$ and
$
\bF(z) = 1 - 
F(0)(1 - \frac{\bl}{\mu - \lambda} (1 - e^{-(\mu - \lambda)z}),\quad z > 0.
$
\end{corollary}
We provide a comparison of the mean response time for the $\pi(d,\infty,0)$ policy under deterministic slowdown and \iid exponential service times in Fig.~\ref{Fig:STVsAR-Ind-RISS}.
We observe that for low arrival rates, the performance is comparatively worse when the service times are identical but not independent.
However, the performance of both models converge for higher arrival rates as the chances of secondary replicas getting admitted at the servers diminish with increase in arrival rate.
\begin{figure}[h!]
\centerline{\scalebox{0.65}{
\begin{tikzpicture}

\definecolor{color0}{rgb}{0.83921568627451,0.152941176470588,0.156862745098039}
\definecolor{color1}{rgb}{0.12156862745098,0.466666666666667,0.705882352941177}
\definecolor{color2}{rgb}{0.172549019607843,0.627450980392157,0.172549019607843}
\definecolor{color3}{rgb}{1,0.498039215686275,0.0549019607843137}

\begin{semilogyaxis}[
legend cell align={left},
legend style={fill opacity=0.8, draw opacity=1, text opacity=1, at={(0.05,0.8)}, anchor=north west, draw=white!80!black},
tick align=outside,
tick pos=left,
x grid style={white!69.01960784313725!black,   dotted},
xlabel={Arrival rate $\lambda$},
xmajorgrids,
xmin=-0.0375, xmax=1.0075,
xminorgrids,
xtick style={color=black},
y grid style={white!69.01960784313725!black,   dotted},
ylabel={Conditional mean response time $\tau$},
ymajorgrids,
ymin=0.08, ymax=100,
yminorgrids,
ytick style={color=black}
]
\addplot [color0, mark=asterisk, mark size=1, mark options={solid}, line width=1.4cm, very thick]
table {%
0.01	1.00
0.06	1.003531757
0.11	1.0254375886
0.16	1.0765969667
0.21	1.1598058474
0.26	1.2740128164
0.31	1.4173449074
0.36	1.5886545464
0.41	1.7883172913
0.46	2.0188034919
0.51	2.2853718171
0.56	2.5972125458
0.61	2.9695352429
0.66	3.4275966667
0.71	4.0150028181
0.76	4.8124960556
0.81	5.9865126038
0.86	7.941902016
0.91	11.9838399467
0.96	25.9443012645
};

\addlegendentry{Det. Slowdown}
\addplot [color1, mark=+, mark size=2.4, mark options={solid}, line width=1.4cm, very thick]
table {%
0.01	0.2627717791
0.06	0.3357484905
0.11	0.4251382773
0.16	0.5304752497
0.21	0.6505330885
0.26	0.7845724458
0.31	0.9328314626
0.36	1.0967439936
0.41	1.2791505919
0.46	1.4846552862
0.51	1.7202736039
0.56	1.9966001794
0.61	2.3299442892
0.66	2.7464088116
0.71	3.2902383758
0.76	4.0426450174
0.81	5.1703964304
0.86	7.0785590272
0.91	11.0724280157
0.96	24.9840122765
};
\addlegendentry{$\emph{i.i.d.}$}


\end{semilogyaxis}

\end{tikzpicture}}}
\caption{\small{For the deterministic slowdown with slowdown factor $s =1$ and \iid service times under $\pi(d,\infty,0)$ policy with number of replicas $d=4$, service rate $\mu=1$ and number of servers $N = 20$, conditional mean response time $\tau$ as function of arrival rate $\lambda$.}}
\label{Fig:STVsAR-Ind-RISS}
\end{figure}

\end{document}